\documentclass[aps,prl,preprint,nopacs,superscriptaddress]{revtex4}
\usepackage{wrapfig}
\usepackage{makecell}
\usepackage{graphicx}
\usepackage{verbatim}
\usepackage{relsize}
\usepackage{mathrsfs}
\usepackage{color}
\usepackage{braket}
\usepackage{ulem}
\usepackage{gensymb}

\usepackage{amsmath,amsfonts,amssymb}
\usepackage{graphicx}

\definecolor{mygray}{gray}{0.6}
\usepackage{mathtools}
\usepackage{array}
\usepackage{graphicx}

\usepackage{bbding}
\usepackage{pifont}
\usepackage{wasysym}
\usepackage{amssymb}

\usepackage{amsmath,amsfonts,amssymb}

\def \beq {\begin{equation}}
\def \eeq {\end{equation}}
\usepackage{braket}
\usepackage{bbold}
\usepackage{multirow}

\renewcommand{\thetable}{\arabic{table}}

\begin{document}
%Discovery of the Axion induction by optical Axion electrodynamics\\Discovery of the Axion induction in a 2D antiferromagnetic topological insulator\\Observation of the Axion induction and the Axion circular dichroism\\Observation of the Axion induction by the Axion circular dichroism\\

\title{Axion optical induction of antiferromagnetic order}
%\title{Observation of the Axion optical induction in a 2D antiferromagnetic Axion insulator}

\author{Jian-Xiang Qiu}\affiliation{\footnotesize Department of Chemistry and Chemical Biology, Harvard University, MA, USA}
\author{Christian Tzschaschel}\affiliation{\footnotesize Department of Chemistry and Chemical Biology, Harvard University, MA, USA}
\author{Junyeong Ahn}\affiliation{\footnotesize Department of Physics, Harvard University, Cambridge, MA, USA}

\author{Anyuan Gao}\affiliation{\footnotesize Department of Chemistry and Chemical Biology, Harvard University, MA, USA}

\author{Houchen Li}\affiliation{\footnotesize Department of Chemistry and Chemical Biology, Harvard University, MA, USA}

\author{Xin-Yue Zhang}\affiliation{\footnotesize Department of Physics, Boston College, Chestnut Hill, MA, USA}

\author{Barun Ghosh}\affiliation{\footnotesize Department of Physics, Northeastern University, Boston, MA, USA}

\author{Chaowei Hu}\affiliation{\footnotesize Department of Physics and Astronomy and California NanoSystems Institute, University of California, Los Angeles, Los Angeles, CA, USA}

\author{Yu-Xuan Wang}\affiliation{\footnotesize Department of Physics, Boston College, Chestnut Hill, MA, USA}

\author{Yu-Fei Liu}\affiliation{\footnotesize Department of Chemistry and Chemical Biology, Harvard University, MA, USA}\affiliation{\footnotesize Department of Physics, Harvard University, Cambridge, MA, USA}

\author{Damien B\'erub\'e}\affiliation{\footnotesize Department of Chemistry and Chemical Biology, Harvard University, MA, USA}
\author{Thao Dinh}\affiliation{\footnotesize Department of Physics, Harvard University, Cambridge, MA, USA}\affiliation{\footnotesize Department of Chemistry and Chemical Biology, Harvard University, MA, USA}
\author{Zhenhao Gong}\affiliation{\footnotesize Shenzhen Institute for Quantum Science and Engineering and Department of Physics, Southern University of Science and Technology (SUSTech), Shenzhen, China}
\affiliation{\footnotesize Quantum Science Center of Guangdong-Hong Kong-Macao Greater Bay Area (Guangdong), Shenzhen, China}
\affiliation{\footnotesize Shenzhen Key Laboratory of Quantum Science and Engineering, Shenzhen, China}
\affiliation{\footnotesize International Quantum Academy, Shenzhen, China}

\author{Shang-Wei Lien}\affiliation{\footnotesize Department of Physics, National Cheng Kung University, Tainan, Taiwan}\affiliation{\footnotesize Center for Quantum Frontiers of Research and Technology (QFort), Tainan, Taiwan}\affiliation{\footnotesize  Physics Division, National Center for Theoretical Sciences, Taipei, Taiwan}
\author{Sheng-Chin Ho}\affiliation{\footnotesize Department of Chemistry and Chemical Biology, Harvard University, MA, USA}
\author{Bahadur Singh}\affiliation{\footnotesize Department of Condensed Matter Physics and Materials Science, Tata Institute of Fundamental Research, Colaba, Mumbai, India}

\author{Kenji Watanabe}\affiliation{\footnotesize Research Center for Functional Materials, National Institute for Materials Science, 1-1 Namiki, Tsukuba, Japan}

\author{Takashi Taniguchi}\affiliation{\footnotesize International Center for Materials Nanoarchitectonics, National Institute for Materials Science,  1-1 Namiki, Tsukuba, Japan}

\author{David C. Bell}\affiliation{\footnotesize Harvard John A. Paulson School of Engineering and Applied Sciences, Harvard University, Cambridge, MA, USA}\affiliation{\footnotesize Center for Nanoscale Systems, Harvard University, Cambridge, MA, USA}

\author{Hai-Zhou Lu}\affiliation{\footnotesize Shenzhen Institute for Quantum Science and Engineering and Department of Physics, Southern University of Science and Technology (SUSTech), Shenzhen, China}
\affiliation{\footnotesize Quantum Science Center of Guangdong-Hong Kong-Macao Greater Bay Area (Guangdong), Shenzhen, China}
\affiliation{\footnotesize Shenzhen Key Laboratory of Quantum Science and Engineering, Shenzhen, China}
\affiliation{\footnotesize International Quantum Academy, Shenzhen, China}

\author{Arun Bansil}
\affiliation{\footnotesize Department of Physics, Northeastern University, Boston, MA, USA}
\author{Hsin Lin}
\affiliation{\footnotesize Institute of Physics, Academia Sinica, Taipei, Taiwan}

\author{Tay-Rong Chang}\affiliation{\footnotesize Department of Physics, National Cheng Kung University, Tainan, Taiwan}\affiliation{\footnotesize Center for Quantum Frontiers of Research and Technology (QFort), Tainan, Taiwan}\affiliation{\footnotesize  Physics Division, National Center for Theoretical Sciences, Taipei, Taiwan}

\author{Brian B. Zhou}\affiliation{\footnotesize Department of Physics, Boston College, Chestnut Hill, MA, USA}

\author{Qiong Ma}\affiliation{\footnotesize Department of Physics, Boston College, Chestnut Hill, MA, USA}\affiliation{\footnotesize Canadian Institute for Advanced Research, Toronto, Canada}

\author{Ashvin Vishwanath}\affiliation{\footnotesize Department of Physics, Harvard University, Cambridge, MA, USA}

\author{Ni Ni$^{*}$}\affiliation{\footnotesize Department of Physics and Astronomy and California NanoSystems Institute, University of California, Los Angeles, Los Angeles, CA, USA}

\author{Su-Yang Xu\footnote{Corresponding authors (emails): suyangxu@fas.harvard.edu and nini@physics.ucla.edu }}\affiliation{\footnotesize Department of Chemistry and Chemical Biology, Harvard University, MA, USA}

\date{\today}

\maketitle

\textbf{Using circularly-polarized light to control quantum matter is a highly intriguing topic in physics, chemistry and biology. Previous studies have demonstrated helicity-dependent optical control of spatial chirality and magnetization $M$. The former is central for asymmetric synthesis in chemistry and homochirality in bio-molecules, while the latter is of great interest for ferromagnetic spintronics. In this paper, we report the surprising observation of helicity-dependent optical control of fully-compensated antiferromagnetic (AFM) order in 2D even-layered MnBi$_2$Te$_4$, a topological Axion insulator with neither chirality nor $M$. We further demonstrate helicity-dependent optical creation of AFM domain walls by double induction beams and the direct reversal of AFM domains by ultrafast pulses. The control and reversal of AFM domains and domain walls by light helicity have never been achieved in any fully-compensated AFM. To understand this optical control, we study a novel type of circular dichroism (CD) proportional to the AFM order, which only appears in reflection but is absent in transmission. We show that the optical control and CD both arise from the optical Axion electrodynamics, which can be visualized as a Berry curvature real space dipole. Our Axion induction provides the possibility to optically control a family of $\mathcal{PT}$-symmetric AFMs such as Cr$_2$O$_3$, CrI$_3$ and possibly novel states in cuprates. In MnBi$_2$Te$_4$, this further opens the door for optical writing of dissipationless circuit formed by topological edge states.}

%\cite{huck1996dynamic,de2017quantized,xu2020spontaneous,barron2020false}By shining circularly-polarized light while cooling across the N{\'e}el temperature, we found that light helicity and wavelength can directly control the fully-compensated AFM order parameter. 

\vspace{0.5cm}
\textbf{Main}

%A primary example is the helicity-dependent optical control of chiral matter. By shining circularly-polarized light, researchers managed to control the formation of chiral molecules  \cite{huck1996dynamic} and the development of gyrotropic electronic domains \cite{xu2020spontaneous}, which are of great interest for asymmetrical synthesis in chemistry and novel charge orders in condensed matter physics. Another example is the helicity-dependent optical control of ferromagnets (FM). By shining circularly-polarized light, researchers managed to control the magnetization ($M$) domains \cite{kirilyuk2010ultrafast}, which has important implications in spintronics. 

%The concept of using light helicity to control materials is both intellectually stimulating and also practically useful.  Therefore,Optical control is non-contact, flexible, has good spatial resolution and further allows for ultrafast manipulation. It also enables fundamental understanding of the interaction of photons with quantum geometry, charges, spins, and lattice.
There is tremendous interest in finding innovative ways to control and manipulate complex quantum materials \cite{basov2017towards}. Antiferromagnets (AFMs) have zero net $M$, so AFM domains are immune to perturbing magnetic field. This leads to the prospect of robust magnetic storage \cite{jungwirth2016antiferromagnetic,nvemec2018antiferromagnetic}. However, this robustness also means that manipulating fully-compensated AFM order is extremely difficult \cite{kirilyuk2010ultrafast, nvemec2018antiferromagnetic, manz2016reversible, higuchi2016control} (discussion in SI.V.2). As such, controlling AFM order has been recognized a key challenge toward the AFM spintronics \cite{jungwirth2016antiferromagnetic,nvemec2018antiferromagnetic}. One known approach is to use the parallel $E$ and $B$ fields \cite{fiebig2005revival, jiang2018electric, gao2021layer}. Compared to such electrical approach, optical control is non-contact, flexible, has good spatial resolution and further allows for ultrafast manipulation. It also enables fundamental understanding of the interaction of photons with charges, spins, lattice, and quantum geometry. 

In this paper, we explore the novel possibility of controlling fully-compensated AFM order by circularly-polarized light, which has never been achieved. We got inspirations from (1) discoveries of helicity-dependent optical control of chiral materials and magnetization $M$ \cite{kirilyuk2010ultrafast, huck1996dynamic,xu2020spontaneous} and (2) previous experiments reporting novel circular dichroism (CD) proportional to the AFM order in Cr$_2$O$_3$ \cite{krichevtsov1993spontaneous, krichevtsov1996magnetoelectric} and the pseudo-gap state of cuprates \cite{xia2008polar}. We report helicity-dependent optical control of fully-compensated AFM order induced by the optical Axion electrodynamics in even-layered MnBi$_2$Te$_4$. \color{black}

%Here, we discover that circularly-polarized light can directly control the fully-compensated antiferromagnetic (AFM) domains in 2D even-layered MnBi$_2$Te$_4$. Moreover, by shining two close-by light beams with opposite helicity, we optically create an AFM domain wall. Furthermore, using ultrafast pulses with circular polarization, we achieve direct reversal of AFM domain. We show that the underlying microscopic origin is the optical Axion electrodynamics \cite{ahn2022theory}. Given the simple experimental procedure and the fact that the optical Axion electrodynamics can be applied to systems with proper symmetry, our discovery can be applied to a wide range of $\mathcal{PT}$-symmetric AFMs such as Cr$_2$O$_3$ and CrI$_3$ and even potentially the pseudo-gap state of cuprates \cite{xia2008polar, orenstein2011optical,varma2014gyrotropic}.

\begin{figure*}[t]
\includegraphics[width=14cm]{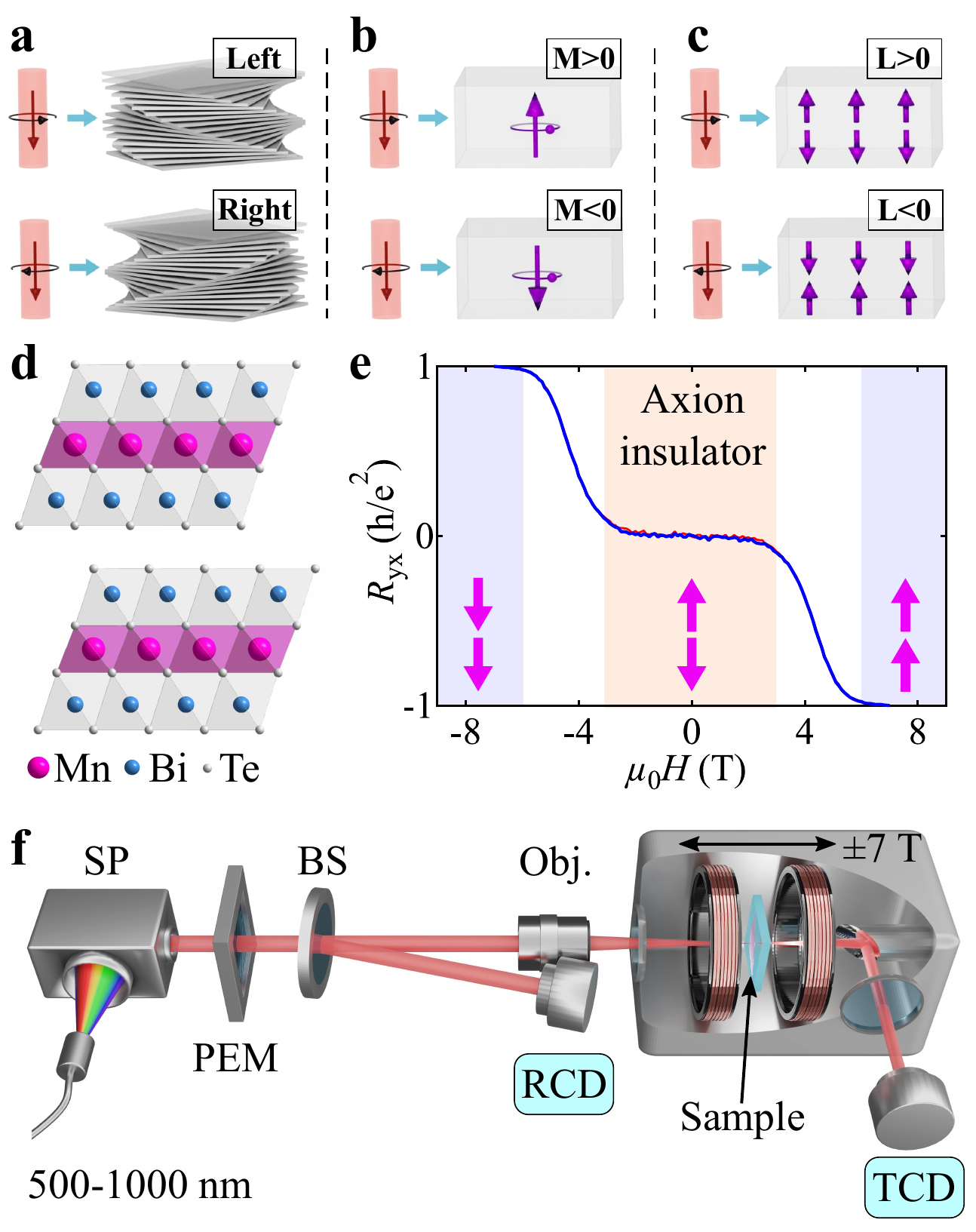}
\caption{ {\bf Helicity-dependent optical control of quantum matter and introduction to MnBi$_2$Te$_4$.} \textbf{a,b,} Previous studies have demonstrated helicity-dependent optical control of spatial chirality and magnetization \cite{huck1996dynamic,xu2020spontaneous,kirilyuk2010ultrafast}. \textbf{c,} We report the surprising observation of helicity-dependent optical control of fully-compensated antiferromagnetic (AFM) order. \color{black} \textbf{d,} Lattice structure of MnBi$_2$Te$_4$. \textbf{e,} Hall resistivity of our 6SL MnBi$_2$Te$_4$  with the layered magnetic state shown by the pink arrows. For each state, two out of six layers are pictured for simplicity. The Axion insulator state is realized by the fully-compensated AFM near $B=0$.  \textbf{f,} Our circular dichroism set up. SP, PEM, BS, Obj. are spectrometer, photoelastic modulator, beam-splitter, and objective, respectively. }
\label{Fig1}
\end{figure*}

MnBi$_2$Te$_4$, the first intrinsic magnetic topological insulator recently synthesized in 2019, has attracted great interest as it bridges three primary fields in quantum condensed matter: topology, magnetism and 2D van der Waals (vdW) materials \cite{Otrokov2019a, Otrokov2019unique,Li2019a,Zhang2019a,Zhang2020, Liu2020, Deng2020, Liu2020a, deng2021high, yang2021odd, Ovchinnikov2020,gao2021layer,cai2021electric, li2021nonlocal,tai2021polarity}. MnBi$_2$Te$_4$'s lattice consists of septuple layers (SL) separated by vdW gaps. Its magnetic ground state is layered AFM, which can be further tuned into a ferromagnetic state by a large $B$ field (Fig.~\ref{Fig1}\textbf{e}).  Previous theoretical works have comprehensively studied the electronic, magnetic and topological properties of MnBi$_2$Te$_4$ bulk and thin films (SI.V.3) \cite{Otrokov2019a, Otrokov2019unique, Li2019a,Zhang2019a,Zhang2020, Liu2020}. \color{black}The 2D magnetic and topological ground states can be classified into two kinds. The first kind has an obvious, nonzero static $M$ hosting the Chern insulator state \cite{Deng2020, Liu2020a, deng2021high, yang2021odd, Ovchinnikov2020,gao2021layer,cai2021electric, li2021nonlocal}. It includes odd-layered MnBi$_2$Te$_4$ near $B=0$ as well as odd-layered and even-layered MnBi$_2$Te$_4$ under large $B$ fields. By contrast, the second, more special kind is the even-layered AFM MnBi$_2$Te$_4$, which will be our focus. It is expected to host fully-compensated AFM with an Axion insulator state \cite{Otrokov2019unique,Li2019a, Liu2020, Deng2020, Liu2020a} near $B=0$.
%These FM or FM-like systems host the Chern insulator state \cite{Deng2020, Liu2020a, deng2021high, yang2021odd, Ovchinnikov2020,gao2021layer,cai2021electric, li2021nonlocal} (see Fig.~\ref{Fig1}\textbf{e}). In this case, there is an equal number of spin-up and spin-down layers. 

Our magneto-optical setup (Fig.~\ref{Fig1}\textbf{f}) allows us to investigate the interaction between circularly-polarized light and quantum materials by probing CD, the difference between $\sigma^+$ and $\sigma^-$ light. Importantly, our setup can measure CD both in the reflection and transmission channels and has a supercontinuum light source with tunable wavelength ($500$ nm to $1000$ nm). These capabilities are crucial for our findings, including unique helicity-dependence, wavelength-dependence, and reflection and transmission properties. The measurement temperature is $2$ K unless noted otherwise. 

%, where Mn spins within each SL are ferromagnetically aligned out-of-plane but Mn spins between adjacent SLs are anti-parallel

\vspace{0.5cm}
\textbf{Optical induction in a 2D topological antiferromagnet}

\begin{figure*}[t]
\includegraphics[width=12.5cm]{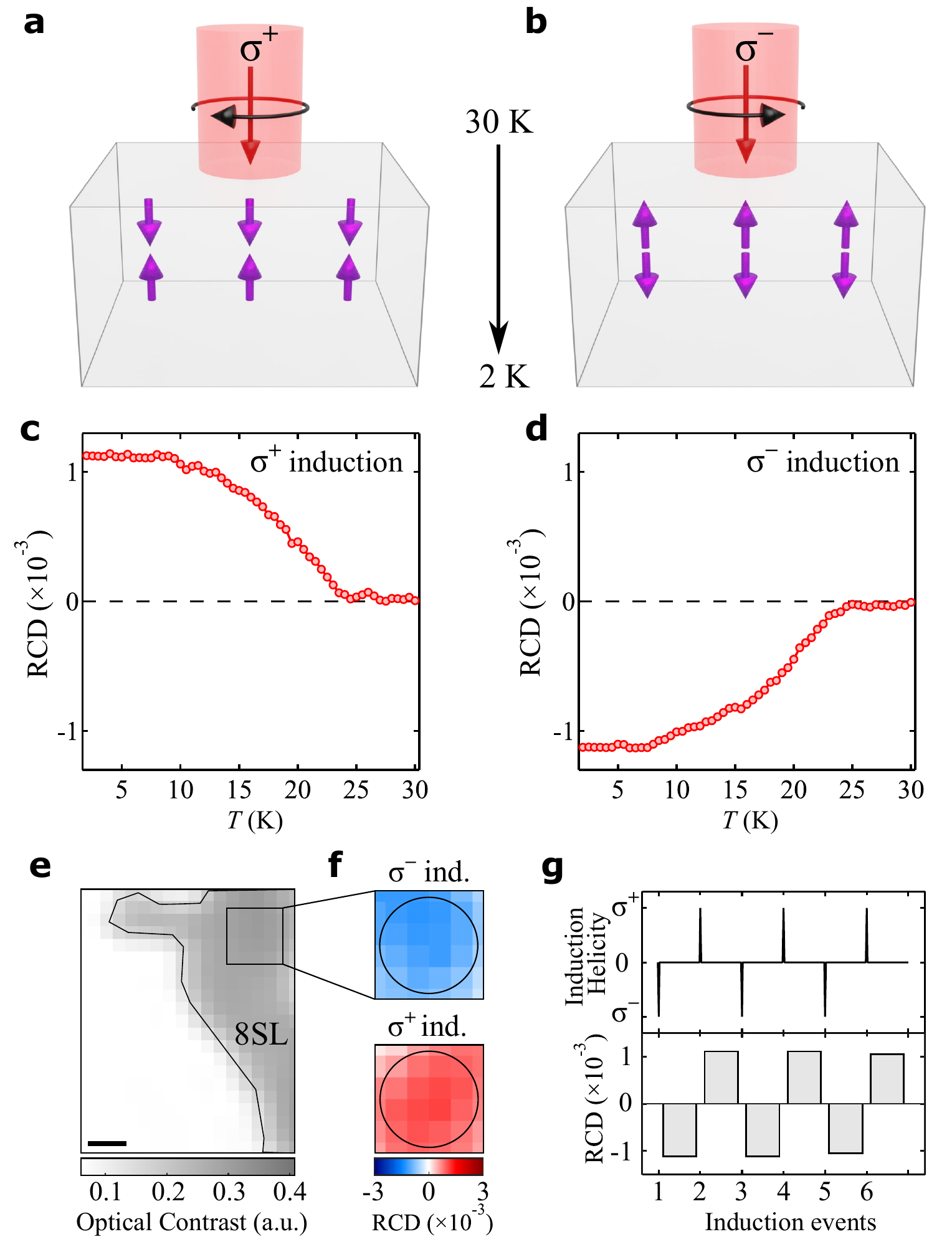}
\vspace{-5mm}
\caption{{\bf Optical induction in a 2D topological antiferromagnet.} \textbf{a,} We shine $\sigma^+$ circularly-polarized induction light ($\lambda_{\textrm{induction}}=840$ nm, $P_{\textrm{induction}}\simeq1$ mW) on the sample S1 (8SL MnBi$_2$Te$_4$ flake on a sapphire substrate) while lowering its temperature from $T=30$ K to $2$ K. \textbf{c,} Upon reaching $2$ K, we turn off the induction light and measure the reflection CD (RCD) with the detection light at $\lambda_{\textrm{detection}}=946$ nm and $P_{\textrm{detection}}\simeq30$ $\mu$W. Surprisingly, we observe a significant RCD at $B=0$. Upon warming up, the RCD vanishes above $T_{\textrm{N}}$. \textbf{b,d,} Same as panels (\textbf{a,c}) except that we shine $\sigma^-$ circularly-polarized induction light on the sample while lowering its temperature from $T=30$ K to $2$ K. \textbf{e,} Spatial mapping of the optical contrast near the 8SL flake. Scale bar: $2$ $\mu$m. \textbf{f,} RCD signal after induction with opposite helicity. The circle marks the spot subject to the induction light while cooling. \textbf{g,} RCD signal at the center of the circle in panel (\textbf{f}) after consecutive induction processes with opposite helicity.}
\label{Fig2}
\end{figure*}

In this section, we show the observation of optical induction at specific wavelengths. Systematic wavelength dependences will be presented later. As shown in Fig.~\ref{Fig2}\textbf{a}, starting from $T=30$ K, we shine $\sigma^{+}$ circularly-polarized light ($\lambda_{\textrm{induction}}=840$ nm, $P_{\textrm{induction}}\simeq1$ mW) onto a spot on a 8SL MnBi$_2$Te$_4$ flake (sample-S1, see Fig.~\ref{Fig2}\textbf{e}) while lowering its temperature. Upon reaching $2$ K, we turn off the induction light and measure the reflection CD (RCD) with the detection light ($\lambda_{\textrm{detection}}=946$ nm and $P_{\textrm{detection}}\simeq30$ $\mu$W). We observe significant anomalous RCD at $B=0$ (Fig.~\ref{Fig2}\textbf{c}). We then measure the anomalous RCD while warming up. The RCD vanishes above $T_{\textrm{N}}$. From $T=30$ K, we repeat the same induction process (Fig.~\ref{Fig2}\textbf{b}) only changing the induction helicity to $\sigma^{-}$. We turn off the induction light at $2$ K and repeat the measurements. Remarkably, the global sign of RCD data is reversed (Fig.~\ref{Fig2}\textbf{d}). We repeated the induction nine consecutive times (Figs.~\ref{Fig2}\textbf{e-g} and Extended Data Fig.~\ref{Induction_temporal_reproducibility}). We find that the RCD at $2$ K is consistently controlled by the induction light helicity. On the other hand, when cooling down without induction light, we still observe the anomalous RCD. Only the sign is random (SI.Fig. S17).

\vspace{0.5cm}
\textbf{Understanding the optical induction by investigating the anomalous CD}

Because the anomalous RCD correlates with the AFM order, the data above hint an exciting possibility that induction helicity can control the AFM order in 8SL MnBi$_2$Te$_4$. To understand this optical induction, we first investigate the anomalous CD, because it serves as the experimental indicator of the AFM order. Here, we focus on sample-S2 on a diamond substrate, which consists of four connected flakes of $5-8$ SLs (Extended Data Fig.~\ref{Fig3}\textbf{a}). We performed systematic RCD measurements (Extended Data Figs.~\ref{Fig3}\textbf{c-p}). In 5SL and 7SL, we observed the conventional magnetic CD proportional to $M$. In 6SL and 8SL, we observed the anomalous RCD. We have further confirmed the reproducibility of the anomalous CD in more than 10 samples. 

%Interestingly, the anomalous CD in even-layered flakes is quite strong: the amplitude of the anomalous CD in 6SL and that of the magnetic CD in 7SL are similar. 
 
There are two possibilities for the anomalous RCD. (1) It can be the magnetic CD proportional to $M$ (we will explain the origin of $M$ below); (2) Actually, in the absence of any $M$, there can be an AFM CD unrelated to $M$ but proportional to the AFM order $L$ in $\mathcal{PT}$ symmetric AFMs as reported in Cr$_2$O$_3$ \cite{krichevtsov1993spontaneous, krichevtsov1996magnetoelectric} (this is symmetry allowed see SI.III), which is also likely the origin of the CD observed in the pseudogap of cuprates \cite{xia2008polar, orenstein2011optical,varma2014gyrotropic}. One may think that (1) and (2) can be easily discerned because they are proportional to different order parameters. But in reality this is often not feasible, because the $M$ in an AFM is typically coupled with the AFM order $L$. For example, suppose our sample is subject to a fixed vertical, static electric field $E_z$ due to substrate, which in turn generates an $M$ due to the static ME coupling $\alpha$, i.e., $M=\alpha E_z$. Because the two AFM states have opposite $\alpha$, if one flips the AFM order $L$, $M$ will also flip. In fact, the induced $M$ turned out to be the dominant mechanism for RCD in even-layered CrI$_3$ \cite{jiang2018electric, huang2018electrical}. Therefore, new measurements beyond the RCD are crucial to distinguish the above two possibilities.
%nonzero RCD alone is insufficient to discern the two possibilities. N
%: When one carefully thinks about possible $M$ in a nominally fully-compensated AFM, one finds that such
\color{black}

We now proceed to show that CD in transmission, i.e., TCD, provides the decisive new measurement, as proposed in Ref. \cite{canright1992ellipsometry, ahn2022theory}. Magnetic CD is known to also occur in transmission channel (just like the Faraday effect). By contrast, the AFM CD has $\textrm{TCD} =0$ because of $\mathcal{PT}$ symmetry. Extended Data Fig.~\ref{Symmetry}\textbf{a} describes a conceptual experiment with $\sigma^-$ light transmitting through sample. Upon $\mathcal{PT}$ inversion, the even-layered MnBi$_2$Te$_4$ remains invariant and light path also stays the same, but light helicity is reversed. As such, $\mathcal{PT}$ enforces the transmission coefficients for $\sigma^{\pm}$ light to be identical, which means $\textrm{TCD} =0$ (similar analysis can show that RCD is allowed, Extended Data Fig.~\ref{Symmetry}\textbf{b}). Therefore, what truly distinguishes the AFM CD from the magnetic CD is $\mathrm{TCD=0}$.

\begin{figure*}[t]
\includegraphics[width=14cm]{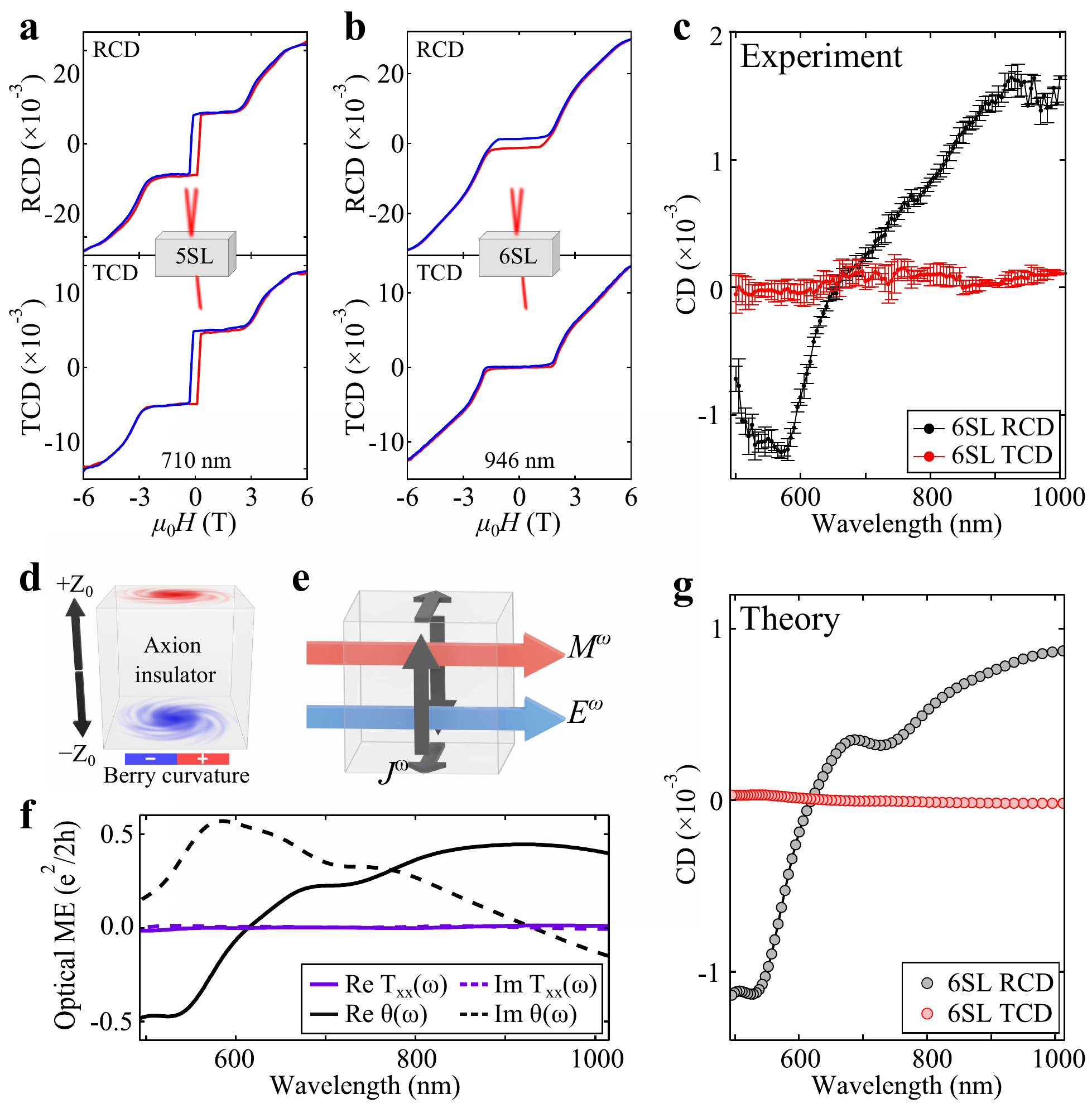}
\vspace{-6mm}
\caption{{\bf Unique reflection and transmission properties and observation of the Axion CD.} \textbf{a,b,} Simultaneous RCD and TCD measurements of the 5SL (panel \textbf{a}) and 6SL (panel \textbf{b}) in sample-S3. \textbf{c,} Wavelength dependence of RCD and TCD at $B=0$ for 6SL. \textbf{d,} Schematic illustration of the Berry curvature real-space dipole. Electrons at opposite positions ($\pm Z$) in even-layered MnBi$_2$Te$_4$ have opposite Berry curvature. \textbf{e,} the Axion optical ME coupling can be visualized an itinerant electron circulation in response to electric field as a consequence of the Berry curvature real-space dipole. \textbf{f,} Optical Axion $\theta(\omega)$ and gyrotropic birefringence ($T_{xx}$) of 6SL MnBi$_2$Te$_4$ calculated from first-principles band structures. \textbf{g,} Theoretically computed RCD and TCD, which are obtained directly from the calculated $\alpha({\omega})$ (see expressions in SI.IV.1). The TCD is strictly zero with perfect $\mathcal{PT}$ symmetry. The calculated TCD here takes account into the weak $\mathcal{PT}$ breaking due to asymmetric dielectric environment (hBN and diamond, see details in SI.IV.1.(5)).}
\label{Fig4}
\end{figure*}

As such, we study TCD and RCD simultaneously in sample-S3, which consists of 5SL and 6SL on diamond (Fig.~\ref{Fig4}). In 5SL, the magnetic CD indeed shows up prominently in both reflection and transmission. We now turn to 6SL. At $946$ nm where significant anomalous RCD was observed at $B=0$ (Fig.~\ref{Fig4}\textbf{b}), the TCD, by contrast, is zero. Continuous wavelength dependence (Fig.~\ref{Fig4}\textbf{c}) shows that, strikingly, TCD is negligibly small over the entire spectrum. Also, we have repeated the RCD $vs.$ TCD experiments in sample-S1 (Extended Data Fig.~\ref{8SL_sapphire}), on which the induction experiments were performed. In SI.II.1, we show additional data to further substantiate this. Therefore, we showed that the anomalous RCD in even-layered MnBi$_2$Te$_4$ only appears in reflection but is absent in transmission. Such unique reflection and transmission characters, although has been long proposed in theory \cite{canright1992ellipsometry}, have never been observed before, allowing us to rule out the magnetic CD due to uncompensated $M$. As such, the anomalous RCD in even-layered MnBi$_2$Te$_4$ is the AFM CD. Below, we show that the AFM CD can be further categorized by the microscopic mechanisms and our results provide the first demonstration of the optical Axion mechanism. %\color{red} Apart from AFMs, optical nonreciprocity with nominally zero magnetization was also observed in the superconducting states of UPt$_3$ and UTe$_2$ \cite{schemm2014observation,hayes2021multicomponent}. \color {black}
%In particular, (1) sample-S1 has an 8SL flake, and 8SL also shows nonzero RCD but zero TCD; (2) sample-S1 is on a sapphire substrate, so we proved that our observations are not due to a particular substrate. Lastly, to minimize the vertical $E$ field due to top and bottom dielectric asymmetry, we studied a hexagonal boron nitride (hBN)-encapsulated 6SL flake (sample S4, SI.XXX), and find the same consistent results.
%}. The GB leads to a nonreciprocal linear birefringence \cite{

\vspace{0.5cm}
\textbf{CD arising from the optical Axion electrodynamics}

 The AFM CD arises from the diagonal optical ME coupling $\alpha(\omega)_{ii}$ \cite{krichevtsov1993spontaneous, canright1992ellipsometry}, but the optical ME coupling has different components corresponding to different microscopic mechanisms. Specifically, the traceless part of $\alpha(\omega)_{ii}$ is known as the gyrotropic birefringence (GB) [$\textrm{GB}=\frac{1}{3}(\alpha(\omega)_{xx}-\alpha(\omega)_{zz})$] \cite{malashevich2010band, graham1997macroscopic}; while the trace part is the Axion contribution $\theta(\omega)=\frac{1}{3}\sum_{i}\alpha(\omega)_{ii}=\frac{1}{3}[2\alpha(\omega)_{xx}+\alpha(\omega)_{zz}]$ (we have applied $\alpha(\omega)_{xx}=\alpha(\omega)_{yy}$ because of MnBi$_2$Te$_4$'s $C_{3z}$ symmetry). However, for a long time, only the GB (traceless part) was theoretically derived \cite{malashevich2010band}. Only very recently, the theory of Axion electrodynamics at optical frequencies was developed \cite{ahn2022theory}, which allows us to compute the Axion optical ME coupling in quasi-2D periodic systems (see Methods for expressions). Importantly, in MnBi$_2$Te$_4$, because its bulk respects inversion symmetry, the GB contribution is expected to be negligible, whereas the Axion contribution dominates. As such, MnBi$_2$Te$_4$ is an ideal system to isolate the Axion optical ME contribution. Figure~\ref{Fig4}\textbf{f} shows the calculated GB and $\theta(\omega)$ of 6SL MnBi$_2$Te$_4$, from which we indeed see that Axion $\theta(\omega)$ strongly dominates. Interestingly, as shown in Fig.~\ref{Fig4}\textbf{e}, the physics of Axion ME coupling can be visualized by a Berry curvature real space dipole (see derivation in the Methods): Because the top and bottom surfaces have opposite Berry curvature, by applying $E$ field, they feature opposite Hall currents. In fact, if one considers the Hall currents on all four facets parallel to $E$ field, one naturally obtains a circulating current, which leads to an $M$. This physical picture works for both static and optical Axion ME effects. We only need the following correspondence: the static $E$ $\leftrightarrow$ optical $E^{\omega}$ and Berry curvature $\leftrightarrow$ inter-band Berry curvature. We note that, in contrast to the static limit, for our photon energy ($500-1000$ nm), the optical transition involves many bands, not just the topological surface states; and the contributions from the higher bands are more significant (SI.IV.2).

Using the calculated $\theta(\omega)$, we can theoretically compute RCD (see expressions in SI.IV.1), and thus compare it with the experimental RCD data. We note that the reflection from a surface with a nonlocal ME contribution is a difficult problem with extensive previous discussions \cite{agranovich1973phenomenological, Halperin_Book, hosur2015kerr}, and we have carefully considered this (see SI.IV.3). \color{black}As shown in Figs.~\ref{Fig4}\textbf{c,g}, we observe good agreement between experimental data and theoretical calculation in terms of the magnitude and the spectral shape. Therefore, by comparing data with calculations, we demonstrated the Axion CD in even-layered MnBi$_2$Te$_4$, i.e., AFM CD arising from the optical Axion electrodynamics.

\vspace{0.5cm}
\textbf{Optical induction arising from the optical Axion electrodynamics}

\begin{figure*}[t]
\includegraphics[width=11.5cm]{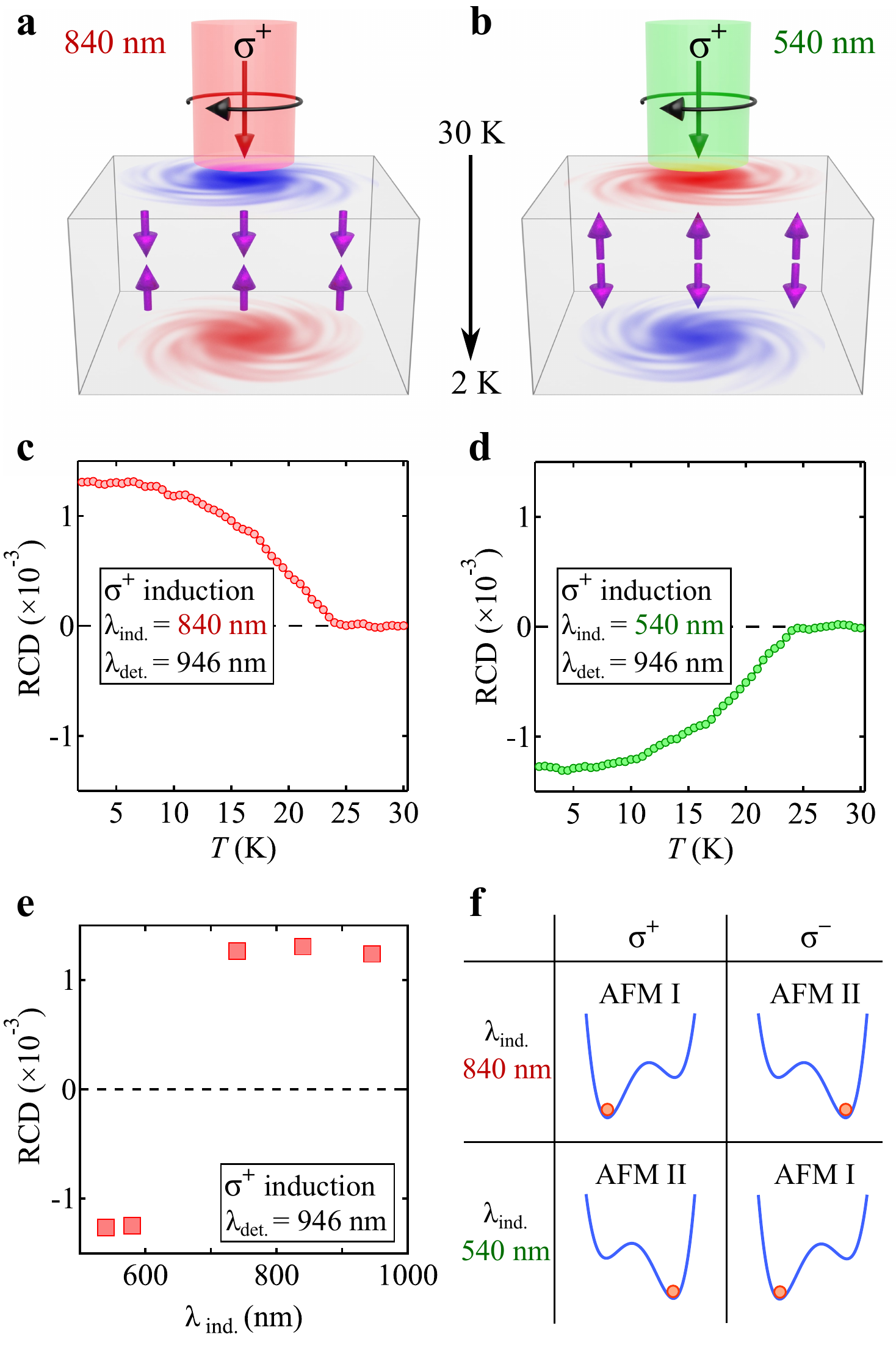}
\caption{{\bf Observation of the Axion induction.} \textbf{a,c,} We shine $\sigma^+$ circularly-polarized induction light ($\lambda_{\textrm{induction}}=540$ nm, $P_{\textrm{induction}}\simeq1$ mW) on the 8SL MnBi$_2$Te$_4$ flake (sample-S1) while lowering its temperature from $T=30$ K to $2$ K (panel \textbf{a}). We turn off the induction light and measure the RCD with $\lambda_{\textrm{detection}}=946$ nm while warming up (panel \textbf{c}). \textbf{b,d,} Same as panels (\textbf{a,c}) except the induction wavelength is $\lambda_{\textrm{induction}}=840$ nm. \textbf{e,} Induction wavelength dependence. To consistently compare how induction wavelength $\lambda_{\textrm{induction}}$ influences the results of induction, we fix all other experimental parameters including induction helicity (fixed at $\sigma^+$) and detection wavelength (fixed at $\lambda_{\textrm{detection}}=946$ nm) and we only vary $\lambda_{\textrm{induction}}$. \textbf{f,} Free energy diagrams summarize the optical control of the AFM order in even-layered MnBi$_2$Te$_4$ with light helicity and wavelength.}
\label{Fig5}
\end{figure*}

%We demonstrated that net $M$ and the resulting magnetic CD are negligibly small in even-layered MnBi$_2$Te$_4$. So
Our simultaneous RCD and TCD measurements demonstrated that the $M$ in even-layered MnBi$_2$Te$_4$ is negligibly small. Instead, circularly-polarized light with opposite helicity couples differently to the opposite AFM domains. To further confirm that this is also the origin of the optical induction, we now investigate its wavelength dependence. In particular, we notice that the RCD data has distinct spectral dependence (Fig.~\ref{Fig4}\textbf{c}): E.g. RCD at $840$ nm and $540$ nm have opposite signs. Therefore, if the induction has the same physical origin as the CD, i.e., the optical Axion electrodynamics, then the induction effects using $\lambda_{\textrm{induction}}=840$ nm and $\lambda_{\textrm{induction}}=540$ nm should be opposite. Specifically, with the same light helicity, the induction using $\lambda_{\textrm{induction}}=840$ nm and $\lambda_{\textrm{induction}}=540$ nm should lead to opposite AFM domains. As such, we carry out the induction with $\lambda_{\textrm{induction}}=540$ nm and 840 nm (note that $\lambda_{\textrm{detection}}$ is fixed to achieve consistent comparison). By directly comparing Fig.~\ref{Fig5}\textbf{a,c} ($\lambda_{\textrm{induction}}=540$ nm) and Fig.~\ref{Fig5}\textbf{b,d} ($\lambda_{\textrm{induction}}=840$ nm), we indeed find that the results are entirely opposite (see free energy analysis in Fig.~\ref{Fig5}\textbf{f}). We further study the induction at other wavelengths. As shown in Fig.~\ref{Fig5}\textbf{e}, our data show that the effect of induction at $540$ nm and $580$ nm is opposite to that of $740$ nm, $840$ nm and $946$ nm. These results are consistent with the sign of the RCD spectra for even-layered MnBi$_2$Te$_4$, which provide strong evidence that the induction and CD share the same physical origin, i.e., the optical Axion electrodynamics. Therefore, we conclude on the observation of the Axion induction, i.e., helicity-dependent control of fully-compensated AFM order based on the optical Axion electrodynamics. 

%We summarize the key evidence in even-layered MnBi$_2$Te$_4$: (1) the exclusion of $M$ and magnetic CD, (2) the reproducible induction results proving the induction as an intrinsic effect, (3) the coupling with circularly-polarized light is dominated by the optical Axion electrodynamics, and (4) the wavelength dependences of CD and induction showing that they share the same origin. Taking these collectively,

\vspace{0.5cm}
  
\textbf{Optical creation of AFM domain wall by double induction}

\begin{figure*}[t]
\includegraphics[width=12.5cm]{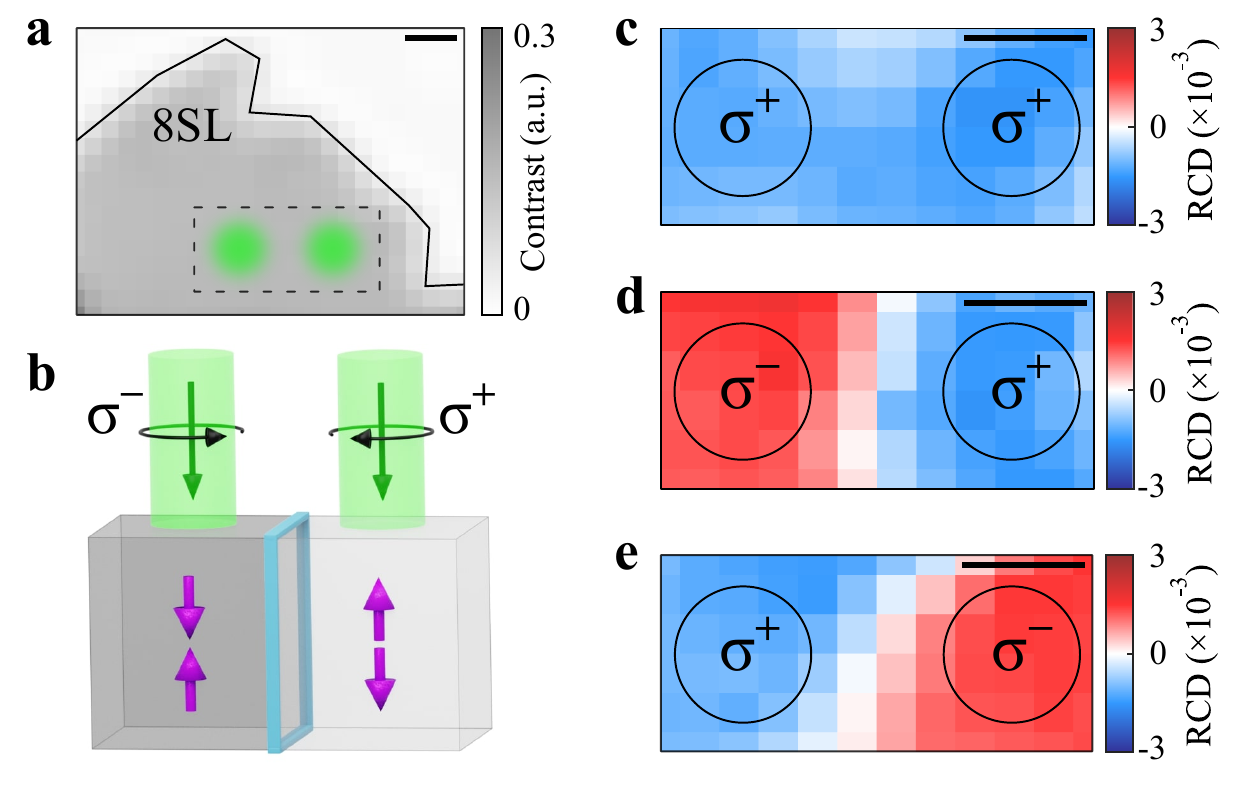}
\caption{ {\bf Optical creation of AFM domain wall by double Axion induction.} \textbf{a,} We shine two close-by circularly-polarized induction light beams on an 8SL MnBi$_2$Te$_4$ flake (sample-S5). Scale bar: 2 $\mu$m. $\lambda_{\textrm{induction}}=540$ nm. \textbf{b,} Schematic illustration of the double induction leading to an AFM domain wall.  \textbf{c-e,} RCD mappings of the area subject to the double induction with ($\sigma^+, \sigma^+$), ($\sigma^-, \sigma^+$), and ($\sigma^+, \sigma^-$), respectively. \color{black}}
\label{Fig6}
\end{figure*}

The control of AFM order with light helicity makes it possible to spatially modulate the AFM domain structure. For instance, one can think of creating AFM domain wall using two close-by light beams of opposite helicity. Here, we demonstrate this possibility in a 8SL flake (sample-S5). As shown in Fig.~\ref{Fig6}\textbf{a}, the two light beams are spatially separated and their polarizations can be controlled separately. When both beams are $\sigma^{+}$ polarized (Fig.~\ref{Fig6}\textbf{c}), the double induction yields one AFM domain, similar to the single induction before. We then change the two beams to $\sigma^{+}$ and $\sigma^{-}$ (Fig.~\ref{Fig6}\textbf{d}). Indeed, the double induction yields opposite AFM domains separated by a domain wall. If we further change the two beams to $\sigma^{-}$ and $\sigma^{+}$, then both AFM domains are flipped and again an AFM domain wall is created. In SI.II.3, we show more systematic data. By double Axion induction, we achieve helicity-dependent optical creation of AFM domain wall for the first time.
\color{black} 

\vspace{0.5cm}
  
\textbf{Direct optical switching of AFM domain by ultrafast pulse}

\begin{figure*}[h]
\includegraphics[width=14cm]{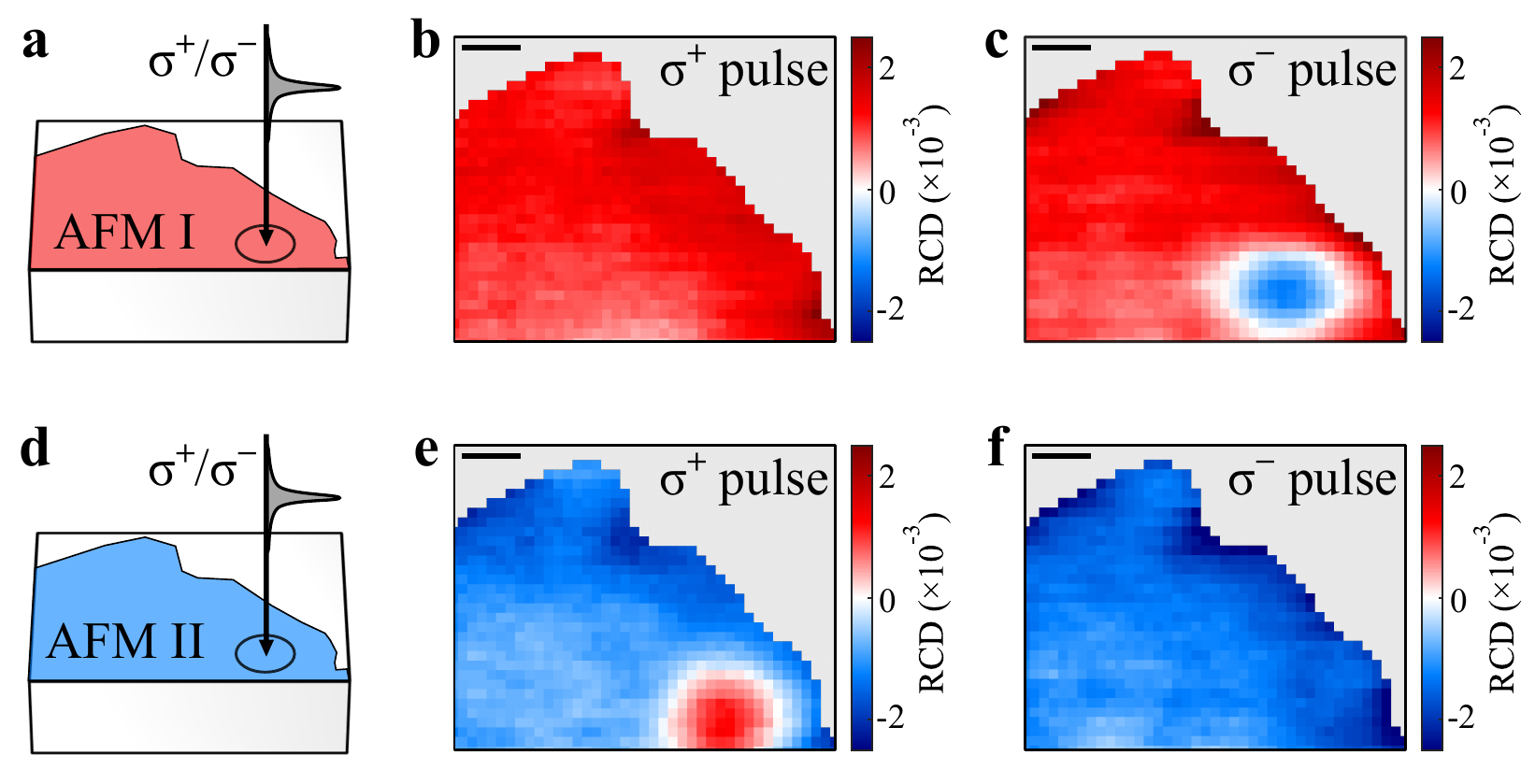}
\vspace{-5.5mm}
\caption{\textbf{Direct optical switch of AFM order by ultrafast pulse with circularly polarization.} \textbf{a,} Schematic illustration of the entire 8SL sample (sample-S1) in the same AFM domain (achieved by sweeping the B field from $+7$ T to $0$ T). We shine circularly-polarized ultrafast pulsed light while keeping the sample at $T=18$ K (below $T_\textrm{N}=25$ K). \textbf{b,c,} RCD maps after shining circularly-polarized ultrafast pulsed light. \textbf{d-f}, The same as panels (\textbf{a-c}) but for the opposite AFM domain prepared by sweeping the $B$ field from $-7$ T to $0$ T. See additional data in SI.II.4. Scale bar: 2 $\mu$m.}
\label{Ultrafast}
\end{figure*}

The optical induction requires warming up the entire sample and then cooling down across $T_\textrm{N}$ with light. To achieve optical writing of complex AFM structures at will, direct optical switching would be highly desirable. We have achieved such direct optical switching of the AFM domain using ultrafast pulsed light with circular polarization. We start from the entire 8SL sample in a single AFM domain (Fig.~\ref{Ultrafast}), while the sample is kept at $T=18$ K (below $T_\textrm{N}$). We shone ultrafast laser pulses with circular polarization, turned off the ultrafast laser, then checked the AFM order by RCD. As shown in Fig.~\ref{Ultrafast}, we indeed directly switch the AFM domain at the ultrafast laser spot with clear helicity dependence. In SI.II.4, we show more systematic data. Direct helicity-dependent optical switching of AFM has never been achieved before. This new result opens a pathway to photolithography for AFM structures. 

\color{black} 
%Moreover, by shining two close-by light beams with opposite helicity, we even optically create an AFM domain wall. The control of AFM domains by light helicity has never been achieved in any fully-compensated AFM. 

\vspace{0.5cm}
\textbf{Discussions} 

 Our results have demonstrated a new type of helicity-dependent optical control (Extended Data Fig.~\ref{Three_Classes}): It has been previously known that the rotating electric field of circularly-polarized (CP) light serves as an effective $\mathbf{B}$ field [$(\mathbf{E}^*\times \mathbf{E})$ has the same symmetry as $\mathbf{B}$], while the rotating electric field multiplies the light propagation vector leads to an effective chiral force [$(\mathbf{E}^*\times \mathbf{E})\cdot\hat{q}$ has the same symmetry as chirality]. Therefore, CP light can control magnetization $M$ and chirality \cite{kirilyuk2010ultrafast, xu2020spontaneous}. In our work, we discovered that CP light can control the AFM order. Such new control can be visualized by the picture that CP light provides an effective Axion $\mathbf{E}\cdot\mathbf{B}$ field [$(\mathbf{E}^*\times \mathbf{E})\cdot\hat{z}$ has the same symmetry as $\mathbf{E}\cdot\mathbf{B}$], where the rotating electric field of CP light serves as an effective $B_z$ field and the sample surface normal as an effective $E_z$. Looking forward, we highlight the following future directions: First, our simultaneous RCD and TCD measurements realize a novel symmetry probe for both $\mathcal{T}$ and $\mathcal{PT}$, which is valuable to investigate novel $\mathcal{T}$-breaking phases in unconventional superconductors and charge orders. For instance, optical nonreciprocity (Kerr rotation) with nominally zero magnetization was also observed in unconventional superconductors such as UPt$_3$ \cite{xia2006high, schemm2014observation,hayes2021multicomponent}. A finite Kerr signal means $\mathcal{T}$-breaking. Whether this state preserves/breaks $\mathcal{PT}$ symmetry is unknown, which can be learnt by simultaneous transmission experiments. Interestingly, theory predicts exotic $\mathcal{PT}$-symmetric topological superconductivity \cite{kawabata2018parity}. Second, we note that the optical Axion $\theta(\omega)$ electrodynamics is quantum geometrical (i.e., it depends on the geometrical properties of Bloch wavefunction such as Berry curvature) but not topological. This is in contrast to the static $\theta$ \cite{qi2008topological, essin2009magnetoelectric}, which can lead to topological quantized effects with exciting experimental progress \cite{wu2016quantized, xiao2018realization, Liu2020a, nenno2020axion, mogi2022experimental}. This means that the optical $\theta(\omega)$ cannot be used to discern topology at photon energies larger than the band gap. On the flip side, it also makes this novel physics more widely applicable in other $\mathcal{PT}$-symmetric AFMs without mirror planes, including Cr$_2$O$_3$ and CrI$_3$ and even the pseudo-gap state of cuprates \cite{xia2008polar}. Third, the direct switching by ultrafast pulses (Fig.~\ref{Ultrafast}) is potentially on the ultrafast timescale. So future pump probe experiments to directly demonstrate ultrafast AFM reversal would be highly desirable. Finally, for MnBi$_2$Te$_4$, because the AFM order is directly coupled to the sign of static $\theta$ angle (a topological invariant) as well as the half-quantized surface Hall conductivity \cite{Otrokov2019unique,Li2019a,Liu2020}, our definitive, versatile optical control of AFM domains and domain walls also leads to an optical writing of ballistic circuits of topological chiral edge states.\color{black}

%, one can achieve heightened clarity and understanding. As shown in Extended Data Fig.~\ref{Three_Classes}, chiral crystals, ferromagnets and $\mathcal{PT}$-symmetric, fully-compensated AFMs feature distinctly different kinds of nontrivial interactions with circularly-polarized light, which are distinguished by their reflection and transmission properties as we achieved here for the AFM case. Also, we note that the nontrivial interactions with circularly-polarized light all have distinct geometrical origins (Extended Data Fig.~\ref{Three_Classes}\textbf{e-f}). Specifically in our AFM MnBi$_2$Te$_4$, we showed that the interaction the is the Berry curvature real space dipole, enabled by the optical $\theta(\omega)$. 

\vspace{0.5cm}

\bibliographystyle{naturemag}
\bibliography{Axion_optical_control_of_antiferromagnetic_order}

\vspace{0.5cm}
\textbf{Methods}

\noindent\textbf{Crystal growth:} Bulk crystals were grown by the flux method \cite{Yan2019Crystal}. Elemental Mn, Bi and Te were mixed at a molar ratio of $15:170:270$, and sealed in a quartz tube with argon environment. The ampule was first heated to $900^{\circ}$C for $5$ hours. It was then moved to another furnace where it slowly cooled from $597^{\circ}$C to $587^{\circ}$C and stayed for one day at $587^{\circ}$C. Finally, MnBi$_2$Te$_4$ were obtained by centrifuging the ampule to separate the crystals from the Bi$_2$Te$_3$ flux. %In the solid-state reaction method, elemental form of Mn, Bi, Te and I$_2$ were first mixed at a stoichiometric ratio of $1.5:2:4:0.5$ and sealed in a quartz ampoule under vacuum. The sample was heated to $900^{\circ}$C in 24 hours in a box furnace and stayed at the temperature for over 5 hours to ensure a good mixture. The ampoule was then air quenched and moved to another furnace preheated at $597^{\circ}$C, where it then slowly cooled to $587^{\circ}$C in 72 hours and stayed at the final temperature for two weeks.% before it was taken out and air quenched again. The product was one chunk of MnBi$_2$Te$_4$ from which single crystals can be cleaved and isolated after the excessive MnI$_2$ was rinsed with water. 

    \vspace{1.0ex}
\noindent\textbf{Sample fabrication:} To preserve the intrinsic properties of 2D MnBi$_2$Te$_4$ flakes, the entire device fabrication process was performed without exposure to air, chemicals, or heat in an Ar-filled glovebox with O$_2$ and water levels below $0.01$ ppm. First, thin flakes were mechanically exfoliated on a $300$-nm SiO$_2$/Si wafer. The number of layers was determined based on the optical contrast shown in Ref. \cite{gao2021layer}. Second, we picked up a desired MnBi$_2$Te$_4$ flake and transferred it onto a diamond, sapphire, or hBN substrate by the cryogenic pickup method developed in Ref. \cite{Zhao2021BSCCO}, where a thin piece of PDMS (polydimethylsiloxane) was cooled to $-110^{\circ}$ by liquid nitrogen to achieve the pickup. Third, a $20-50$ nm hBN flake was transferred onto the MnBi$_2$Te$_4$ flake. A 200 nm layer of PMMA (poly(methyl methacrylate)) was spin-coated onto the sample to further protect it before transferring it from the glovebox to a cryostat.  

%The established pick-up method using the polymer known as ``PC'' poly(Bisphenol A carbonate) requires heating $>150$ $^{\circ}$C (Fig.~\ref{Pickup_comparison}\textbf{a}), at which 2D MnBi$_2$Te$_4$ can be unstable even inside the glovebox. 

%
% Because heating will damage the quality of MnBi2Te4, commonly used polymers such as poly(Bisphenol A carbonate) (PC) and poly(propylene carbonate) (PPC) cannot been used. To transfer the thin flakes of MnBi2Te4 to other substrate without heating, we used a special exfoliation and transfer method. 
%First, diamonds, sapphire, and 300-nm SiO2/Si substrates were baked at 240 ? for more than 12 hours to remove water absorbed on their surfaces. Second, MnBi2Te4 thin flakes were mechanically exfoliated on a 300-nm SiO2/Si wafer. The number of layers was determined based on the optical contrast which has been proven by previous works (Nature 595, 521Ð525 (2021)). Third, a thin PDMS was cooled down to -110 ? by liquid Nitrogen was used to transfer the target MnBi2Te4 flake from the 300-nm SiO2/Si wafer to diamond, sapphire, or BN substrates (arXiv preprint arXiv:2108.13455). 
%To further protect the MnBi2Te4 samples, we also directly exfoliated BN on PDMS and transferred BN onto MnBi2Te4 at room temperature. Afterwards, 200 nm PMMA was spin-coated on the flakes before transferring the sample from the glovebox to a cryostat.  

    \vspace{1.0ex}
\noindent\textbf{Circular dichroism and optical induction:} Optical CD measurements were performed in the closed-loop magneto-optical cryostat OptiCool by Quantum Design (base temperature $\sim2$ K and $B$ field $\pm 7$ T) using a supercontinuum laser SuperK-EXR20 by NKT photonics (wavelength $500$ nm to $2500$ nm, pulse width $\sim12$ ps at 1064 nm).   We focused on $500-1000$ nm due to constraints of the photodetector, lens, objective and beam splitter. \color{black} A spectrometer SpectraPro-300i by Acton Research was used to select the wavelength. The beam went through a photoelastic modulator (PEM200, Hinds instruments) operating at $\frac{\lambda}{4}$ retardation with a frequency of $\mathrm{50\ kHz}$. After an optical chopper ($1000\ \mathrm{Hz}$) and a broadband plate beam-splitter (near normal, Thorlabs BSW26), the beam was focused onto the sample by a 50X Mitutoyo Plan Apochromat Objective (MY50X-825). The reflected beam went through the cryostat's top window, was collimated by the same objective, and was collected by a Si Avalanche Photodetector (APD410A, Thorlabs). The transmitted beam was collimated by a parabolic mirror (37-282 Edmund Optics) inside the cryostat and passed through a side window to reach the APD. The corresponding reflection and transmission APD signals were analyzed by two lock-in amplifiers at $\mathrm{50\ kHz}$ (the PEM frequency) and $\mathrm{1000 \ Hz}$ (the chopper frequency), respectively. The RCD and TCD were the ratio of the $\mathrm{50\ kHz}$ and $\mathrm{1000\ Hz}$ signals. Spatial imaging were achieved using a galvo scanning mirror system.  The background CD were obtained by performing the same measurement at a location immediately next to MnBi$_2$Te$_4$ flake (SI.II.1). In order to reduce the background CD, the beam splitter (BSW26, Thorlabs) was intentionally used at near normal incidence (Fig.~\ref{Fig1}\textbf{f}) \color{black}

Induction experiments were performed using the same supercontinuum laser. The induction light shared the same beam path. When conducting induction experiments, the PEM was turned off. An achromatic $\frac{\lambda}{4}$ waveplate (AQWP10M-580, Thorlabs) was installed before the objective, which generates $\sigma^{\pm}$ polarization. After the induction was completed, the induction light was then turned off, and the $\mathrm{\lambda/4}$ waveplate was removed from the beam path, allowing us to measure the CD using the PEM.  In order to check if the phase of the signal was definitive and consistent, we deliberately turned off and on the PEM multiple times and took the identical measurements. Every time, the phase (sign) of the signal was consistent. We show the data and explain this based on the PEM instrumentation in SI.II.1.

The direct switching was achieved with the sample kept at $T=18$ K (below $T_\textrm{N}=25$ K). The pulsed light was generated by an amplified Yb:KGW laser (Pharos, LightConversion) with pulse duration $168$ fs, wavelength $1030$ nm, repetition rate $100$ kHz. The power applied on the sample was $0.04$ mW ($= 0.4$ nJ per pulse). We shone the ultrafast light for $1$min, turned it off, and then checked the AFM by RCD. \color{black}%The single AFM domain was prepared by sweeping B field to 0T from +7T or -7T, respectively.

% At the DC or low-frequency limit, the $\theta$ term has been theoretically formulated about a decade ago \cite{qi2008topological, essin2009magnetoelectric}. This led to the proposal of ``Axion electrodynamics'' \cite{qi2008topological, essin2009magnetoelectric, qi2009inducing} with exciting experimental progress \cite{wu2016quantized, xiao2018realization, Liu2020a, nenno2020axion, mogi2022experimental}, which set the foundation for realizing elusive electromagnetic phenomena arising from Berry phase and topology including the quantized magnetoelectric (ME) effect \cite{qi2008topological, essin2009magnetoelectric} and the Witten effect \cite{rosenberg2010witten, qi2009inducing}. At the 
%The NV ensemble is created by 15N implantation with 45 keV energy and 10^12 ions/cm^2 dose. MBT flakes are transferred onto the diamond surface inside a glovebox and encapsulated by hBN and PMMA prior to zero-field cooling to 4 K inside a low-temperature confocal microscope. 

% is determined by converting the stray field image into a magnetization image. The detection limit is then taken to be twice the standard deviation of the magnetization distribution in the areas surrounding where the 8-SL and 6-SL MBT flake borders the bare NV sample.

     \vspace{1.0ex}
\noindent\textbf{NV center magnetometry:} NV center magnetic imaging was performed using a diamond sample containing a near-surface ensemble of NV centers. A green laser ($515$ nm, $100$ $\mu$W power, beam spot FWHM $400$ nm) was used to probe the optically-detected magnetic resonance across the NV ensemble \cite{thiel2019}. A pulsed electron spin resonance measurement ($500$ ns pulse length) was performed on the $|0\rangle$ to $|1\rangle$ NV ground-state transition at a background field of $141$ mT along the NV axis ($\sim1.08$ GHz). To determine the stray field dB due to the flake, a linear plane-fit background is subtracted from the raw field image. The NV detection limit was about 2$\mu_{\textrm{B}}/{\textrm{nm}}^2$.

   \vspace{1.0ex}
\noindent\textbf{Optical Axion Electrodynamics:} 

\begin{itemize}
    \vspace{-1.5ex}
%     \item The detailed expressions for $\theta(\omega)$, $\alpha(\omega)$, RCD and TCD are presented in SI.IV.1.   
%     \vspace{-1.0ex}
    \item $\theta$ and $\mathbf{E}\cdot\mathbf{B}$ have identical symmetry properties. They require the breaking of $\mathcal{P}$, $\mathcal{T}$ and all mirrors. Note that there are $\mathcal{PT}$-symmetric phases with mirror symmetry \cite{kimura2020imaging}. They do not support the Axion optical ME coupling because mirror forces $\theta=0$ but they can support other novel optical effect such as the nonreciprocal directional dichroism.  
    \vspace{-1.0ex}
    \item  By adding $\theta(\omega)\frac{e^2}{2\pi hc}\mathbf{E}^\omega\cdot\mathbf{B}^\omega$ into the Lagrangian, the modified Maxwell's equations read
\begin{align}
\mathbf{\nabla}\cdot \mathbf{E}^{\omega} &= \rho - \frac{e^2}{2\pi hc}\mathbf{\nabla}\theta(\omega)\cdot \mathbf{B}^{\omega} \\
\mathbf{\nabla}\times \mathbf{B}^{\omega} &= \partial_t\mathbf{E}^{\omega}+\mathbf{j}^{\omega} + \frac{e^2}{2\pi hc}(\mathbf{\nabla}\theta(\omega)\times\mathbf{E}^{\omega}+\partial_t\theta(\omega)\mathbf{B}^{\omega} )
\end{align} The other equations (the Gauss's law for magnetism and the Faraday's law) are unchanged. 

    \vspace{-1.0ex}
    \item The low frequency limit is defined as frequencies below the magnetic gap at the surface Dirac point, which is typically  $\sim10$ meV in magnetic topological insulators. Therefore, according to this definition, terahertz light is in the low frequency limit. 
 
  \vspace{-1.0ex}
     \item  According to Ref.~\cite{ahn2022theory}, $\theta(\omega)$ is given by
%\small
\normalsize
\begin{align}
 \theta(\omega) &= \pi \frac{2h}{e^2}\frac{1}{3}(2\alpha(\omega)_{xx}+\alpha(\omega)_{zz}), \label{eq:theta-definition}
\end{align}
\begin{align}
\alpha_{xx}(\omega) &= \frac{e^2 }{\hbar L}\sum_{\textrm{o,u}}\int d^2\textbf{k}\frac{\varepsilon_{\textrm{uo}}}{\varepsilon_{\textrm{uo}}-\hbar\omega}\ {\rm Im}[\frac{\hbar^2\braket{\textrm{o}| \hat{v}^x|\textrm{u}}\braket{\textrm{u}|-\frac{1}{2}\left(\hat{v}^y\hat{r}^z+\hat{r}^z\hat{v}^y\right)+\hat{m}^{\rm s}_x|\textrm{o}}}{\varepsilon_{\textrm{uo}}^2}], \label{ME_xx}\\
\alpha_{zz}(\omega) &= \frac{e^2 }{\hbar L}\sum_{\textrm{o,u}}\int d^2\textbf{k}\frac{\varepsilon_{\textrm{uo}}}{\varepsilon_{\textrm{uo}}-\hbar\omega}\ {\rm Im}[\frac{\frac{\hbar^2}{2}(\braket{\textrm{o}| \hat{r}^z|\textrm{u}}\braket{\textrm{u}| \hat{v}^x\hat{v}^y|\textrm{o}}-\braket{\textrm{o}|  \hat{r}^z\hat{v}^x|\textrm{u}}\braket{\textrm{u}|\hat{v}^y |\textrm{o}}}{\varepsilon_{\textrm{uo}}^2}\nonumber\\
& \qquad\qquad\quad\qquad\qquad\qquad\qquad\qquad\quad\quad\ \ \frac{-(x\leftrightarrow y))+\hbar^2\braket{\textrm{o}|\hat{v}^z|\textrm{u}}\braket{\textrm{u}| \hat{m}^{\rm s}_z |\textrm{o}}}{\varepsilon_{\textrm{uo}}^2}]
\end{align}\normalsize where $L$ is the sample thickness, $\varepsilon_{\textrm{uo}}(\mathbf{k})$ is the energy difference between occupied (o) and unoccupied (u) states, $\hat{v}_x$ and $\hat{v}_y$ are velocity operators, $\hat{r}_z$ is the position operator along $z$, and $\hat{m}^{\rm s}$ is the spin operator. 
 
    \vspace{-1.0ex}

\item   To get the Berry curvature real space dipole, we start from $\alpha(\omega)_{xx}$ (because the traceless part is small, $\theta (\omega)\simeq \pi\frac{2h}{e^2}\alpha(\omega)_{xx}$.). 
   \begin{align}
\alpha(\omega)_{xx} &=\frac{e^2 }{ \hbar L} \sum_{\mathrm{o, u}} \int d^2\textbf{k} \frac{ \varepsilon_{\textrm{uo}}}{\varepsilon_{\textrm{uo}}-\hbar\omega} \ {\rm Im}[\frac{\hbar^2\braket{\textrm{o}| \hat{v}^x|\textrm{u}}\braket{\textrm{u}|\hat{v}^y\hat{r}^z|\textrm{o}}}{\varepsilon_{\textrm{uo}}^2}]\nonumber\\
&=\frac{e^2 }{ \hbar L} \sum_{\mathrm{o, u}} \int d^2\textbf{k} \frac{ \varepsilon_{\textrm{uo}}}{\varepsilon_{\textrm{uo}}-\hbar\omega} \ {\rm Im}[\frac{\hbar^2\braket{\textrm{o}| \hat{v}^x|\textrm{u}}\sum_{\mathrm{p}}\braket{\textrm{u}|\hat{v}^y|\textrm{p}(\textbf{k})}\braket{\textrm{p}(\textbf{k})|\hat{r}^z|\textrm{o}}}{\varepsilon_{\textrm{uo}}^2}]\nonumber\\
&\simeq\frac{e^2 }{ \hbar L} \sum_{\mathrm{o, u}} \int d^2\textbf{k} \frac{ \varepsilon_{\textrm{uo}}}{\varepsilon_{\textrm{uo}}-\hbar\omega} \ \langle\hat{r}^z\rangle_{\textrm{o}}{\rm Im}[\frac{\hbar^2\braket{\textrm{o}| \hat{v}^x|\textrm{u}}\sum_{\mathrm{p}}\braket{\textrm{u}|\hat{v}^y|\textrm{p}(\textbf{k})}\delta_{\textrm{po}}}{\varepsilon_{\textrm{uo}}^2}]\nonumber\\
&=\frac{e^2 }{ \hbar L} \sum_{\mathrm{o, u}} \int d^2\textbf{k} \frac{ \varepsilon_{\textrm{uo}}}{\varepsilon_{\textrm{uo}}-\hbar\omega} \ \langle\hat{r}^z\rangle_{\textrm{o}}{\rm Im}[\frac{\hbar^2\braket{\textrm{o}| \hat{v}^x|\textrm{u}}\braket{\textrm{u}|\hat{v}^y|\textrm{o}}}{\varepsilon_{\textrm{uo}}^2}]\nonumber\\
&=\frac{e^2}{2\hbar L}\sum_{\textrm{o,u}}\int d^2\mathbf{k} \ \frac{\varepsilon_{\textrm{uo}}(\mathbf{k})}{\varepsilon_{\textrm{uo}}(\mathbf{k})-\hbar\omega}\  \langle\hat{r}^z\rangle_{\textrm{o}} \Omega_{\textrm{uo}}
\label{eq:Appro}
\end{align} 
Therefore, the Berry curvature real space dipole is a good approximation when the wavefunction of the electronic states is concentrated in a particular layer ($\braket{\textrm{p}(\textbf{k})|\hat{r}^z|\textrm{o}}\simeq\delta_{\textrm{po}}\langle\hat{r}^z\rangle_{\textrm{o}}$). In MnBi$_2$Te$_4$, because it is a vdW layered material, the interlayer coupling is expected to be relatively weak. Hence, the wavefunction of the electronics states is relatively localized.\color{black}
      \vspace{-1.0ex}
%     
%        \item In MnBi$_2$Te$_4$, the ME coupling only has the Axion contribution because 3D bulk MnBi$_2$Te$_4$ has inversion symmetry. By contrast, for AFMs whose bulk breaks inversion, there is an additional contribution to the ME coupling, the traceless contribution \cite{ahn2022theory, malashevich2010band, graham1997}. 
%      
%\color{black}
    \end{itemize}

   \vspace{1.0ex}
\noindent\textbf{Free energy analysis:} Similar to previous works \cite{kirilyuk2010ultrafast,xu2020spontaneous}, we expand the system's free energy in the presence of light. Here we assume the light propagates along $\mathbf{\hat{z}}$. 

 \vspace{-7mm}
 \begin{align}
\delta F=\beta \mathbf{M} \cdot[ \mathbf{E}^*\times \mathbf{E}] +   \gamma \Phi_{\textrm{chiral}} [ \mathbf{E}^*\times \mathbf{E}] \cdot\mathbf{\hat{q}} + \xi L_z [ \mathbf{E}^*\times \mathbf{E}] \cdot\mathbf{\hat{z}},
\label{FE_ME} 
\end{align} where $\mathbf{E}$ and $\mathbf{\hat{q}}$ are the electric field and unit wavevector of light; $\mathbf{M}$, $\Phi_{\textrm{chiral}}$ and $\mathbf{L}$ are the order parameters for FM, chiral crystals and AFM, respectively; $\beta$, $\gamma$ and $\xi$ are the corresponding coupling tensors. First, we explain how each term is constructed. The guiding principle \cite{toledano1987} is that a valid free energy term must be invariant under all symmetries. For instance, $\mathbf{M}$ is odd under $\mathcal{T}$ but even under $\mathcal{P}$; one can check that the same is true for $[ \mathbf{E}^*\times \mathbf{E}]$, so that $\mathbf{M} \cdot[ \mathbf{E}^*\times \mathbf{E}]$ is invariant under both $\mathcal{T}$ and $\mathcal{P}$. Similarly, the spatially-chiral order $\Phi_{\textrm{chiral}}$ is odd under $\mathcal{P}$ but even under $\mathcal{T}$, and the same is true for $[ \mathbf{E}^*\times \mathbf{E}] \cdot\mathbf{\hat{q}}$. The AFM order $L_z$ (as in even-layered MnBi$_2$Te$_4$) is odd under both $\mathcal{P}$ and $\mathcal{T}$, and the same is true for $[ \mathbf{E}^*\times \mathbf{E}] \cdot\mathbf{\hat{z}}$.

Next, we explain the physical meaning of each term. Importantly, one can check that $\mathbf{E}^*\times \mathbf{E}$, $[ \mathbf{E}^*\times \mathbf{E}] \cdot\mathbf{\hat{q}}$, and $[ \mathbf{E}^*\times \mathbf{E}] \cdot\mathbf{\hat{z}}$ all flip sign upon reversing light helicity (while keeping the propagation direction invariant). The first term is the energy coupling between $M$ and circularly-polarized light, which is responsible for the helicity-dependent optical control of magnetization observed in FMs  \cite{kirilyuk2010ultrafast}. The second term is the energy coupling between spatial chirality and circularly-polarized light, which is responsible for the helicity-dependent optical control of spatial chirality observed in asymmetrical chemical reactions \cite{huck1996dynamic} and gyrotropic electronic order \cite{xu2020spontaneous}. The last term is the energy coupling between the fully-compensated AFM order and circularly-polarized light, which is responsible for the helicity-dependent optical control of the fully-compensated AFM order, achieved for the first time in even-layered MnBi$_2$Te$_4$ here. The coupling constant $\xi$ directly arises from the optical Axion electrodynamics, as we demonstrated from the data above.

    \vspace{1.0ex}
\noindent\textbf{First-principles calculations:} First-principles band structure calculations were performed using the projector augmented wave method as implemented in the VASP package within the generalized gradient approximation (GGA) schemes. $9\times9\times1$ Monkhorst-Pack $k$-point meshes with an energy cutoff of $400$ eV were adapted for the Brillouin zone integration. Experimentally determined lattice parameters were used. In order to treat the localized Mn $3d$ orbitals, we follow previous first-principles works \cite{Otrokov2019a, Otrokov2019} on MnBi$_2$Te$_4$ and used an onsite $U~=~5.0$ eV. The Wannier model for the few-layered MnBi$_2$Te$_4$ was built using the Bi $p$, Te $p$ and Mn $d$ orbitals. All optical response functions were calculated based on the Wannier model.

\vspace{0.5cm}

\textbf{Data availability:} The data that support the plots within this paper and other findings of this study are available from the corresponding author upon reasonable request.

\textbf{Acknowledgement:} 
 We gratefully thank Xiaodong Xu and Tiancheng Song for sharing experience on CD setup and Manfred Fiebig for providing the Cr$_2$O$_3$ bulk crystals. We also thank Yang Gao, Bertrand I. Halperin, Pavon Hosur, and Philip Kim for helpful discussions. Work in the SYX group was supported through NSF Career (Harvard fund 129522) DMR-2143177. SYX acknowledge the Corning Fund for Faculty Development. SYX, JA, QM, AV acknowledge support from the Center for the Advancement of Topological Semimetals, an Energy Frontier Research Center funded by the U.S. Department of Energy (DOE) Office of Science, through the Ames Laboratory under contract DE-AC0207CH11358. QM was also supported through the CIFAR Azrieli Global Scholars programme. CT acknowledges support from the Swiss National Science Foundation under project P2EZP2\_191801. YFL, SYX, and DCB were supported by the STC Center for Integrated Quantum Materials (CIQM), NSF Grant No. DMR-1231319. This work was performed in part at the Center for Nanoscale Systems (CNS) Harvard University, a member of the National Nanotechnology Coordinated Infrastructure Network (NNCI), which is supported by the National Science Foundation under NSF award no.1541959. Work at UCLA was supported by the U.S. Department of Energy (DOE), office of Science, office of Basic Energy Sciences under Award Number DE-SC0021117. The work at TIFR Mumbai is supported by the Department of Atomic Energy of the government of India under Project No. 12-R\&D-TFR-5.10-0100. The work at Northeastern University was supported by the Air Force Office of Scientific Research under award number FA9550-20-1-0322, and it benefited from the computational resources of Northeastern University's Advanced Scientific Computation Center (ASCC) and the Discovery Cluster. H.L. acknowledges the support by the National Science and Technology Council (NSTC) in Taiwan under grant number MOST
111-2112-M-001-057-MY3. T.-R.C. was supported by the 2030 Cross-Generation Young Scholars Program from the National Science and Technology Council (NSTC) in Taiwan (Program No. MOST111-2628-M-006-003-MY3), National Cheng Kung University (NCKU), Taiwan, and National Center for Theoretical Sciences, Taiwan. This research was supported, in part, by Higher Education Sprout Project, Ministry of Education to the Headquarters of University Advancement at NCKU. HZL was supported by the National Key R\&D Program of China (2022YFA1403700), the National Natural Science Foundation of China (11925402), Guangdong province (2016ZT06D348, 2020KCXTD001), the Science, Technology and Innovation Commission of Shenzhen Municipality (ZDSYS20170303165926217, JCYJ20170412152620376, KYTDPT20181011104202253), and Center for Computational Science and Engineering of SUSTech. K.W. and T.T. acknowledge support from JSPS KAKENHI (Grant Numbers 19H05790, 20H00354 and 21H05233). X-YZ, Y-XW, and BBZ acknowledge support from the NSF award No. ECCS-2041779.

\textbf{Author contributions:}   JXQ performed the optical measurements and analyzed data with help from CT and HCL. JXQ performed the transport measurements with help from AG. AG fabricated the 2D MnBi$_2$Te$_4$ samples with help from JXQ, HCL, YFL, DB, TD, SCH, and QM. CH and NN grew the bulk MnBi$_2$Te$_4$ single crystals. JA and AV developed the theory of optical Axion electrodynamics with discussions with SYX. BG and JA made first-principles calculations with helps BS from under supervisions of HS, TRC, AB, and AV. ZHG and HZL did model simulations. DCB performed transmission electron microscope measurements. X-YZ, Y-XW, and BBZ performed nitrogen-vacancy center magnetometry experiments. KW and TT grew the bulk hBN single crystals. SYX, JXQ and QM wrote the manuscript with input from all authors. JXQ came up with the idea of optical induction of AFM. SYX supervised the project and was responsible for the overall direction, planning and integration among different research units.

%AG fabricated the devices with help from YFL, JXQ, DB, CF, KSB and QM. JXQ came up with the idea of optical induction. 

%, SCH, DB, TD and QMYFL, JXQ, DB, CF, KSB and QM

\textbf{Competing financial interests:} The authors declare no competing financial interests.

\clearpage

%RCD and TCD are reflection circular dichroism and transmission circular dichroism. 

\clearpage

%\textbf{d,e,} Same as panels (\textbf{a,b}) but for the 5SL flake in sample-S3. \textbf{f,} Wavelength dependence for TCD at $B=0$ for the 5SL and 6SL flakes. 
%\addtocounter{figure}{-1}
%\begin{figure*}[t!]
%\caption{wavelength $\lambda_{\textrm{induction}}$ influences the results of induction, we fix all other experimental parameters including induction helicity (fixed at $\sigma^+$) and detection wavelength (fixed at $\lambda_{\textrm{detection}}=946$ nm) and we only vary $\lambda_{\textrm{induction}}$. \textbf{f,} Free energy diagrams summarize the optical control of the AFM order in even-layered MnBi$_2$Te$_4$ with light helicity and wavelength.}
%\end{figure*}
\clearpage

%after induction at $\lambda_{\textrm{induction}}=840$ nm as a reference. 
%The induction ability (the $y$ axis) is defined as the RCD after induction at a particular wavelength normalized to the reference RCD. Therefore, if the induction leads to the same AFM state as the reference induction, then the induction ability will be $\simeq+1$; if the induction leads to the opposite AFM state as the reference induction's, then the induction ability will be $\simeq-1$.

%; if the induction has no effect, then we perform the induction 6 consecutive inductions and take the averaged RCD value, and he induction ability will be $\simeq0$
%measure the RCD's detection wavelength dependence (panel \textbf{e}) at $2$ K. We fix $\lambda_{\textrm{detection}}=946$ nm and 

\clearpage

\setcounter{figure}{0}
\renewcommand{\figurename}{\textbf{Extended Data Fig.}}
\begin{figure*}[t]
\includegraphics[width=17cm]{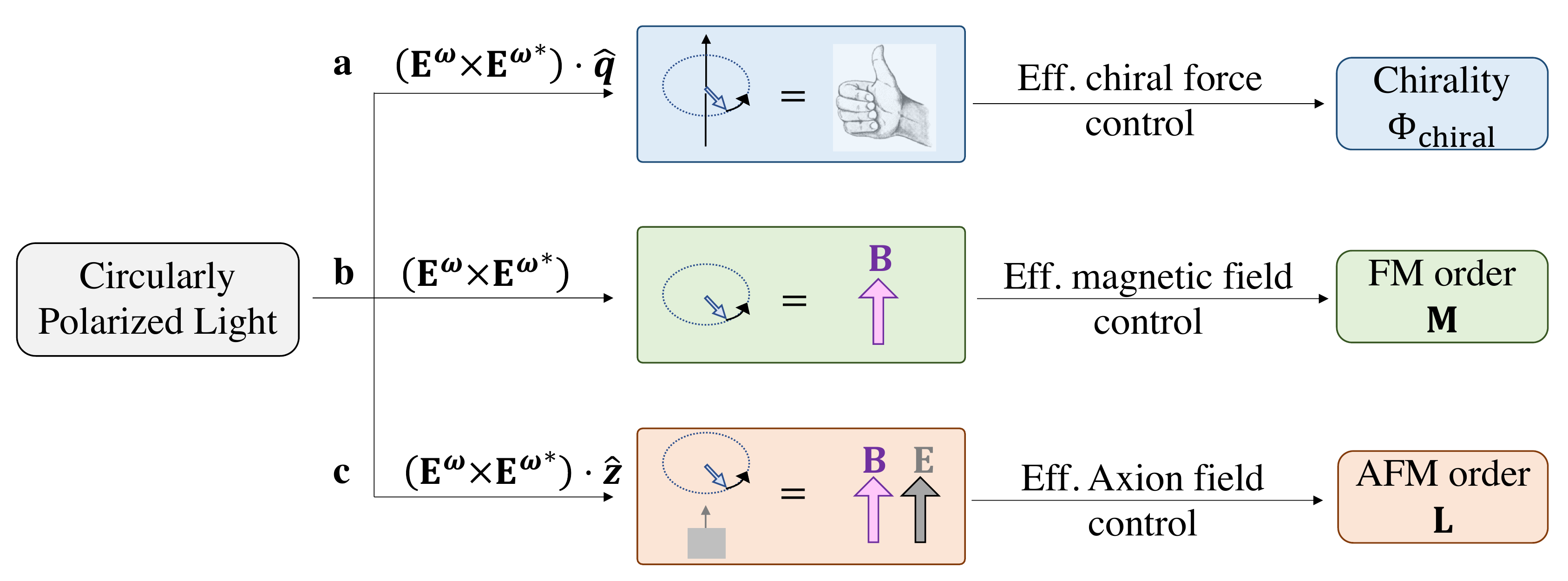}
\vspace{-5.5mm}
\caption{\textbf{Distinct mechanism for the three classes of helicity-dependent optical control.} $(\mathbf{E}^\omega\times {\mathbf{E}^\omega}^*) \cdot\mathbf{\hat{q}}$, $(\mathbf{E}^\omega\times {\mathbf{E}^\omega}^*)$, and $(\mathbf{E}^\omega\times {\mathbf{E}^\omega}^*) \cdot\mathbf{\hat{z}}$ are symmetry-equivalent to chirality, $\mathbf{B}$ and $\mathbf{E}\cdot\mathbf{B}$, respectively.}
\label{Three_Classes}
\end{figure*}

\begin{figure*}[t]
\includegraphics[width=15cm]{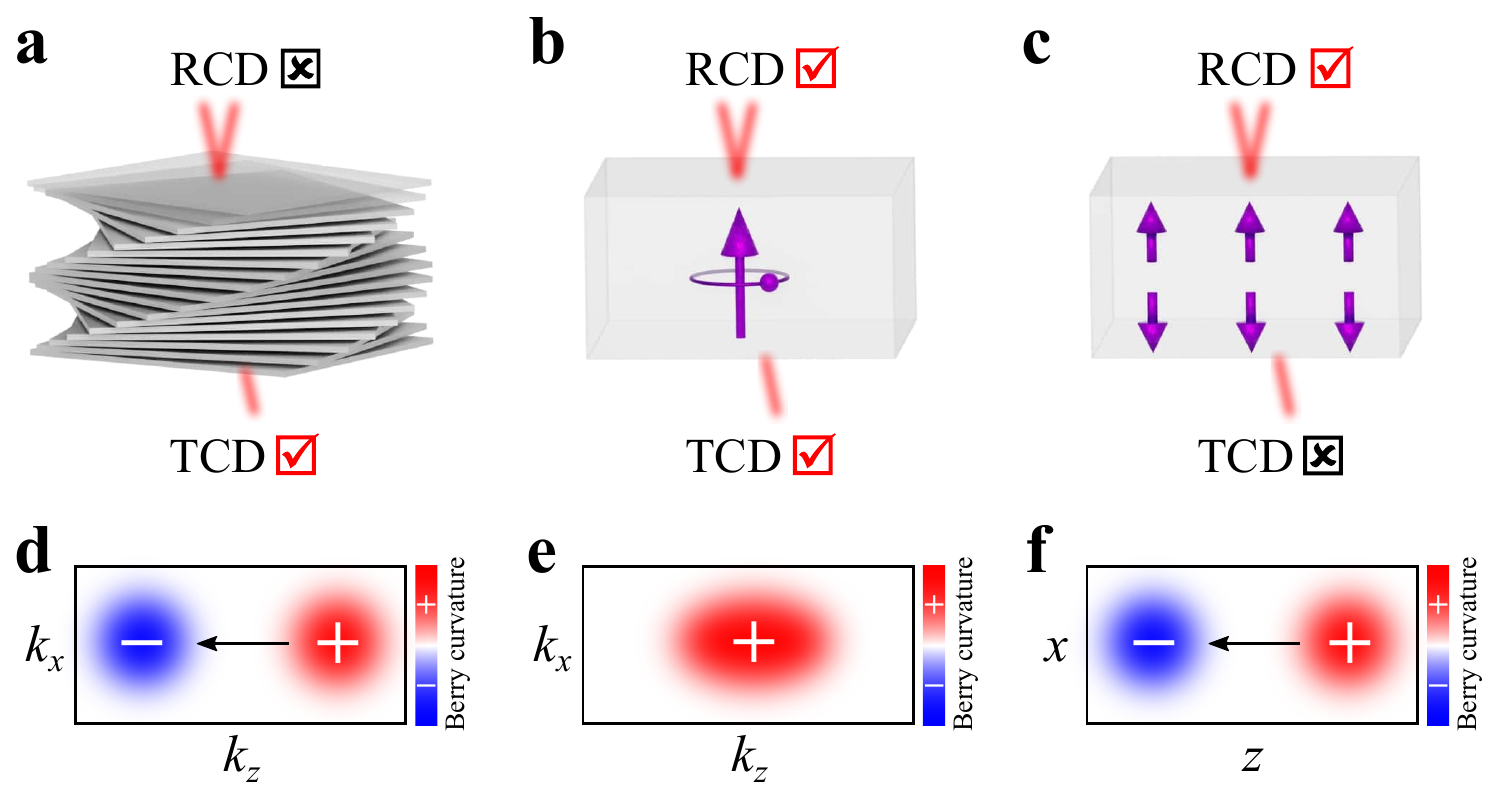}
\vspace{-5.5mm}
\caption{\textbf{Three classes of CD in chiral crystals, ferromagnets and $\mathcal{PT}$-symmetric AFM.} \textbf{a-c}, Chiral crystals, ferromagnets and $\mathcal{PT}$-symmetric AFM feature distinct nontrivial interactions with circularly-polarized light. They can only be distinguished by their transmission and reflection properties. \textbf{d-f}, Equally importantly, when bands with nontrivial topology or giant Berry curvature occur in these three classes of materials, the respective CD can dominantly arise from the Berry curvature properties, namely the Berry curvature $k$ space dipole, the total Berry curvature, and the Berry curvature real space dipole, respectively.}
\label{Three_Classes}
\end{figure*}

%\textbf{Three classes of CD in chiral crystals, ferromagnets and $\mathcal{PT}$-symmetric AFM.} \textbf{a-c}, Chiral crystals, ferromagnets and $\mathcal{PT}$-symmetric AFM feature distinct nontrivial interactions with circularly-polarized light. They can only be distinguished by their transmission and reflection properties. \textbf{d-f}, Equally importantly, when bands with nontrivial topology or giant Berry curvature occur in these three classes of materials, the respective CD can dominantly arise from the Berry curvature properties, namely the Berry curvature $k$ space dipole, the total Berry curvature, and the Berry curvature real space dipole, respectively.

\begin{figure*}[h]
\includegraphics[width=13cm]{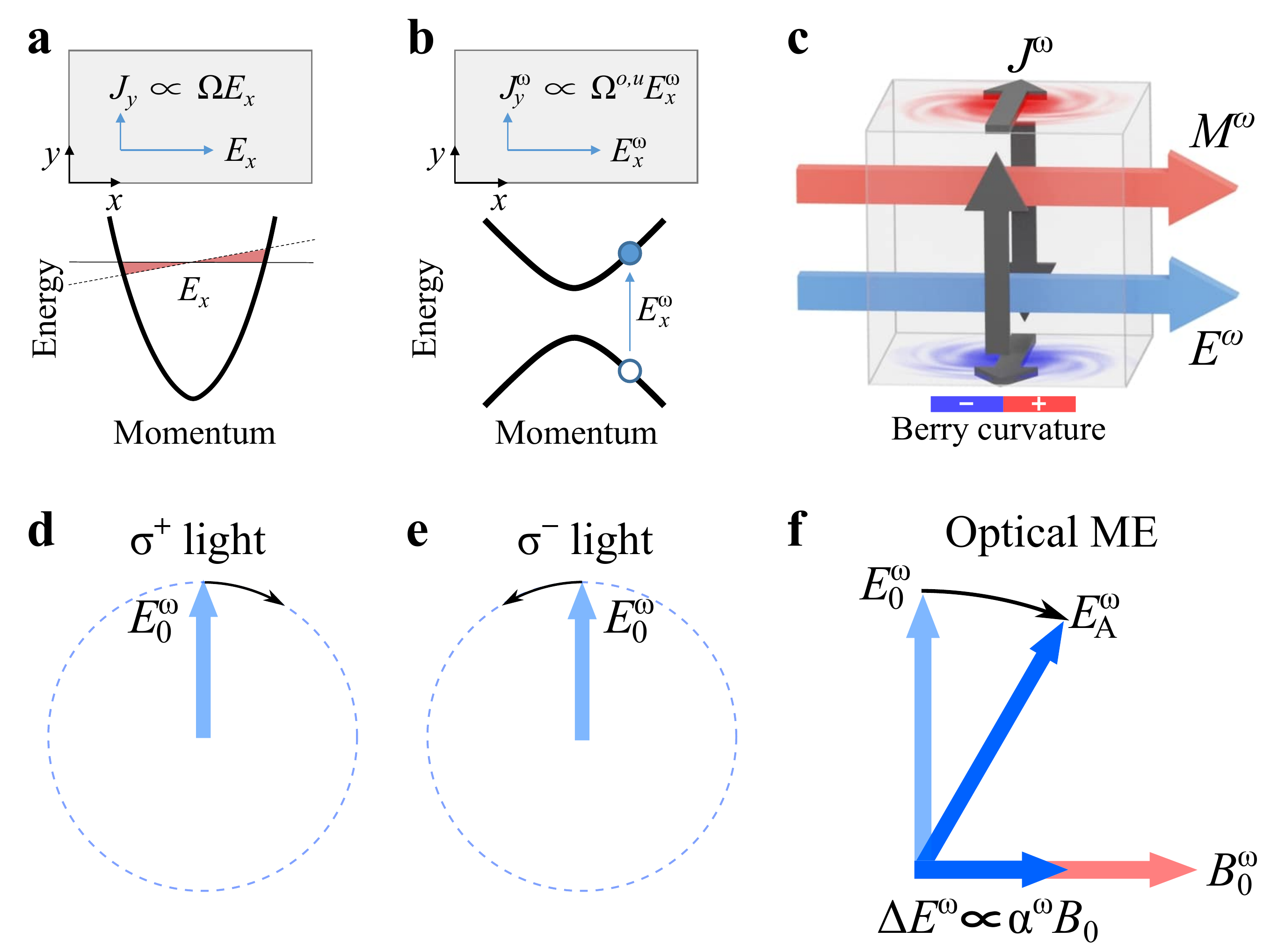}
\vspace{-5.5mm}
\caption{\textbf{a,} The Berry curvature causes transverse electron motion in response to an external DC $E$ field. \textbf{b,} Analogously, the inter-band Berry curvature causes a transverse electron motion in response to light's $E^{\omega}$ field upon an interband transition. Ref.~\cite{ahn2021} provides a detailed theoretical analysis for the clear geometrical origin of the inter-band Berry curvature. \textbf{c,} The Berry curvature induced optical ME coupling can be visualized an itinerant electron circulation in response to electric field. Specifically, upon the application of an electric field $E^\omega$, Berry curvature leads to transverse electron motion. Because of the Berry curvature at $\pm Z$ is opposite, the transverse motions of electrons at $\pm Z$ are in opposite directions. This in turn leads to an itinerant electron circulation $J^\omega$, which is equivalent to magnetization $M^\omega$.  \textbf{d,e,} Rotation of electric field $E^{\mathrm{\omega}}_{\mathrm{0}}$ for $\sigma^{\pm}$ light. \textbf{f,} Light's magnetic field $B^{\omega}$ can lead to an electric polarization $P_0^{\omega}=\alpha(\omega)B_0^{\omega}$ through the optical ME effect, which rotates light's electric field $E^{\omega}$. Suppose this rotation is in the clockwise direction. Then, this additional clockwise rotation would be along the same direction as the intrinsic rotation (also clockwise) for $\sigma^+$ light but opposite to the intrinsic rotation (counter-clockwise) for $\sigma^-$ light. This provides an intuitive picture for why $\sigma^{\pm}$ light can behave differently  in an Axion insulator, a prerequisite for nonzero CD. }
\label{TopoME}
\end{figure*}

%\begin{figure*}[t]
%\includegraphics[width=15cm]{Ext_Data_Fig/Three_Classes_v5.pdf}
%\vspace{-5.5mm}
%\caption{\textbf{Three classes of CD in chiral crystals, ferromagnets and $\mathcal{PT}$-symmetric AFM.} \textbf{a-c}, Chiral crystals, ferromagnets and $\mathcal{PT}$-symmetric AFM feature distinct nontrivial interactions with circularly-polarized light. They can only be distinguished by their transmission and reflection properties. \textbf{d-f}, Equally importantly, when bands with nontrivial topology or giant Berry curvature occur in these three classes of materials, the respective CD can dominantly arise from the Berry curvature properties, namely the Berry curvature $k$ space dipole, the total Berry curvature, and the Berry curvature real space dipole, respectively.}
%\label{Three_Classes}
%\end{figure*}

\begin{figure*}[t]
\includegraphics[width=12cm]{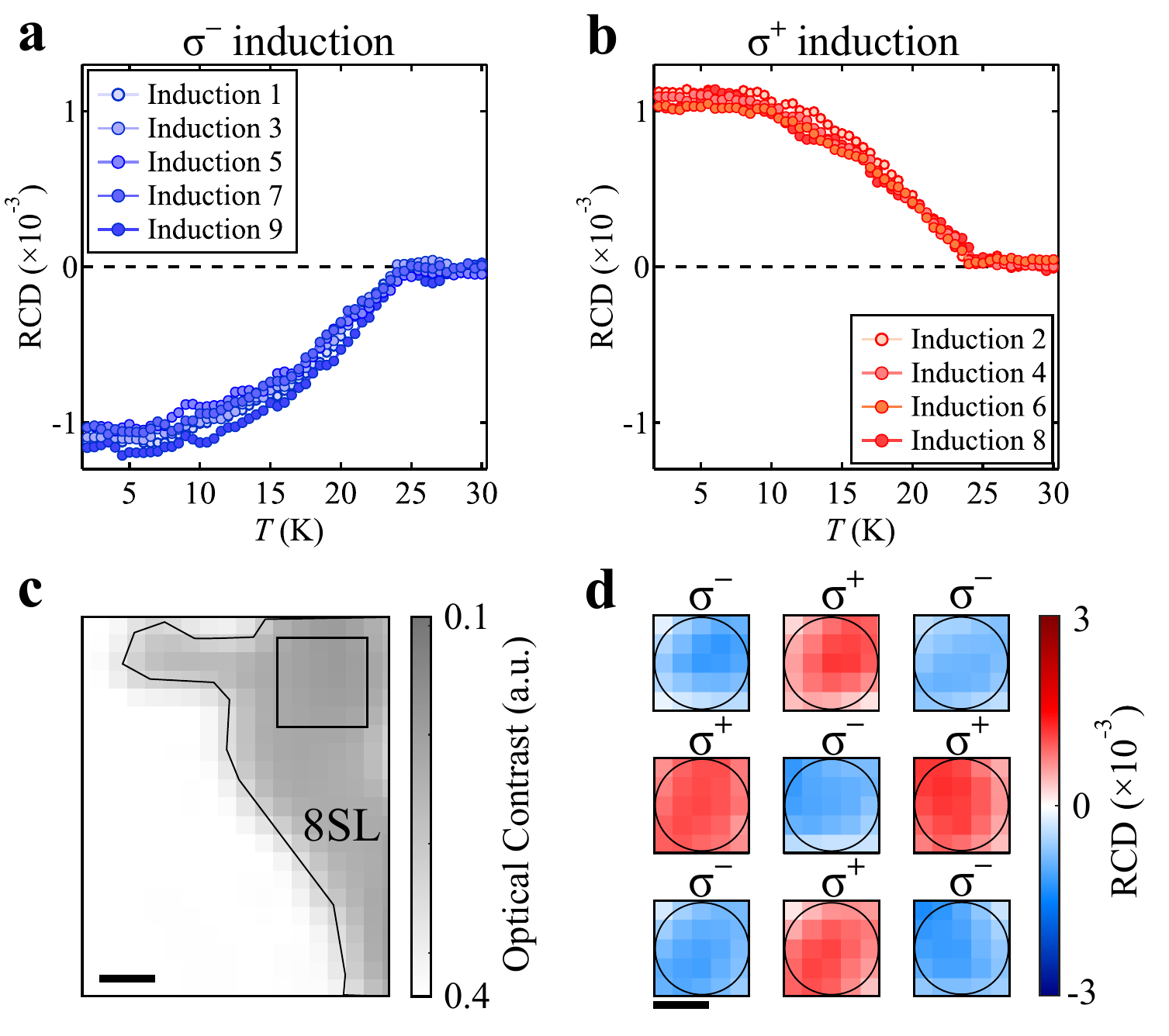}
\caption{\textbf{Reproducible RCD measurements for nine consecutive inductions with alternating induction helicities. a,} RCD as a function of temperature while warming up after induction with $\sigma^-$ light.  \textbf{b,} Same as panel (\textbf{a}) but after induction with $\sigma^+$ light.  \textbf{c,} Spatial mapping of the optical contrast near the 8SL MnBi$_2$Te$_4$ flake. The square box marks the region for induction experiments. \textbf{d,} RCD map after induction with opposite helicity. The circle marks the spot that is subject to the induction light while cooling; The $\sigma^+$ and $\sigma^-$ on each small panel denotes the helicity of the induction light. 
%\textbf{e,} We take the averaged RCD after four $\sigma^+$ inductions as well as the averaged RCD five $\sigma^-$ inductions. The $\Delta$RCD is the difference between these two averaged values.
Experimental parameters used for data in this figure: $\lambda_{\textrm{induction}}=840$ nm, $P_{\textrm{induction}}\simeq1$ mW; $\lambda_{\textrm{detection}}=946$ nm, $P_{\textrm{detection}}\simeq30$ $\mu$W. Scale bars for panels (\textbf{e,f}) are $2$ $\mu$m.
}
\label{Induction_temporal_reproducibility}
\end{figure*}

\clearpage
\begin{figure*}[t]
\includegraphics[width=15cm]{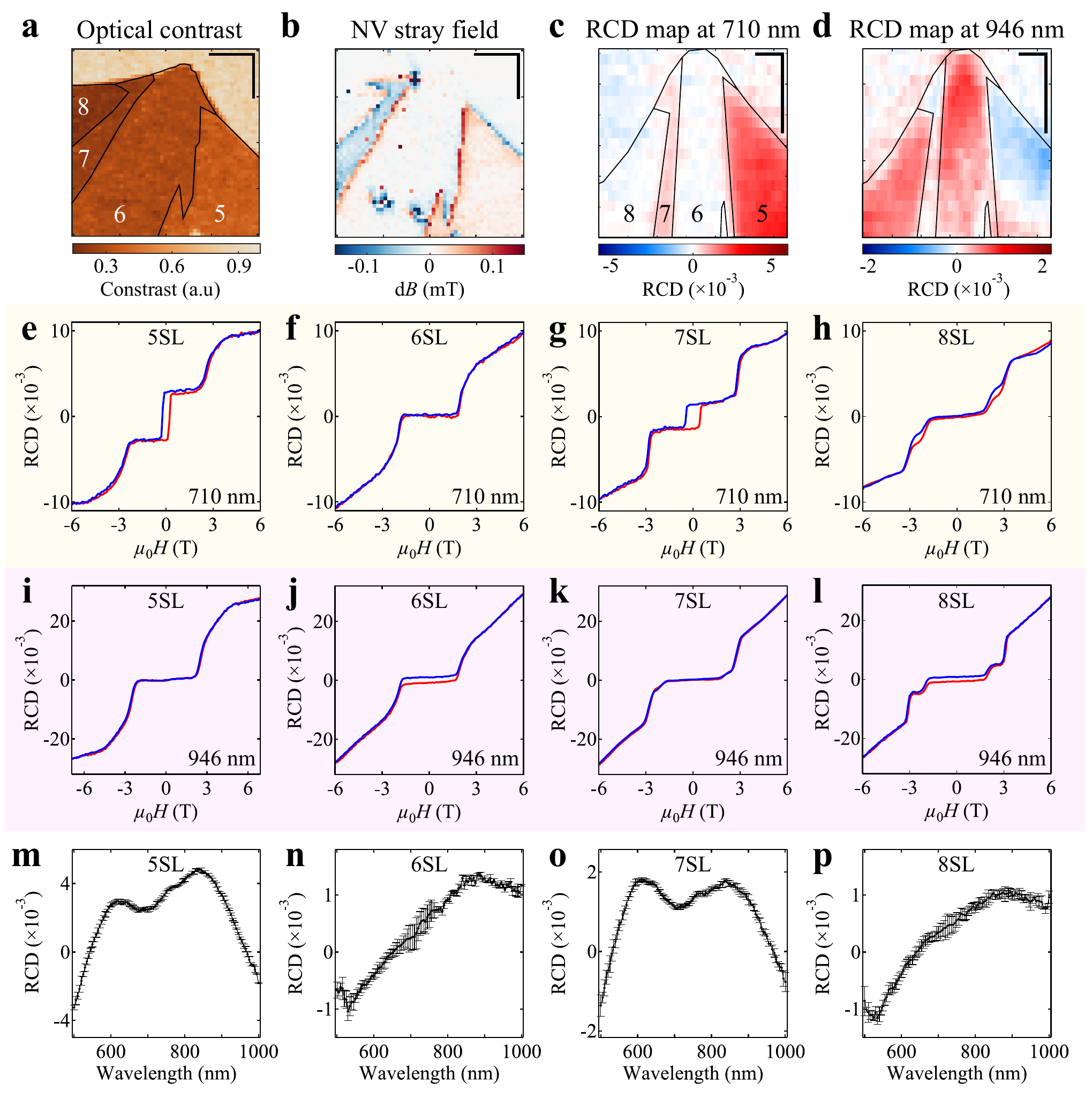}
\caption{{\bf Circular dichroism in 2D MnBi$_2$Te$_4$.} \textbf{a,} Optical contrast of sample-S2 on diamond structure, which consists of four connected flakes of 5SL, 6SL, 7SL and 8SL. \textbf{b,} Nitrogen vacancy center measured stray magnetic field of sample S2. \textbf{c,} RCD spatial mapping at $B=0$ using $\lambda_{\textrm{detection}}=710$ nm. The sample was cooled down with a finite $B$ field and the $B$ field was ramped to zero before the measurements. \textbf{d,} Same as panel \textbf{c} but using $\lambda_{\textrm{detection}}=946$ nm with. No optical induction was performed in panels \textbf{c,d}. Scale bars (horizontal and vertical lines) are all $5$ $\mu$m. We note that the NV and CD measurements were performed in different setups. Hence the spatial mappings (\textbf{b} and \textbf{c,d}) are rotated with respect to each other. \textbf{e-h,} Magnetic hysteresis of RCD for 5SL, 6SL, 7SL and 8SL measured at $\lambda_{\textrm{detection}}=710$ nm. \textbf{i-l,} Same as panels (\textbf{e-h}) but measured at $\lambda_{\textrm{detection}}=946$ nm. \textbf{m-p,} RCD spectra at $B=0$ for 5-8SL. The spectrum strongly depends on the evenness or oddness of the layer number, consistent with the different physical origins of the CDs in even and odd layers. It is interesting to note that $\sim700$ nm is the symmetric (antisymmetric) point for odd (even) layers. }
\label{Fig3}
\end{figure*}

\begin{figure*}[h]
\includegraphics[width=15cm]{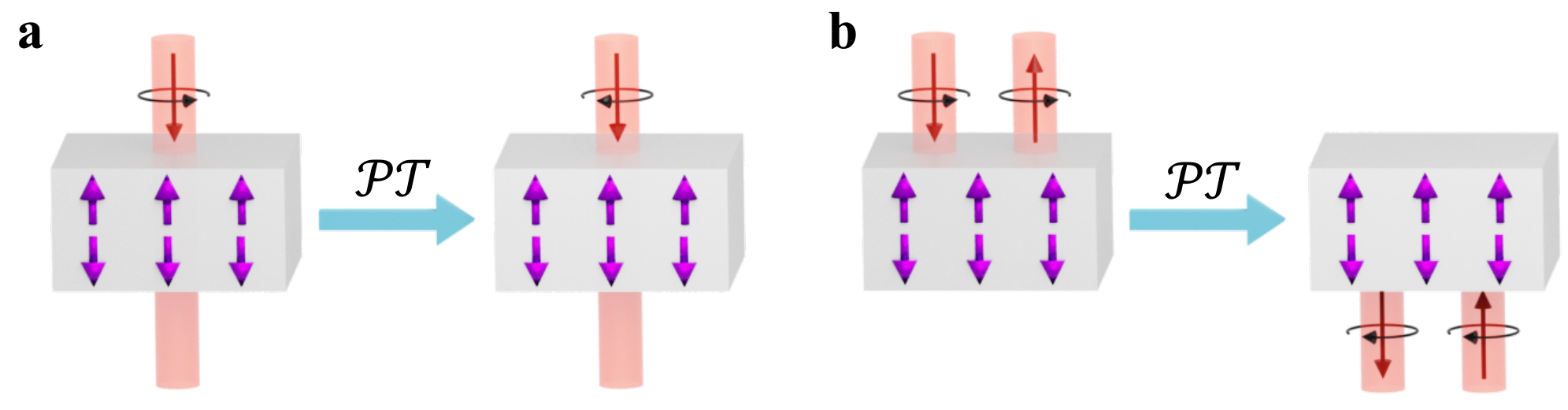}
\caption{{\bf Symmetry analysis for CD in $\mathcal{PT}$-symmetric AFM. a,} $\sigma^-$ light transmitting through a sample. Upon $\mathcal{PT}$ inversion, the AFM remains invariant and the light path also stays the same, but light helicity is reversed. As such, $\mathcal{PT}$ enforces the transmission for $\sigma^{\pm}$ to be identical, which means $\textrm{TCD} =0$. \textbf{b,} $\sigma^-$ light reflecting off a sample. Upon $\mathcal{PT}$ inversion, the AFM remains invariant but the reflection is changed to the bottom surface. This means that $\mathcal{PT}$ does not impose any constraint on RCD experiments, because RCD compares the reflections of $\sigma^{\pm}$ lights from the same side of the sample. Carrying out the same analysis exhaustively confirms that there is no symmetry that can relate the reflections of light with opposite helicity while keeping the AFM invariant. Therefore, RCD is allowed. }
\label{Symmetry}
\end{figure*}
%\begin{figure*}[h]
%\includegraphics[width=12cm]{Ext_Data_Fig/hBN_encapsulation}
%\vspace{-5.5mm}
%\caption{{\bf Comparison between sample-S2 on diamond and sample-S4 encapsulated by hBN.} \textbf{a-c,} For the sample-S2 on diamond substrate. \textbf{a,} Sample schematic, \textbf{b,} RCD magnetic hysteresis at $946$ nm, \textbf{c,}  RCD spectrum at $B=0$. \textbf{d-f,} Same as panels (\textbf{a-c}) but for sample-S4 encapsulated by hBN. }
%\label{hBN_encapsulation}
%\end{figure*}

\begin{figure*}[h]
\includegraphics[width=8.5cm]{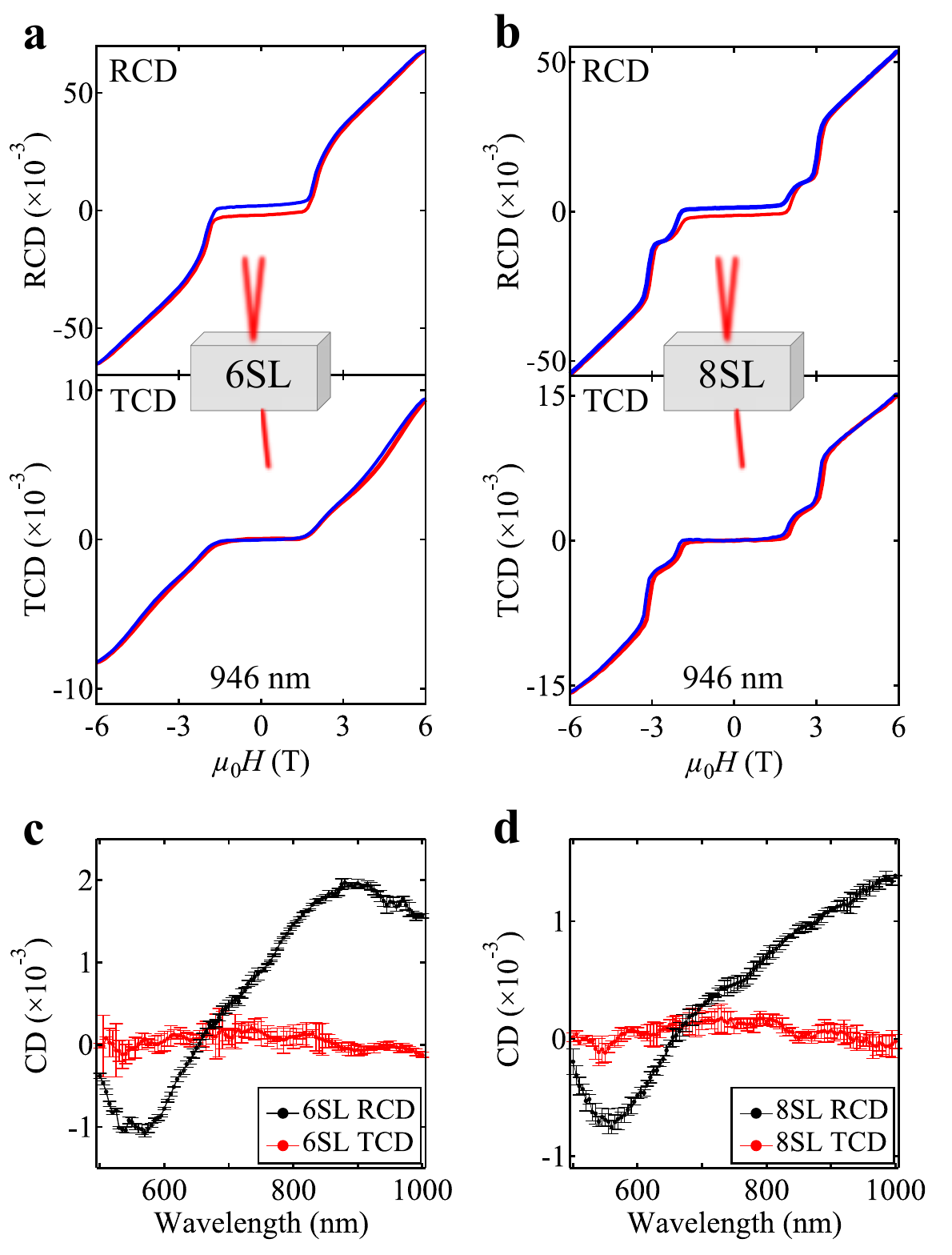}
\vspace{-5.5mm}
\caption{{\bf RCD $vs.$ TCD measurements of the 6SL and 8SL MnBi$_2$Te$_4$ in sample-S1.} \textbf{a,b,} RCD and TCD magnetic hysteresis measurements at $946$ nm. \textbf{c,} RCD and TCD spectra at $B=0$.}
\label{8SL_sapphire}
\end{figure*}

\setcounter{figure}{0}
\renewcommand{\thefigure}{S\arabic{figure}}
\renewcommand{\theequation}{S\arabic{equation}}
\renewcommand{\thetable}{S\arabic{table}}
\renewcommand{\figurename}{\textbf{Fig.}}

%\begin{figure*}[h]
%\includegraphics[width=17.5cm]{Ext_Data_Fig/free_energy_control.pdf}
%\vspace{-3.5mm}
%\caption{\textbf{a,} Opposite helicity can favor opposite AFM states. \textbf{b,} With a fixed helicity, different light wavelength can favor opposite AFM states.}
%\label{free_energy_control}
%\end{figure*}

%\begin{figure*}[h]
%\includegraphics[width=12cm]{Ext_Data_Fig/Induction_spectra.pdf}
%\vspace{-5.5mm}
%\caption{\textbf{a,c,} We shine  $\sigma^-$ induction light ($\lambda_{\textrm{induction}}=840$ nm, $P_{\textrm{induction}}\simeq1$ mW) on the 8SL MnBi$_2$Te$_4$ flake (sample-S1) while lowering its temperature from $T=30$ K to $2$ K (panel \textbf{a}).  Upon reaching $2$ K, we turn off the induction light, and measure the RCD's spectra (panel \textbf{c}) at $2$ K. \textbf{b,d,} Same as panels (\textbf{a,c}) except that we perform induction with $\sigma^+$. \textbf{e-h,} Same as panels (\textbf{a-d}) except that we change $\lambda_{\textrm{induction}}$ to $\lambda_{\textrm{induction}}=540$ nm.}
%\label{Induction_spectra}
%\end{figure*}

\clearpage

\newpage

\vspace{3cm}
\Large
\begin{center}
Supplemental Information for

 \textbf{Axion optical induction of antiferromagnetic order}
\end{center}
\vspace{0.45cm}
\textbf{
\begin{center}
{Table of contents:\\}
\end{center}
}
\vspace{1cm}
\normalsize
\begin{tabular}{l l}
\underline{I.} & Addressing alternative origins for CD in even-layered MnBi$_2$Te$_4$ \\
\enspace  I.1. & Protection layers (PMMA/hBN) induced CD  \\
\enspace  I.2. & Uncompensated magnetization induced magnetic CD  \\
\enspace  I.3. & Light attenuation induced residual CD \\
\enspace  I.4. & Higher order effects \\
\underline{II.} & Additional data \\
\enspace  II.1. & Additional CD data\\
\enspace  II.2. & Additional induction data\\
\enspace  II.3. & Additional double induction data\\
\enspace  II.4. & Additional ultrafast switching data\\
\enspace  II.5. & Additional electrical transport data\\

\underline{III.} & Symmetry analysis of the CD \\
\enspace  III.1. & General principles\\
\enspace  III.2. & Three classes of materials and their symmetries\\
\enspace  III.3. & Application to AFM and other systems\\
\enspace  III.4. & Math derivation for symmetry analysis\\

\underline{IV.} & Theoretical studies \\
\enspace  IV.1. & Theoretical expressions for calculating the Axion CD \\
\enspace  IV.2. & Band structure of $\mathrm{MnBi_2Te_4}$ and additional calculations\\
\enspace  IV.3. & Mathematical derivation of RCD/Kerr under different symmetry condition\\
%\enspace  IV.3. & XXX \\
\underline{V.} & Additional discussions \\
\enspace  V.1. & Additional discussion about Berry curvature real space dipole\\
\enspace  V.2. & Additional discussion about optical control of AFM \\
\enspace  V.3. & Additional discussion about previous works on MnBi$_2$Te$_4$\\
\enspace  V.4. & Additional discussion about thickness dependence \\
\end{tabular}

\date{\today}

%

%We present additional discussion/clarifications of our data, focusing on its relationship with symmetry, Berry curvature, and topology.

\clearpage
\section*{I. Addressing alternative origins for CD in even-layered MnBi$_2$Te$_4$}

\hspace{4mm}In this section, we carefully consider alternative origins for the observed CD beyond the optical Axion electrodynamics. We enumerate systematic experimental data and theoretical analyses, which allow us to show that these alternative origins are not significant for our experiments.

\subsection*{I.1. Protection layers (PMMA/hBN) induced CD}
\hspace{4mm}We covered hBN and PMMA (poly(methyl methacrylate)) layers on the MnBi$_2$Te$_4$ flakes to protect them from oxidation before taking them out of the glovebox (see Methods section). We can exclude the possibility that the observed CD signals come from these protection layers as follows:

\begin{itemize}
\item PMMA and hBN are non-magnetic. Reflection CD is strictly prohibited under $\mathcal{T}$ (see proof in SI.III.3).\\

\item As shown by the temperature-dependence and $B$-hysteresis measurements (Figs. 2 and 3 in the main text), our observed CD signal clearly arises from the AFM order in MnBi$_2$Te$_4$.

%We have made dual-gated 6SL MnBi$_2$Te$_4$ device to directly test the influence of $E$ field. XXX
\end{itemize}

\subsection*{I.2. Uncompensated magnetization induced magnetic CD}

\hspace{4mm}We now consider the possibility that the AFM is uncompensated with nonzero $M$, which leads to magnetic CD. In even-layered MnBi$_2$Te$_4$, this uncompensated $M$ could arise from sample degradation or a vertical $E$ field. Because this has been carefully addressed in the main text, we only enumerate the key evidence.

\begin{itemize}
\item Most importantly, we observe  $\textrm{RCD} \neq0$ but $\textrm{TCD} =0$. We emphasize that RCD and TCD were measured simultaneously, which means on the same spot of the same flake without changing any other condition ($T$, $B$ field).
\item The NV center magnetometry measurements show no observable $M$.
\item The hBN-encapsulated 6SL MnBi$_2$Te$_4$ sample is expected to minimize the vertical $E$ field. However, as shown in Extended Data Fig.5,  the RCD still persists. 
\item Sample degradation is minimized because the entire device fabrication process was finished in an Ar-filled glovebox without exposure to air, chemicals, or heat. The optical CD data are of high quality. Using a similar fabrication process, we have also fabricated a 6SL MnBi$_2$Te$_4$ device with electrical contacts and performed transport experiments in the same cryostat as the optical CD measurements. In the FM phase at $B=-7$ T, we observed clear topological Chern insulator state with fully quantized $R_{yx}$ and zero $R_{xx}$ (Fig.~\ref{Transport_data_dev87}). The high data quality helps to rule out the degradation possibility.
%We have made dual-gated 6SL MnBi$_2$Te$_4$ device to directly test the influence of $E$ field. XXX
\end{itemize}

Based on these systematic data, we conclude that the uncompensated magnetization induced magnetic CD is not the dominant effect in the even-layered MnBi$_2$Te$_4$.

\subsection*{I.3. Light attenuation induced residual CD}

\begin{figure*}[h]
\centering
\includegraphics[width=10cm]{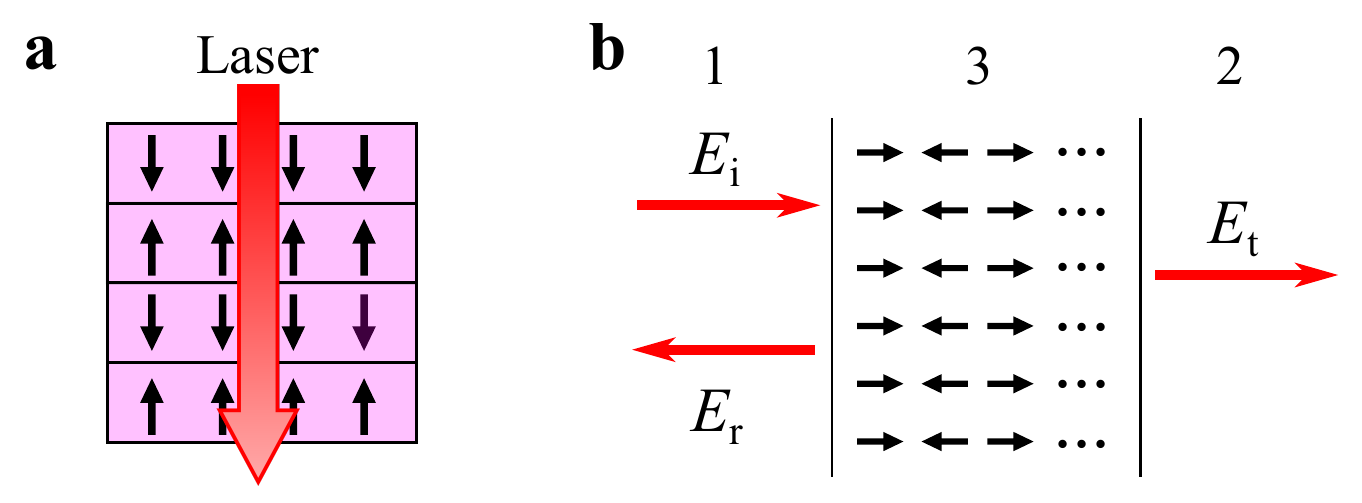}
\caption{\textbf{a,} Schematic illustration for the light attenuation induced residual CD. \textbf{b,} A layout to theoretically compute the light attenuation induced residual CD (see SI.I.3.2).}
\label{Attenuation_schematics}
\end{figure*}

\hspace{4mm} We now consider another important possibility based on the magnetic CD: The even-layered AFM is fully compensated, but the magnetic CD signals coming from opposite spin layers may not cancel, because the light intensity decays due to absorption (Fig.~\ref{Attenuation_schematics}\textbf{a}). We refer to this mechanism as light attenuation induced residual CD.

% Each layer has opposite $M$ and thus gives opposite magnetic CD, but light intensity decays due to absorption. So the total magnetic CD from all layers may not cancel. 

\subsubsection*{I.3.1. Weak absorption in 2D MnBi$_2$Te$_4$ flakes}

\begin{figure*}[h]
\centering
\includegraphics[width=12cm]{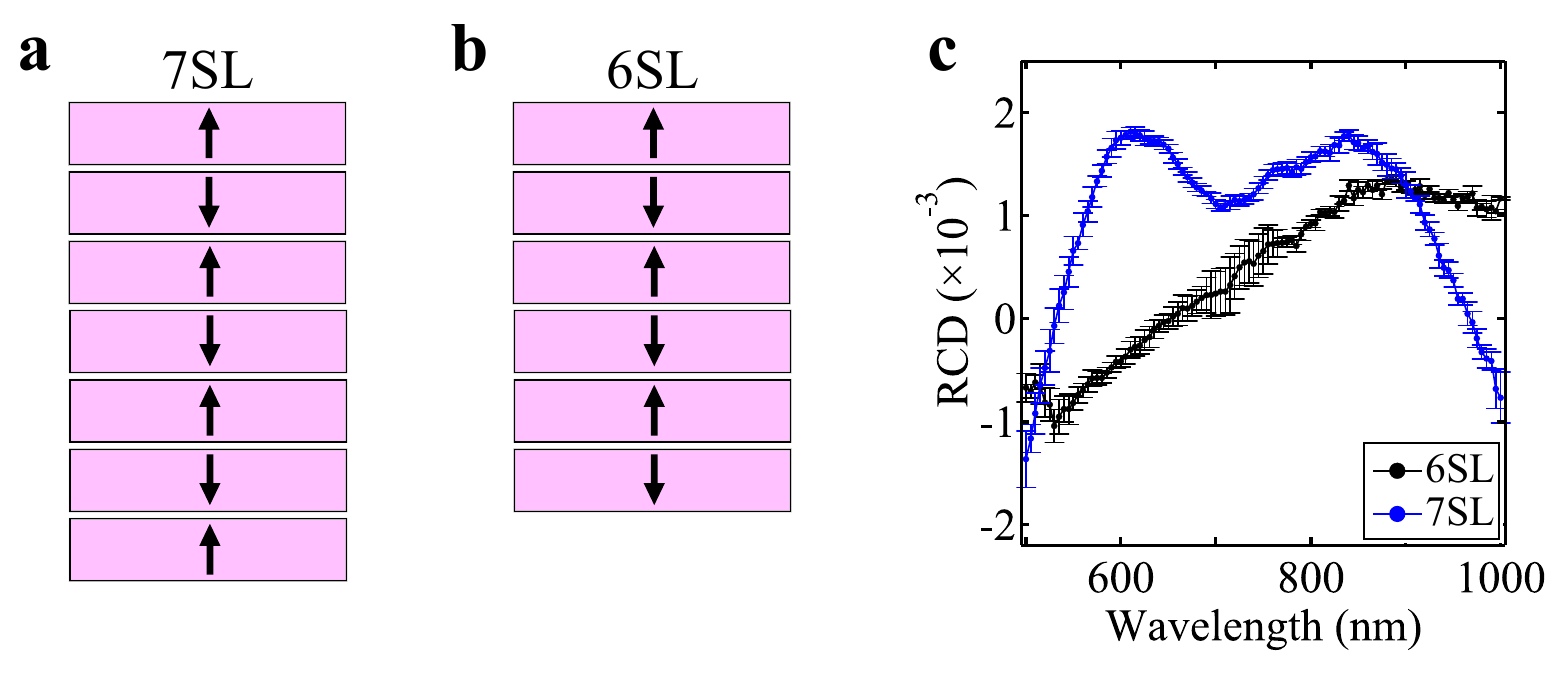}
\caption{Comparison between the RCD in 6SL and 7SL at $B=0$.}
\label{7SL_6SL_comparison}
\end{figure*}

\hspace{4mm} As shown in Fig.~\ref{7SL_6SL_comparison}, the amplitude of the CD in 6SL is similar to that in 7SL. If we purely use the light attenuation induced residual CD mechanism to explain, then the result would only make sense in the very strong absorption limit. For instance, let us assume that the top layer absorbs all light. This means that the CD only probes the top layer, so the number of layers does not matter.

\begin{figure*}[htb]
\centering
\includegraphics[width=16cm]{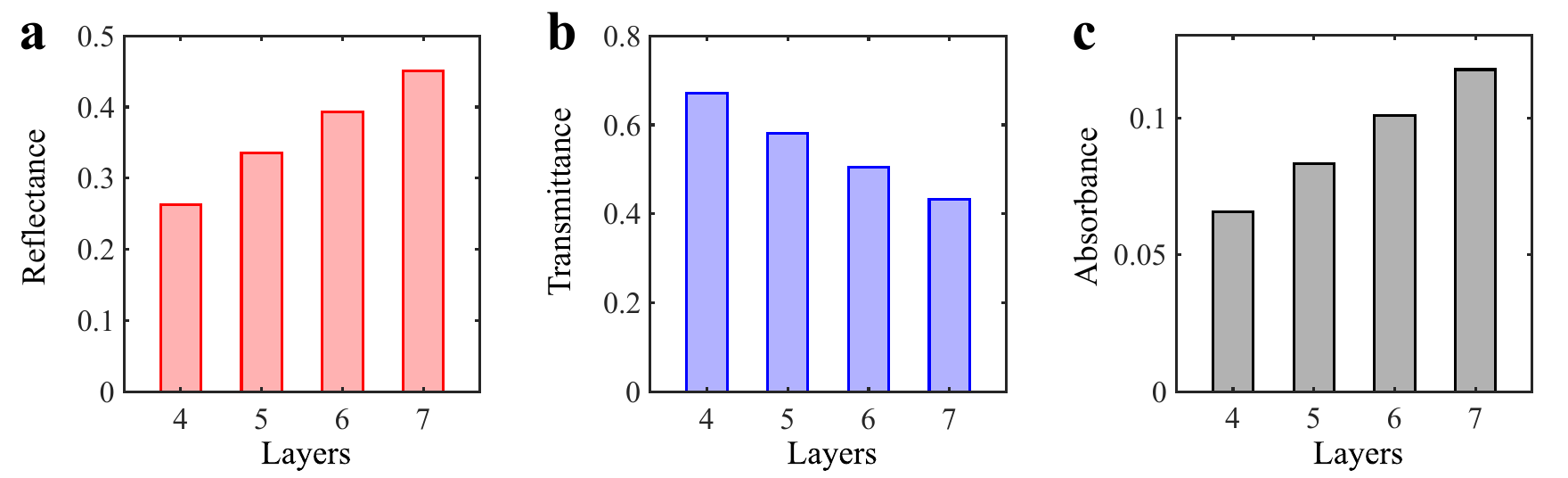}
\caption{Reflectance, transmittance and absorbance of 2D MnBi$_2$Te$_4$ flakes.}
\label{RTA}
\end{figure*}

In sharp contrast, the actual absorption in 2D MnBi$_2$Te$_4$ is weak. Based on our direct measurements (Fig.~\ref{RTA}), each MnBi$_2$Te$_4$ layer absorbs only $\sim1.7\%$ light. This means that light probes the 2D MnBi$_2$Te$_4$ flake as a whole, rather than only the top layer. With a total absorbance of $\sim10\%$ in 6SL or 7SL, it is highly unlikely that the residual CD in 6SL is similarly strong as the magnetic CD in 7SL. 

\begin{figure*}[h]
\centering
\includegraphics[width=12cm]{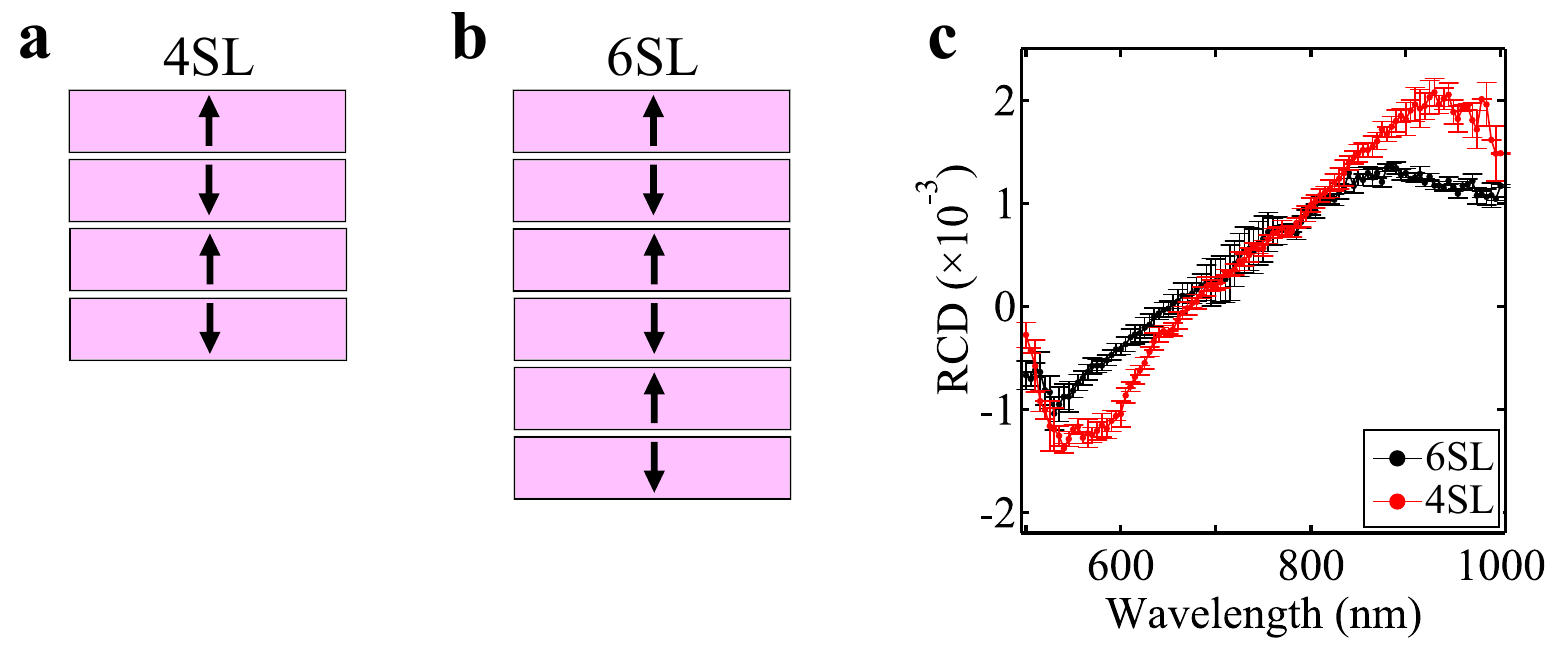}
\caption{Comparison between the RCD in 6SL and 4SL at $B=0$.}
\label{4SL_6SL_comparison}
\end{figure*}

We have further studied the CD in 4SL, which is found to be even stronger than 6SL (Fig.~\ref{4SL_6SL_comparison}) but the total absorbance in 4SL is as low as $\sim6\%$. These studies show that the light attenuation induced residual CD cannot explain our experiments. 

Notes: We determined the absorbance of MnBi$_2$Te$_4$ in two independent ways. First, as shown in Fig.~\ref{RTA}, we measured the reflectance and transmittance  simultaneously in our 2D MnBi$_2$Te$_4$ flakes, and obtain the absorbance. Second, a recent FTIR experiment \cite{xu2021infrared} measured the optical conductivity of bulk MnBi$_2$Te$_4$, based on which we can obtain the light attenuation coefficient and therefore the averaged absorbance per layer. Both methods yield the consistent result , i.e., $\sim1.7\%$ absorbance per layer.

\subsubsection*{I.3.3. Quantitative estimation of the light attenuation induced residual CD}

\hspace{4mm} To make our conclusion above more precise, we aim to quantitatively calculate the magnitude of the light attenuation induced residual CD. In fact, this attenuation induced residual CD effect has been theoretically derived in Ref. \cite{dzyaloshinskii1995nonreciprocal} by expanding the off-diagonal part of the dielectric tensor $\epsilon_{xy}$ to the first order in $\frac{d}{\lambda}$ (the interlayer distance $d\simeq1$ nm) following the standard light scattering theory \cite{landau2013electrodynamics}. $\epsilon_{xy}(z)=\bar{\gamma} d \sum_{n=1}^{N-1}(-)^n\delta(z-nd)$, where $\bar{\gamma}$ represents the rotation power for a single FM layer and $N$ is the number of layers. The $\mathbf{E}$ field in the entire space can be obtained by solving the following equation: $(\epsilon\frac{4\pi^2}{\lambda^2}-\textrm{curl curl})\mathbf{E}=-\frac{4\pi^2}{\lambda^2}\epsilon_{xy}(z)\mathbf{E}_3$, where $\epsilon=\epsilon_{xx}=\epsilon_{yy}$ is the diagonal part of the dielectric tensor, $\mathbf{E}$ and $\mathbf{E}_3$ are the electric fields outside and inside the AFM (see layout in Fig.~\ref{Attenuation_schematics}\textbf{b}). This was solved in Ref. \cite{dzyaloshinskii1995nonreciprocal}. In the thin film limit (number of layer $N$ is small), the magnetic CD and attenuation induced residual CD can be both expressed by the RCD of a single FM layer, 
%i.e., $\textrm{RCD}^{\textrm{1SL}}=\frac{\gamma}{\epsilon-1}$. 
%Using $\textrm{RCD}^{\textrm{1SL}}$ as a reference, Ref. \cite{dzyaloshinskii1995nonreciprocal} has the following conclusions:

%the RCD in odd and even layers are as follows:

\begin{align}
\label{M_CD}\textrm{Magnetic CD} &=\frac{1}{N}\textrm{RCD}^{\textrm{1SL}} \quad\quad N\textrm{ is odd}\\
\label{M_CD_even}&=0 \qquad\qquad\quad\quad N\textrm{ is even}\\
\textrm{attenuation induced residual CD}&=2\pi \frac{d}{\lambda}\textrm{RCD}^{\textrm{1SL}} \quad \textrm{both odd and even}
\label{Attenuation_CD}
\end{align} 

We know that the magnetization $M$ in odd layers is inversely proportional to the layer number $N$, and the $M$ in even layers is zero. Indeed, from From Eqs.~\ref{M_CD} and~\ref{M_CD_even}, we see that the magnetic CD behaves in the same way, consistent with the expectation that the magnetic CD is proportional to $M$. On the other hand, for the attenuation induced residual CD, we see that the relevant quantity is $\frac{d}{\lambda}\sim\frac{1}{1000}$ from Eq.~\ref{Attenuation_CD}, which is very small ($d\simeq1$ nm and typical $\lambda=1000$ nm). In fact, the factor of $\frac{d}{\lambda}$ can be intuitively derived by the following picture: as light propagates and attenuates, each spin layer feels a spatially varying electric field $E(z)=E_0e^{ik_zz}$ ($k_z$ is the light wave-vector). By considering the reflection from each spin layer, the ratio between attenuation induced residual CD and $\mathrm{RCD^{1SL}}$ can be estimated as $(1-e^{2ik_zd}+e^{4ik_zd}-e^{6ik_zd}+\hdots)/(1+e^{2ik_zd}+e^{4ik_zd}+e^{6ik_zd}+\hdots)\propto\frac{d}{\lambda}$. We could now estimate the attenuation induced residual CD based on our own experimental data. Using Eqs.~\ref{Attenuation_CD} and \ref{M_CD}, we can express the light attenuation induced residual CD by the measured magnetic CD in 7SL. By using $d=1$, $\lambda=1000$ nm, and $N=7$, we obtain $\textrm{attenuation induced residual CD}\sim 4\%$ of the magnetic CD in 7SL. By contrast, our 6SL RCD experimental data is similarly large as 7SL (Fig.~\ref{7SL_6SL_comparison}).

\subsection*{I.4. Higher order effects}

\hspace{4mm} We now explore the possibility of CD due to higher order effects  in electromagnetic fields, which in general can arise from many microscopic mechanisms. However, a unifying property is that the CD amplitude due to higher order effects is expected to depend strongly on the optical power. To be more specific, let us consider the following mechanism as an example: The even-layered AFM is fully compensated, but light can generate a nonzero static $M$ (e.g. the inverse Faraday effect), which in turn leads to a magnetic CD. In this case, the CD is expected to be proportional to the laser power. This is quite intuitive: Without light, the material has $M=0$. Light induces $M$. As such, stronger light means larger light-induced $M$, which in turn results in stronger CD.

By contrast, the CD processes we considered in the main text are linear optical effects, i.e., the amplitude of CD is independent of the optical power. This can be understood by the fact that they are direct measures of the material's properties. For example, the Axion CD is proportional to the material's ME coupling; the magnetic CD is proportional to the material's $M$; the natural optical activity is proportional to the material's spatial chirality. Therefore, their magnitude is independent of the optical power. %By contrast, the light-induced phenomenon considered in this section should depend strongly on the laser power. Because these CD signals are direct measures of the material's properties

We have carefully studied how CD depends on the opitcal power. As shown in Fig.~\ref{Power_dep_5SL6SL}, our CD in both 5SL and 6SL MnBi$_2$Te$_4$ is clearly independent of optical power, therefore ruling out the possibility of higher order effects. 

\clearpage
\section*{II. Additonal data}
\subsection*{II.1. Additional CD data}

\begin{figure*}[!htb]
\centering
\includegraphics[width=16cm]{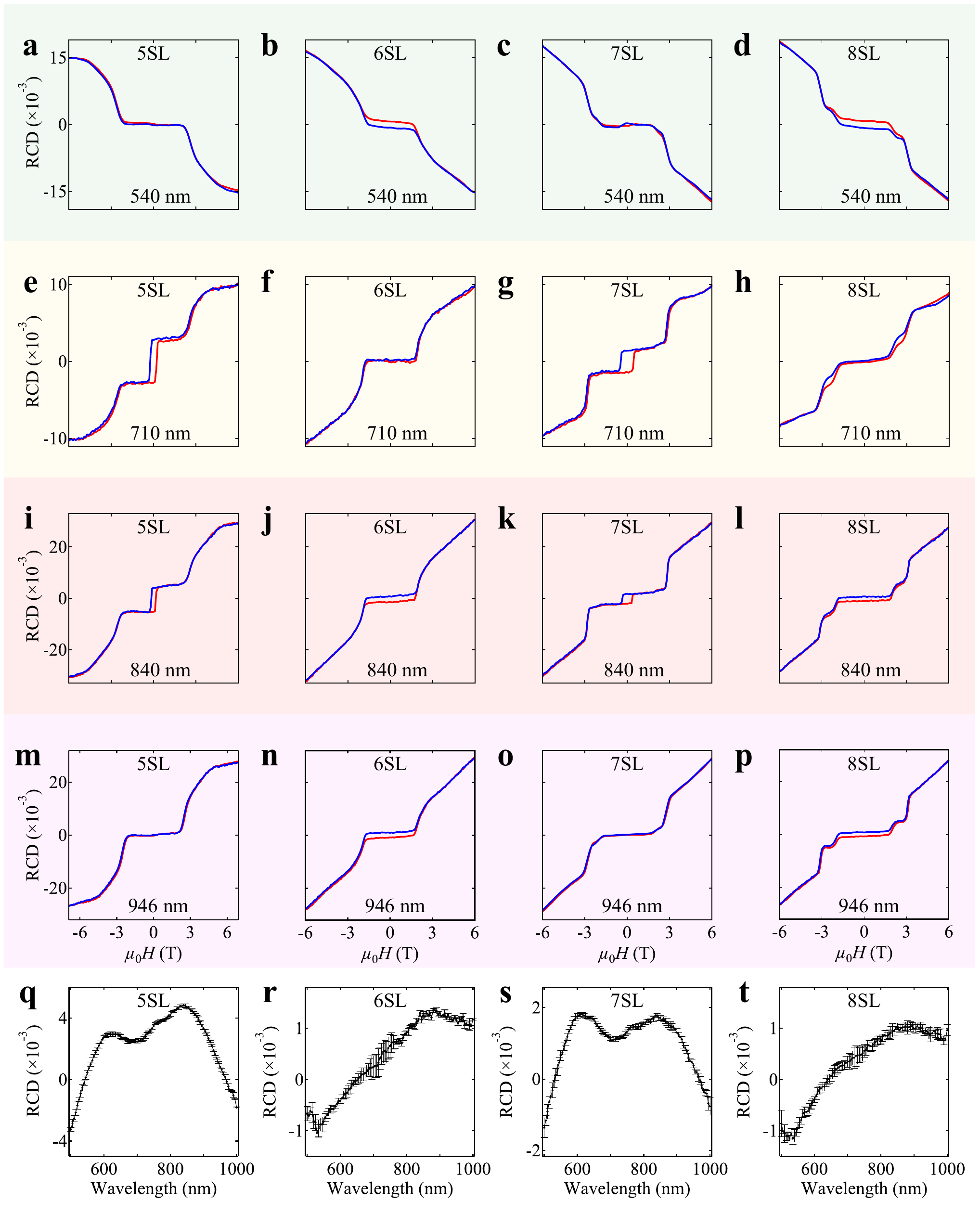}
\caption{\textbf{a-p,} Magnetic hystereses of RCD for 5SL, 6SL, 7SL and 8SL measured using different wavelengths. \textbf{q-t,} RCD spectra for 5SL, 6SL, 7SL and 8SL.}
\label{RCD_540nm_840nm_hysteresis}
\end{figure*}

\hspace{4mm}\textbf{ RCD magnetic hystereses at multiple wavelengths:} To better understand the RCD signals of the $\mathrm{MnBi_2Te_4}$ system, in Figs.~\ref{RCD_540nm_840nm_hysteresis}\textbf{a-p}, we show the RCD magnetic hystereses for 5SL-8SL at multiple wavelengths, including $540$ nm, $710$nm, $840$ nm and $946$ nm. These data are all consistent with the RCD spectra shown in Figs.~\ref{RCD_540nm_840nm_hysteresis}\textbf{q-t}. \\

%\hspace{3mm}\textbf{ RCD map data under zero field cooling:} As shown in Fig.~\ref{RCD_NV_map}\textbf{a}, the NV stray field maps  shows that only 5SL and 7SL have magntization, and their magnetization is opposite under zero field cooling. For consistency, we also measured the RCD map under zero field cooling with $\mathrm{\lambda_{detection}=650\ nm}$ (Fig.~\ref{RCD_NV_map}\textbf{b}), which shows that 5SL and 7SL have opposite RCD signals.  Compared with the RCD map under field cooling shown in Fig.~\ref{RCD_NV_map}\textbf{c}, we can conclude that RCD maps also demonstrates the opposite magnetizations for 5SL and 7SL under zero field cooling.\\

\begin{figure*}[!htb]
\centering
\includegraphics[width=12cm]{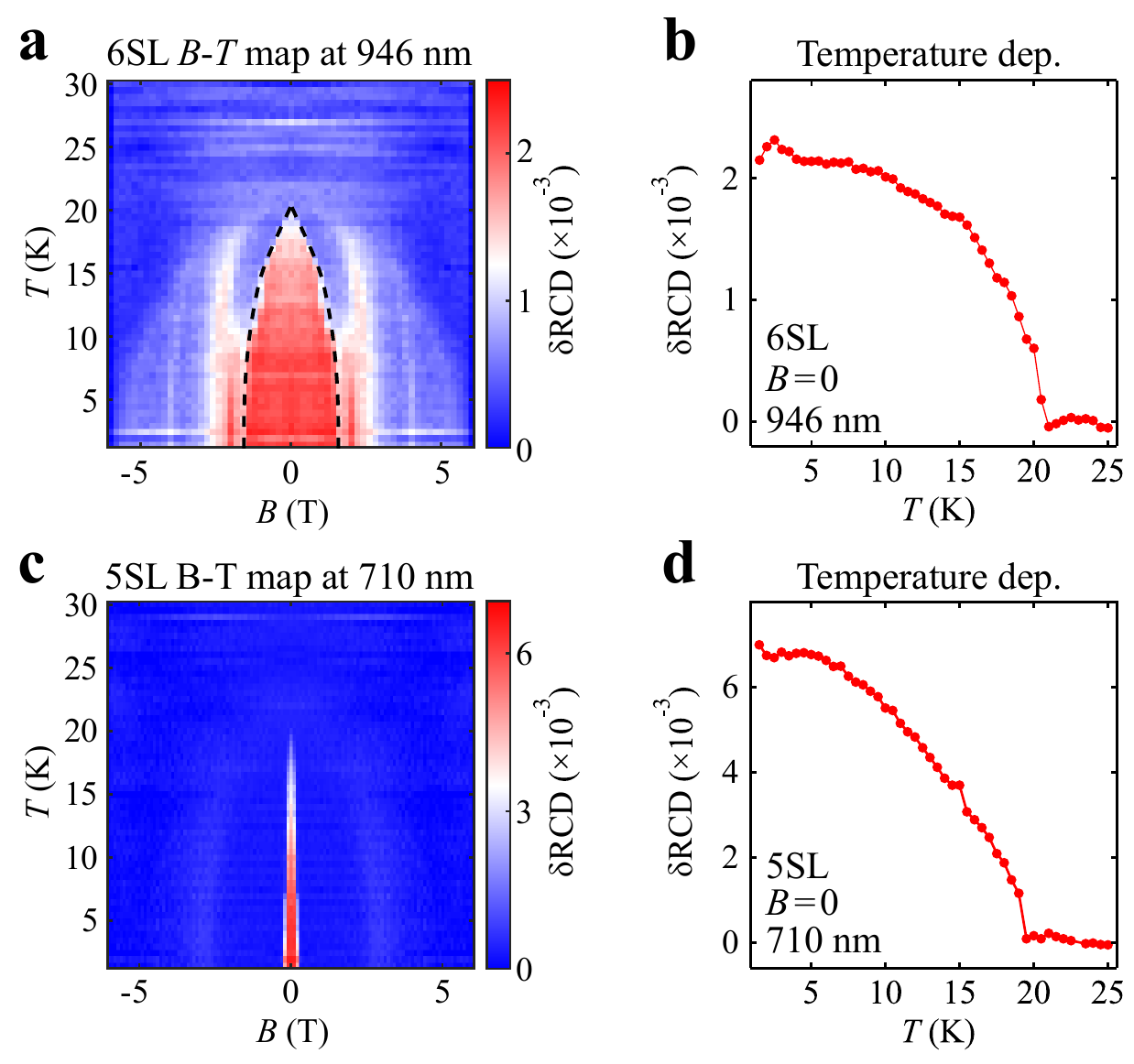}
\caption{$\delta\textrm{RCD}(B)=\textrm{RCD}^{\textrm{Backward}}(B)-\textrm{RCD}^{\textrm{Forward}}(B)$. \textbf{a,b,} Temperature $T$ and magnetic field $B$ dependence of $\delta\textrm{RCD}$ in 6SL at $\lambda_{\textrm{detection}}=946$ nm. The Axion CD near $B=0$ is clearly observed. \textbf{c,d,} Same as panels (\textbf{a,b}) but for 5SL at $\lambda_{\textrm{detection}}=710$ nm. The magnetic CD near $B=0$ is clearly observed.  }
\label{RCD_BT_map_5SL_6SL}
\end{figure*}

\textbf{Temperature dependence of RCD data:}  In order to highlight the nonzero RCD signal at $B=0$, we define $\delta\textrm{RCD}(B)=\textrm{RCD}^{\textrm{Backward}}(B)-\textrm{RCD}^{\textrm{Forward}}(B)$ (Backward/Forward refer to the scanning direction of the magnetic hysteresis). Figures~\ref{RCD_BT_map_5SL_6SL}\textbf{a,b} show the $T$ and $B$ dependence of $\delta\textrm{RCD}$ in 6SL, where the Axion CD near $B=0$ is clearly observed. Figures~\ref{RCD_BT_map_5SL_6SL}\textbf{c,d} show the $T$ and $B$ dependence of $\delta\textrm{RCD}$ in 5SL, where the magnetic CD near $B=0$ is clearly observed. \\

\begin{figure*}[!htb]
\centering
\includegraphics[width=12cm]{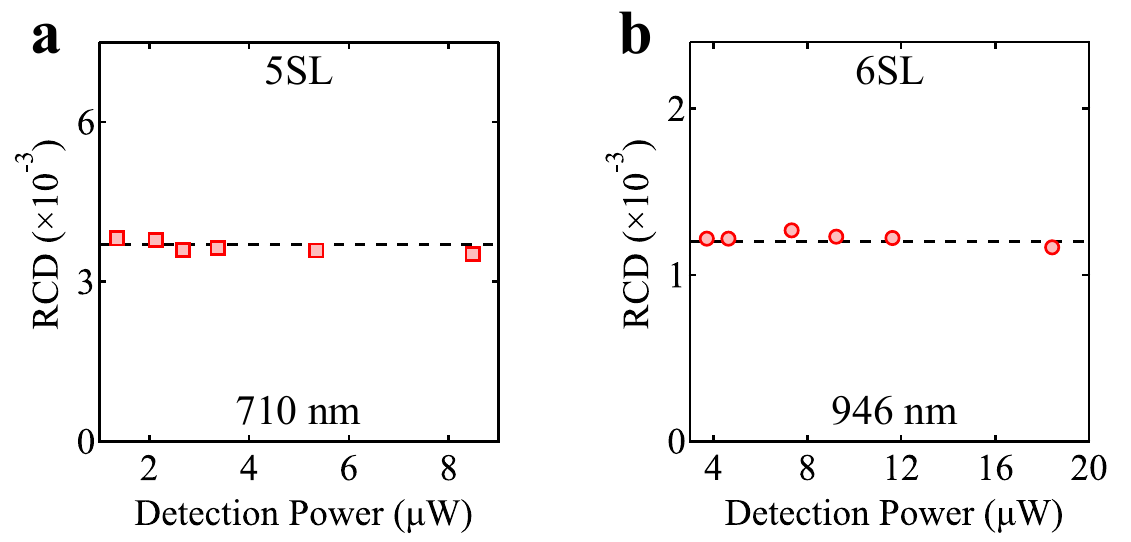}
%\vspace{-5.5mm}
\caption{\textbf{a}, Optical power dependence of the 5SL RCD signal ($\mathrm{\lambda_{detection}=710\ nm}$). \textbf{b}, Detection power dependence of the 6SL RCD signal ($\mathrm{\lambda_{detection}=946\ nm}$). }
\label{Power_dep_5SL6SL}
\end{figure*}

\textbf{Optical power dependence of the RCD data:} As shown in Fig.~\ref{Power_dep_5SL6SL}, the RCD data for both 5SL and 6SL are independent of the detection optical power, which shows that both the magnetic CD and Axion CD are linear optical effects. \\

\begin{figure*}[!htb]
\centering
\includegraphics[width=16cm]{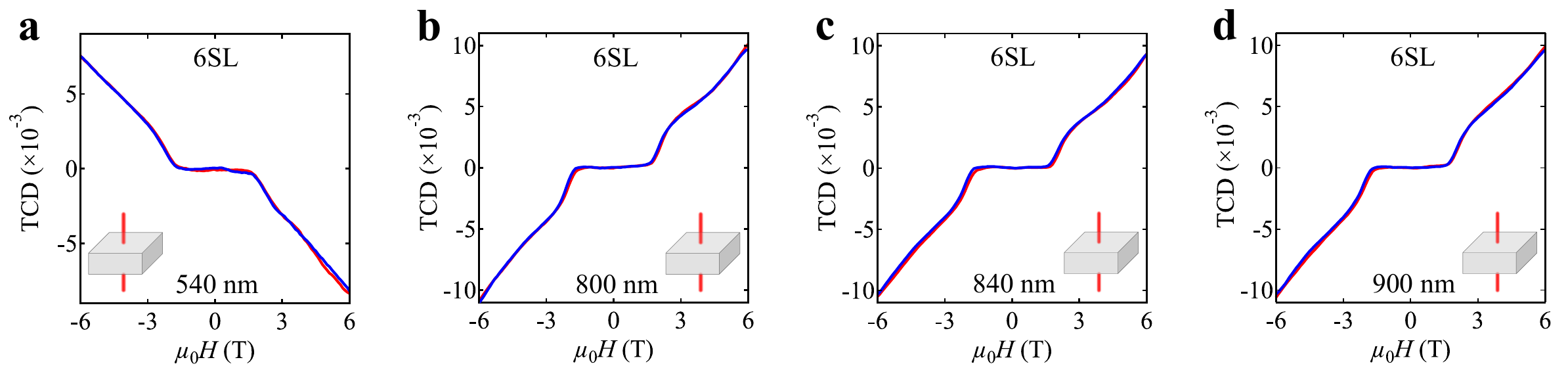}
%\vspace{-5.5mm}
\caption{\textbf{a-d}, Magnetic hystereses of TCD for 6SL measured at 540 nm, 800 nm, 840 nm and 900 nm. }
\label{TCD_6SL_multiple}
\end{figure*}

\begin{figure*}[!htb]
\centering
\includegraphics[width=16cm]{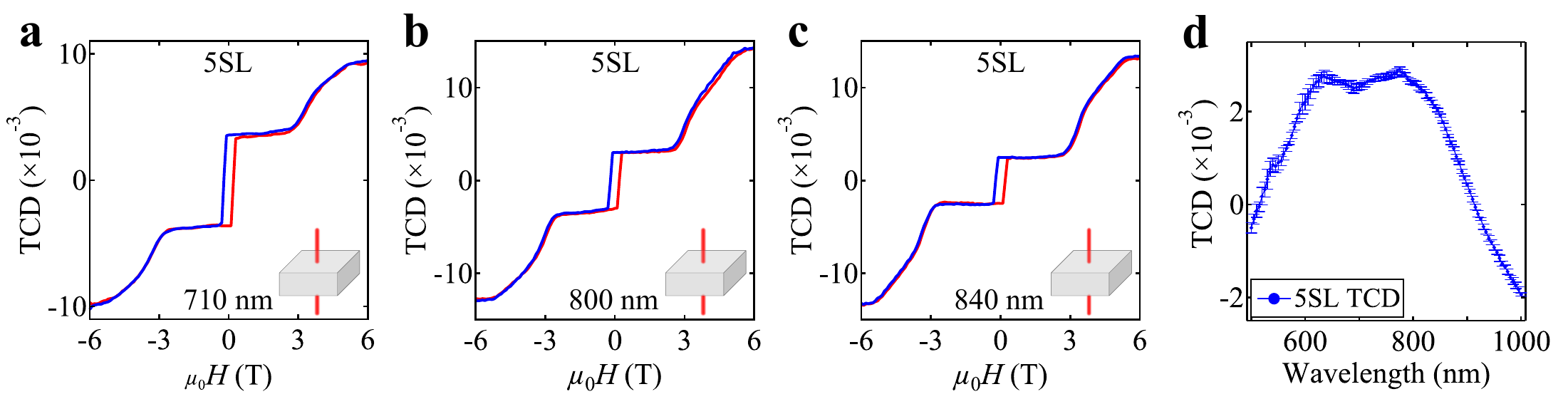}
%\vspace{-5.5mm}
\caption{\textbf{a-c}, Magnetic hystereses of TCD for 5SL measured at 710 nm, 800 nm and 840 nm. \textbf{d}, TCD spectrum at $B=0$ for 5SL. }
\label{TCD_spectra_5SL}
\end{figure*}

\textbf{Additional TCD magnetic hysteresis and spectrum:} In the main text, we showed that the TCD signals of 6SL $\mathrm{MnBi_2Te_4}$ vanish at $B=0$. To substantiate this conclusion, we measured TCD magnetic hystereses for 6SL at multiple wavelengths (540 nm, 800 nm, 840 nm and 900 nm, Fig.~\ref{TCD_6SL_multiple}), which all show vanishing TCD signals at $B=0$.  On the other hand, we also measured TCD magnetic hystereses for 5SL at multiple wavelengths (800 nm, 840 nm and 900 nm, Fig.~\ref{TCD_spectra_5SL}\textbf{a-c}), which further demonstrate that magnetic CD exists in transmission. Additionally,  Fig.~\ref{TCD_spectra_5SL}\textbf{d} shows the 5SL TCD spectrum, which is quite similar to the 5SL RCD spectrum shown in Fig.~\ref{RCD_540nm_840nm_hysteresis}\textbf{q}. \\

\begin{figure*}[!htb]
\centering
\includegraphics[width=14cm]{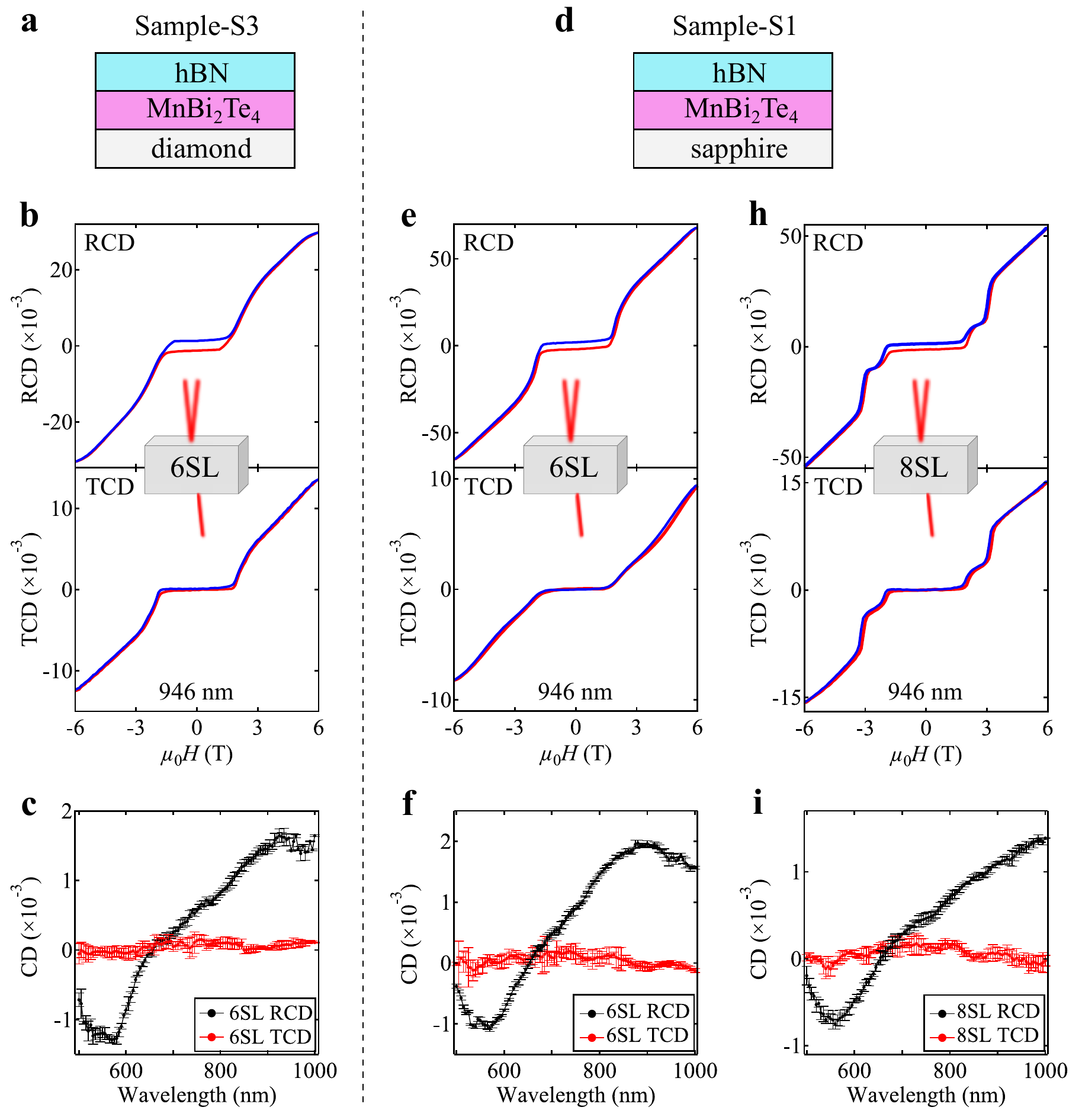}
%\vspace{-5.5mm}
\caption{\textbf{Simultaneous RCD and TCD measurements in both samples-S3 and S1. a-c,} Sample schematic (panel \textbf{a}), simultaneous RCD and TCD magnetic hystereses using $\mathrm{\lambda_{detection}}=946$ nm (panel \textbf{b}), and RCD and TCD spectra at $B=0$ (panel \textbf{c}) for the 6SL flake of sample-S3 on diamond substrate. \textbf{d-i} Same as panels \textbf{a-c} but for the 6SL and 8SL flakes of sample-S1 on sapphire substrate. }
\label{TCD_RCD_data}
\end{figure*}

\textbf{Reproducibility of simultaneous TCD and RCD measurements:} One of our most crucial observations is that the Axion CD only shows up in reflection but is absent in transmission (i.e., nonzero $\textrm{RCD}$ but zero $\textrm{TCD}$). This conclusion was found to be highly reproducible. First, we performed measurements of 6SL flakes on different substrates (see Figs.~\ref{TCD_RCD_data}\textbf{b,e}). Independent of the substrates, we found nonzero $\textrm{RCD}$ but zero $\textrm{TCD}$. Second, on the same substrate, we performed measurements on flakes with different thicknesses (6SL and 8SL, see Figs.~\ref{TCD_RCD_data}\textbf{e,h}). For both 6SL and 8SL, we found  nonzero $\textrm{RCD}$ but zero $\textrm{TCD}$. These systematic results further demonstrate the unique transmission and reflection properties of Axion CD. \\

\begin{figure*}[!htb]
\centering
\includegraphics[width=14cm]{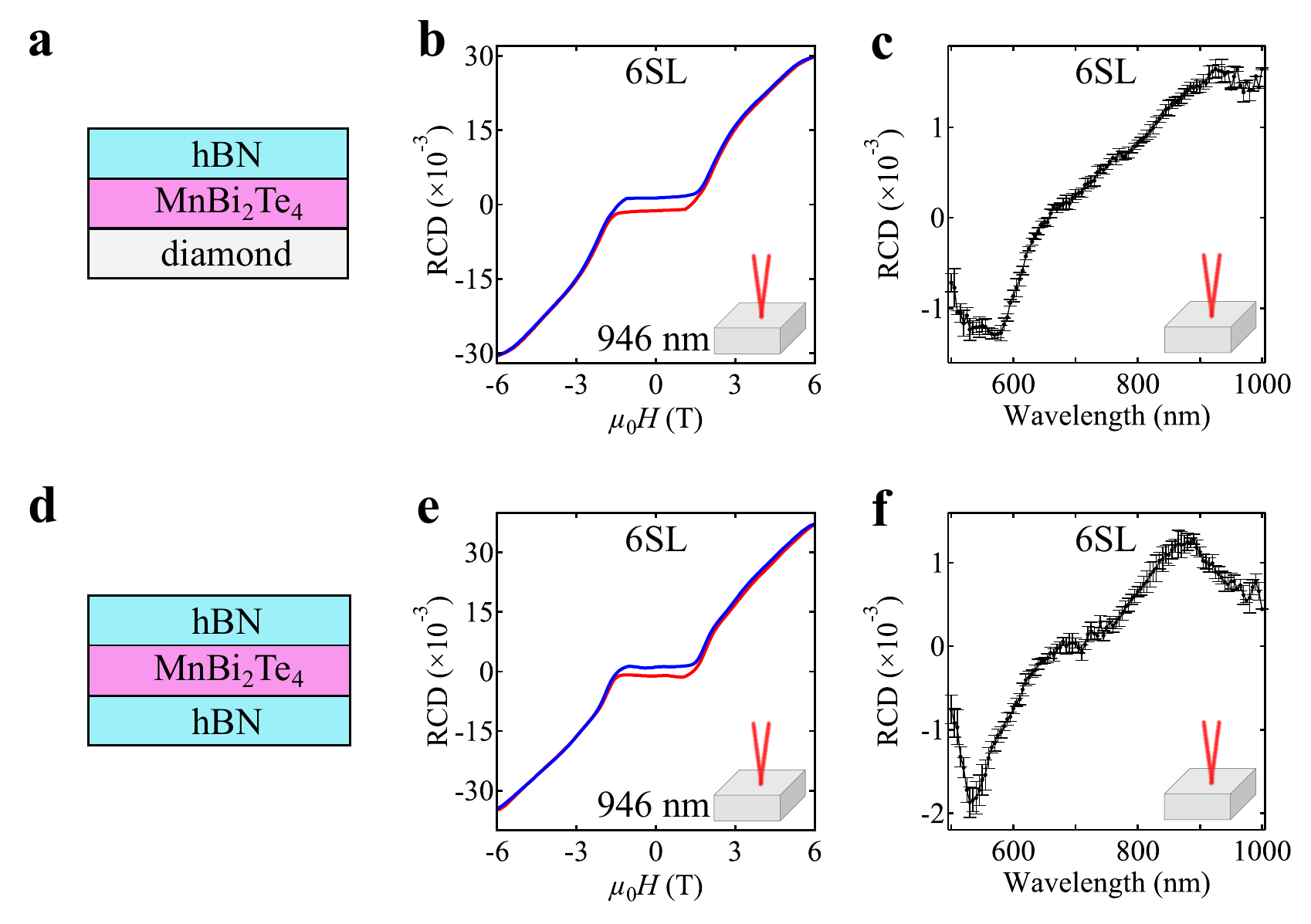}
%\vspace{-5.5mm}
\caption{{\bf Comparison between sample-S2 on diamond and sample-S4 encapsulated by hBN.} \textbf{a-c,} For the sample-S2 on diamond substrate. \textbf{a,} Sample schematic, \textbf{b,} RCD magnetic hysteresis at $946$ nm, \textbf{c,}  RCD spectrum at $B=0$. \textbf{d-f,} Same as panels (\textbf{a-c}) but for sample-S4 encapsulated by hBN. }
\label{hBN_encapsulation}
\end{figure*}

\textbf{RCD signal in hBN-encapsulated $\mathrm{MnBi_2Te_4}$ sample:} In the main text, we discussed the possibility of uncompensated magnetization due to a built-in electric field induced by asymmetric dielectric environment, which is an alternative mechanism for our observed RCD signal. To further rule out this possibility, here we show the RCD signals measured in a hBN-encapsulated sample (S4). The hBN-encapsulated sample is supposed to give minimal built-in electric field. As shown in Fig.~\ref{hBN_encapsulation}, the RCD hysteresis and spectra of hBN-encapsulated sample is very similar to that of sample-S3, which has diamond on bottom and hBN on top. This is contradictory to the possibility of built-in electric field induced magnetization. \\

%\textbf{Layer dependence of reflectance, transmittance and absorbance} Fig.~\ref{RTA} shows the layer dependence of reflectance, transmittance and absorbance for $\mathrm{MnBi_2Te_4}$ flakes, which are all linear to the layer number. This data helps us to rule out the trivial mechanism of light-attenuation induced mechanism as discussed above in SI.I.3. 

\color{black}

\begin{figure*}[!htb]
\centering
\includegraphics[width=14cm]{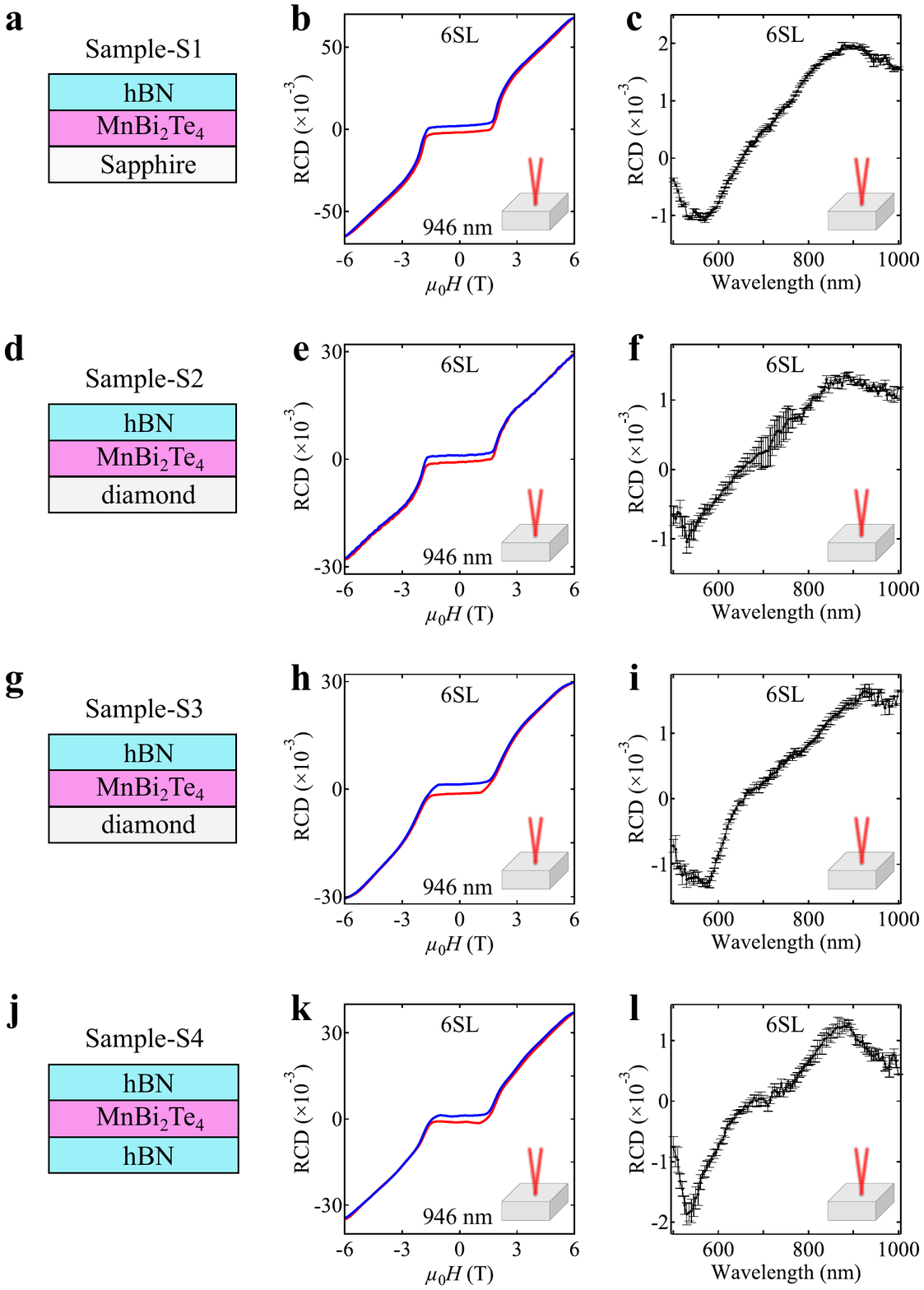}
%\vspace{-5.5mm}
\caption{\textbf{Reproducible observations of the Axion CD in multiple samples on different substrates. a-c}, Sample schematic (panel \textbf{a}), RCD magnetic hysteresis using $\mathrm{\lambda_{detection}}=946$ nm (panel \textbf{b}), and RCD spectrum at $B=0$ (panel \textbf{c}) for Sample-S1. Similar results for sample-S2 (panels \textbf{d-f}), sample-S3 (panels \textbf{g-i}) and sample-S4 (panels \textbf{j-l}). }
\label{all_sample_RCD}
\end{figure*}

\textbf{Reproducibility of Axion CD on different substrates:} In Fig.~\ref{all_sample_RCD}, we summarize all samples mentioned in the main text, which are on different substrates (diamond, sapphire and hBN). Despite the different substrates, all 6SL flakes show highly reproducible Axion CD signals (Figs.~\ref{all_sample_RCD}\textbf{b,e,h,k}). Furthermore, their RCD spectra are quite similar in terms of both amplitude and shape (Figs.~\ref{all_sample_RCD}\textbf{c,f,i,l}). The small variation of the RCD amplitude could come from the different substrate refractive indices. In conclusion, Axion CD is robust and reproducible, and does not rely on a specific type of substrate.\\

\begin{figure*}[!htb]
\centering
\includegraphics[width=14cm]{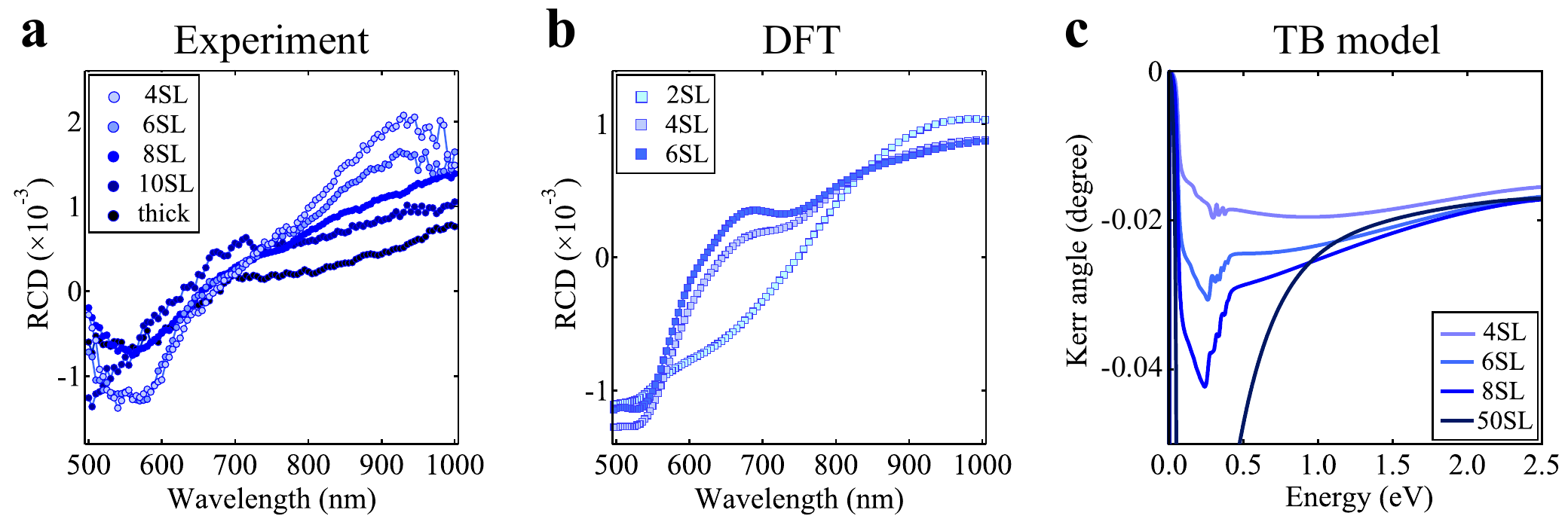}
%\vspace{-5.5mm}
\caption{\textbf{a,} Measured RCD data. \textbf{b,} Calculated RCD based on DFT band structures. \textbf{c,} Calculated complex Kerr angle based on an effective tight-binding model described in Ref. \cite{ahn2022Axion}. }
\label{RCD_thickness}
\end{figure*}

\textbf{Thickness dependence of RCD:} Figure~\ref{RCD_thickness}\textbf{a} shows the measured RCD for 4, 6, 8, 10 SLs and thick samples. RCD were observed in all thicknesses. From our data, we see that the RCD is slightly weaker with increasing thickness and remains finite in the very thick limit. Theoretically, we have computed the RCD for 2, 4, 6 SLs based on DFT band structures. As shown in Fig.~\ref{RCD_thickness}\textbf{b}, the calculated RCD are similarly large. For larger thickness than 6SL, the DFT calculations become too heavy to run. Using a simple tight-binding (TB) effective model \cite{ahn2022Axion}, we can simulate up to 50SL (Fig.~\ref{RCD_thickness}\textbf{c}), which shows that the Kerr effect indeed persists. But we caution that band structures of this TB effective model may not be able to capture the realistic MnBi$_2$Te$_4$ band structure, especially for the high energy states. Nevertheless, the purpose of TB model is to show that optical Axion electrodynamics persists in thick samples.\\

\begin{figure*}[!htb]
\centering
\includegraphics[width=12cm]{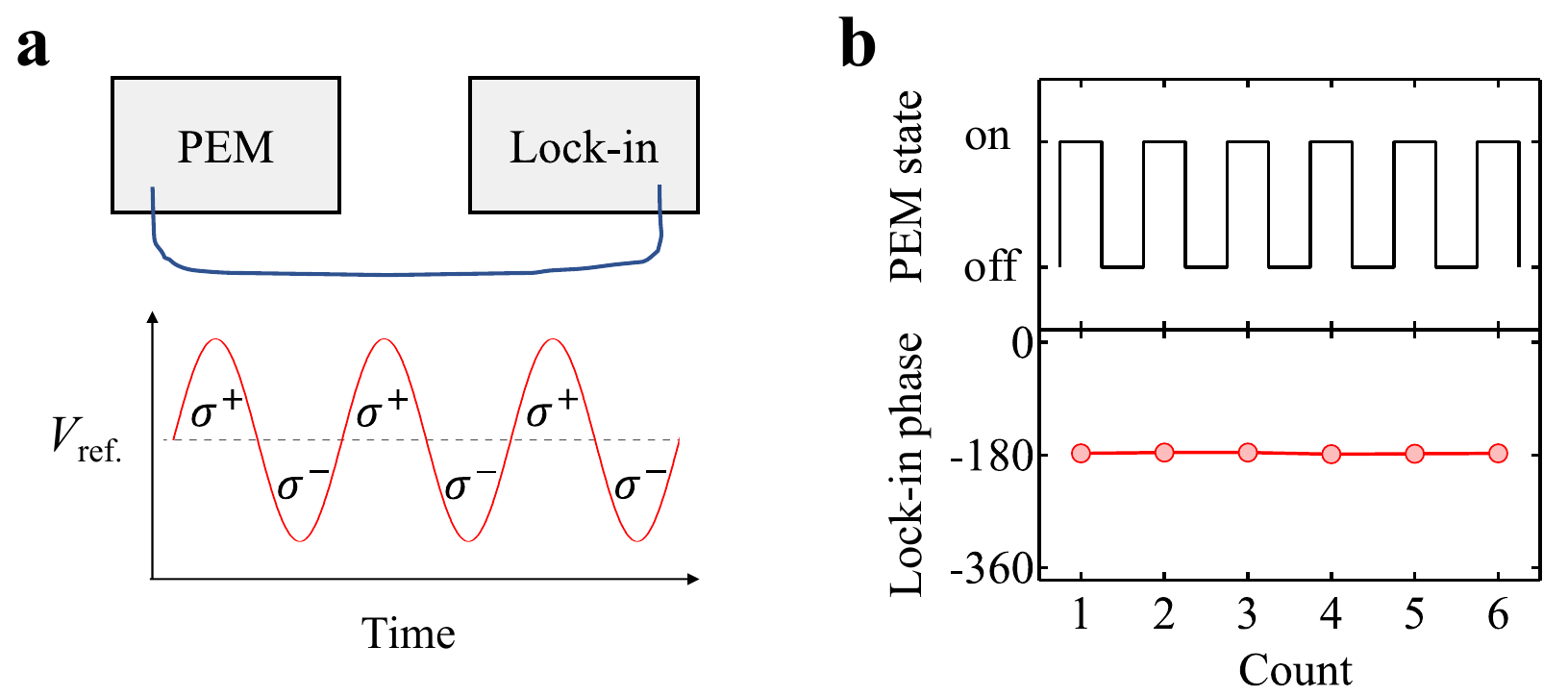}
%\vspace{-5.5mm}
\caption{\textbf{a,} PEM$+$lock-in setup. The PEM reference wave has definitive correspondence with $\sigma{+}$ and $\sigma{-}$ polarizations. \textbf{b,} We parked the beam on a particular spot on an 8SL sample. We turned the PEM off multiple times and measure the lock-in phase (the sign of the CD) every time we turn the PEM back on. }
\label{PEM_phase}
\end{figure*}

\textbf{Consistent sign of the CD:} In order to check if the sign of the CD is consistent every time we turn on the PEM, we have performed the following testing: (1) We measured the CD value of an MnBi$_2$Te$_4$ sample, (2) We turned off the PEM and turned it back on, (3) We redid the same measurement. We repeated the above steps many times. As shown in Fig.~\ref{PEM_phase}\textbf{b}, for every measurement, the sign of the CD (the phase of the lock-in) is invariant. We explain this based on the PEM$+$lock-in setup shown in Fig.~\ref{PEM_phase}\textbf{a}. The PEM output reference wave (also reference for lock-in) definitively corresponds to the $\sigma^+$ and $\sigma^-$ polarizations. This is because the PEM reference wave is synchronized with the piezoelectric voltage, which is used to control the retardation of the optical head. \\

\begin{figure*}[!htb]
\centering
\includegraphics[width=16cm]{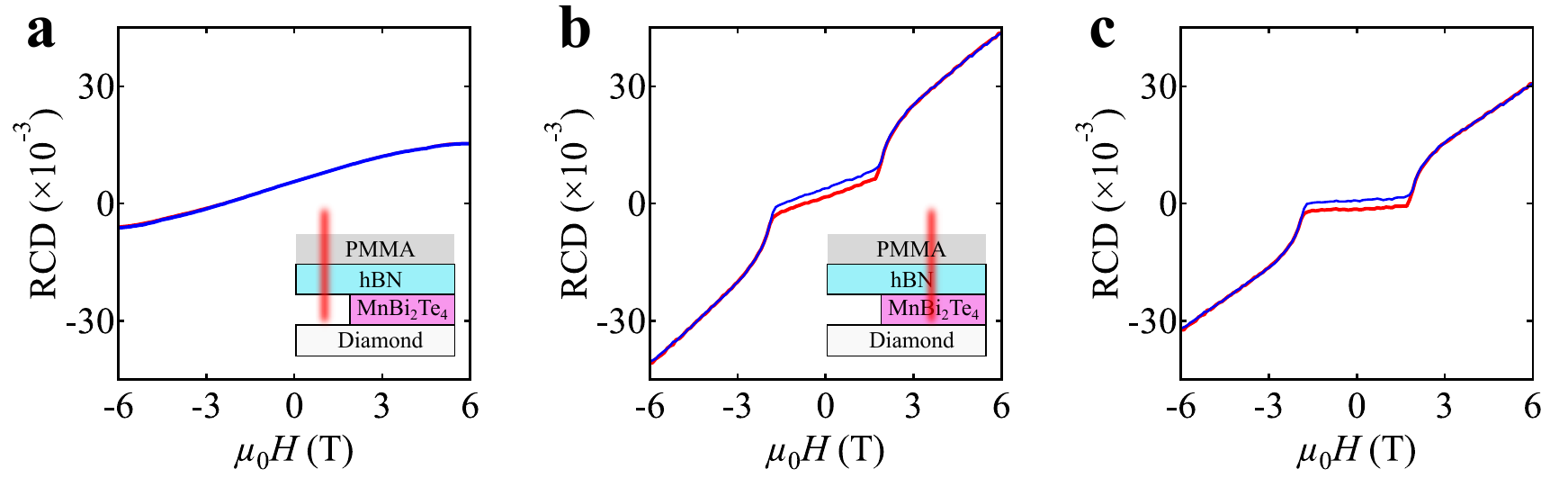}
%\vspace{-5.5mm}
\caption{\textbf{a,b,} Raw CD data when the beam spot is parked on the MnBi$_2$Te$_4$ flake (\textbf{b}) or next to the flake (\textbf{a}). \textbf{c,} CD after we subtract (\textbf{b}) by (\textbf{a}).}
\label{Background}
\end{figure*}

\textbf{Background removal for CD:} It is well-known that beam splitter when used at $\sim 45^{\circ}$ can lead to significant CD offset. In order to minimize the unwanted background, we used a plate beam splitter at near-normal incidence. To check our background level, we measure CD both on the sample (Fig.~\ref{Background}\textbf{b}), and next to the sample (Fig.~\ref{Background}\textbf{a}). We see that the background is in general $\sim5$ times smaller than the signal. We therefore remove the background from the signal by subtracting Fig.~\ref{Background}\textbf{b} with Fig.~\ref{Background}\textbf{a}. \\

\begin{figure*}[!htb]
\centering
\includegraphics[width=6cm]{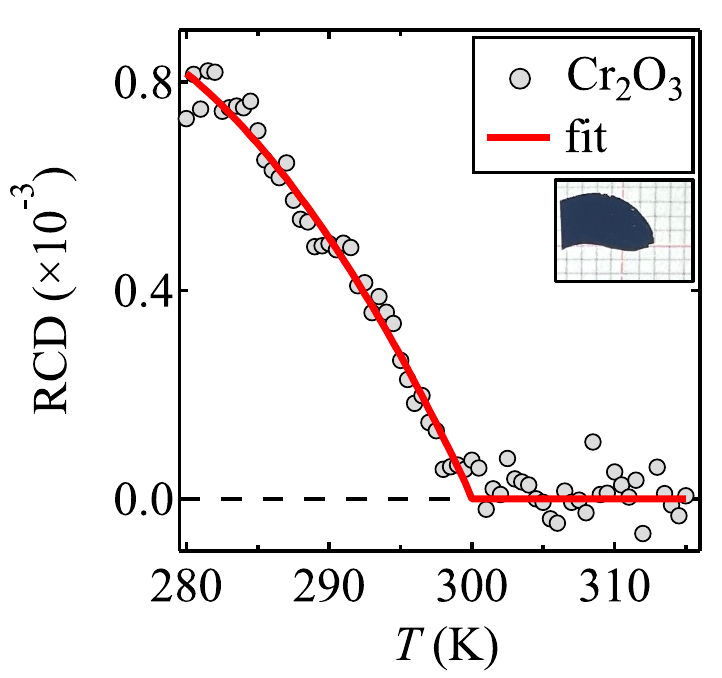}
%\vspace{-5.5mm}
\caption{\textbf{RCD signal of $\mathrm{Cr_2O_3}$ single crystal}. Inset is the image of the $\mathrm{Cr_2O_3}$ crystal (grid: 1 mm). $\lambda_{\textrm{detection}}=540$ nm. }
\label{Cr2O3_CD}
\end{figure*}

\textbf{RCD signal of Cr$_2$O$_3$ single crystal:} In order to substantiate the broad applicability of Axion optical induction, we also measured the RCD signal of $\mathrm{Cr_2O_3}$ (shown in \ref{Cr2O3_CD}). The RCD signal of $\mathrm{Cr_2O_3}$ is consistent with previous pioneering work by Krichevtsov et al\cite{krichevtsov1993spontaneous}.\\

\color{black}

%In our optical experiments (both induction and CD), light is focused on the scale of diffraction limit ($\sim1$ $\mu$m). For the areas that are chosen for induction (see Fig.~\ref{Spatial_dep}), we have carefully studied their RCD signals as a function of $B$-field (by measuring magnetic hysteresis at $2$ K) and $T$ (upon multiple cooldowns without induction light). We found that after cooling down from $30$ K to $2$ K without induction, the RCD always reaches the maximal value; only the sign is random. This suggests that, at least on the length scale of diffraction limit ($\sim1$ $\mu$m), the areas chosen for induction is always in a single-domained state. Therefore, the results of induction are ternary: (1) the induction leads to the AFM state I (positive RCD at $\lambda_{\textrm{detection}}=946$ nm); (2) the induction leads to the AFM state II (negative RCD at $\lambda_{\textrm{detection}}=946$ nm); (3) the induction cannot favor one state, so the results are random just like no induction. 

\pagebreak

\clearpage
\pagebreak
\newpage

\clearpage

%\begin{figure*}[h]
%\centering
%\includegraphics[width=15cm]{./SI_Figures/RTA.pdf}
%\caption{\textbf{Layer dependence of reflectance, transmittance and absorbance of 2D MnBi$_2$Te$_4$ flakes.} \textbf{a-c}, We measured the reflectance ($R$) and transmittance ($T$) simultaneously in our 2D MnBi$_2$Te$_4$ flakes, and obtain the absorbance ($A$) by $A=1-T-R$. }
%\label{RTA}
%\end{figure*}
%\vspace{1cm}

\clearpage

\clearpage
 
\color{black}

\clearpage
\subsection*{II.2. Additional induction data}

\hspace{4mm} \textbf{Default experimental conditions:} In order to systematically study the Axion induction, it is important to keep a consistent set of experimental conditions. In our experiments, we keep the following set of default experimental conditions: $\lambda_{\textrm{induction}}=840$ nm, $P_{\textrm{induction}}\simeq1$ mW; $\lambda_{\textrm{detection}}=946$ nm and $P_{\textrm{detection}}\simeq30$ $\mu$W; induction initial and ending temperatures at $30$ K and $2$ K. When we try to study how the induction depends on a specific parameter, we keep all the other parameters at their default condition and only vary that specific parameter.

%probe of the AFM order. Therefore, in our induction experiments, we fix the detection wavelength $\lambda_{\textrm{detection}}=946$ nm unless noted otherwise.

%\hspace{4mm} \textbf{Detection wavelength is fixed:} In order to systematically study the Axion induction, it is important to have a robust and consistent probe of the AFM order. Therefore, in our induction experiments, we fix the detection wavelength $\lambda_{\textrm{detection}}=946$ nm unless noted otherwise.

\begin{figure*}[!htb]
\centering
\includegraphics[width=12cm]{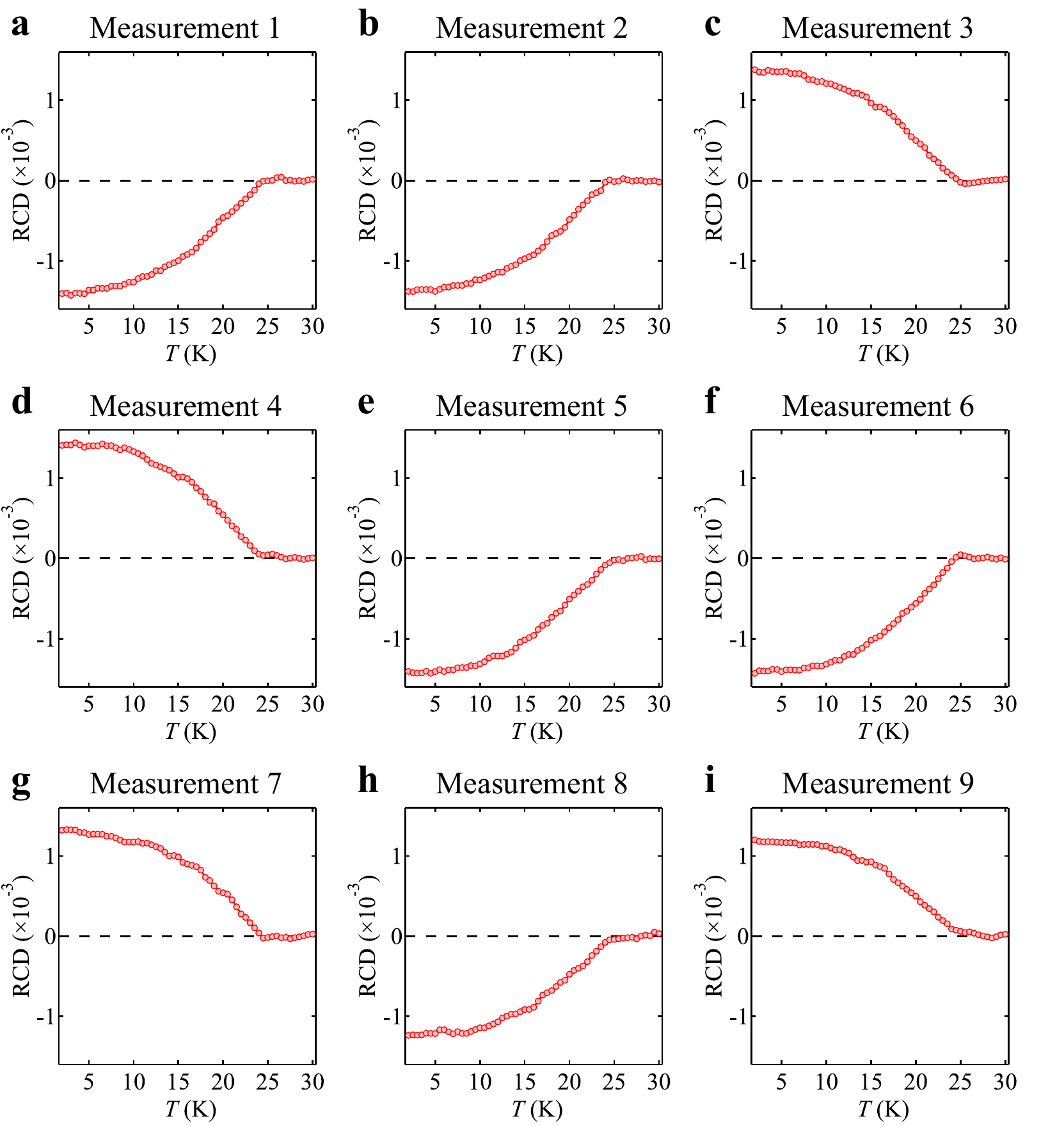}
%\vspace{-5.5mm}
\caption{Consecutive temperature dependent RCD measurements without induction of 8SL MnBi$_2$Te$_4$ in sample-S1 measured at $\lambda_{\textrm{detection}}=946$ nm. These measurements show that cooling down without induction leads to a single-domained state, but the sign is random. }
\label{Cooldown_without_induction}
\end{figure*}

\begin{figure*}[!htb]
\centering
\includegraphics[width=10cm]{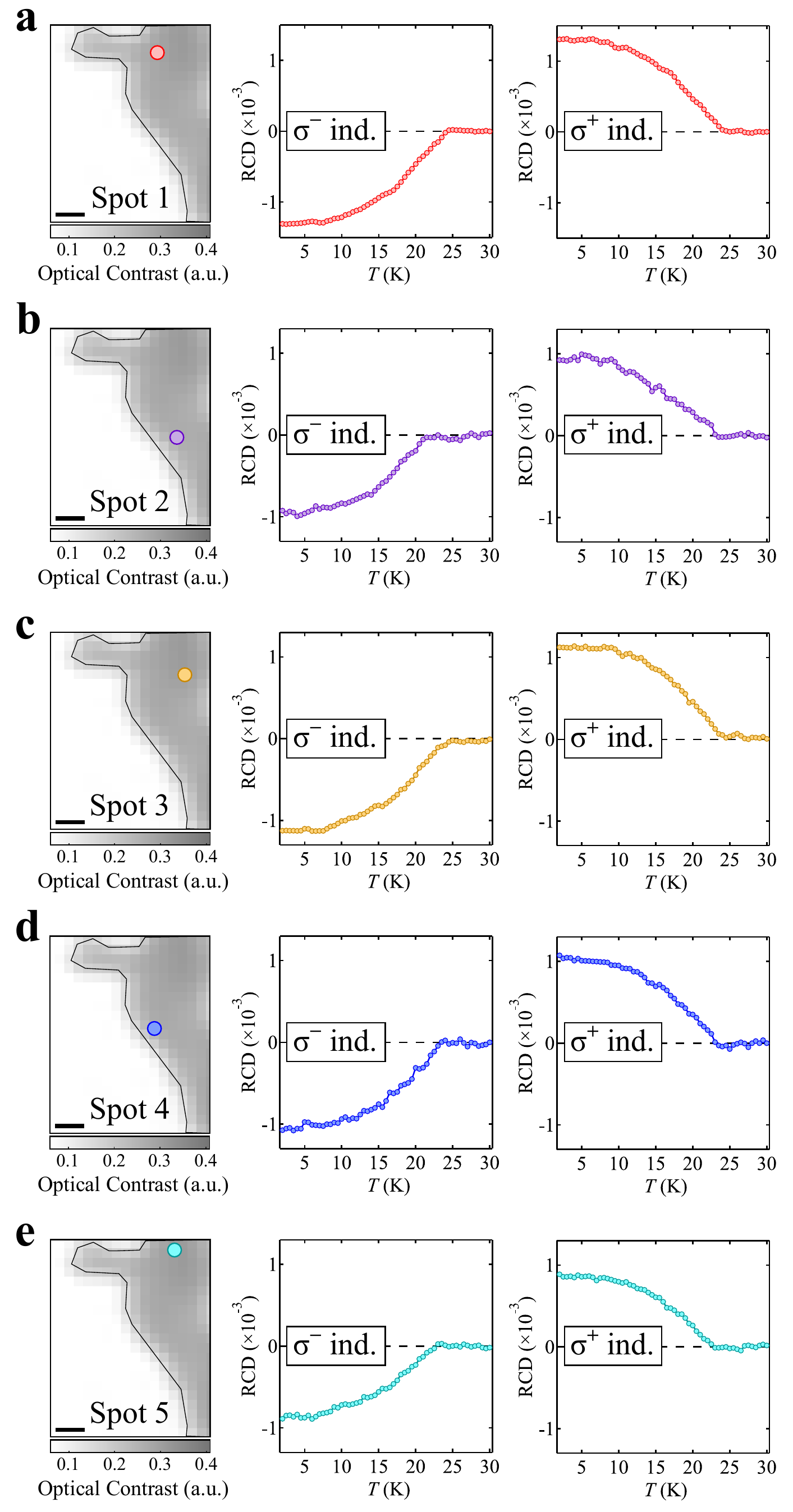}
\caption{\textbf{Spatial reproducibility of the Axion induction}. \textbf{a,} On the left is the optical contrast map of the 8SL MnBi$_2$Te$_4$ flake in sample-S1. The red dot marks the laser spot for induction experiments.  On the right is the RCD signal as a function of temperature while warming up after induction with $\sigma^{\pm}$ light. Scale bar: $2$ $\mathrm{\mu m}$.  \textbf{b-e,} Same as panel (\textbf{a}) but at different spatial locations of the sample. Experimental parameters for data in this figure: $\lambda_{\textrm{induction}}=840$ nm, $P_{\textrm{induction}}\simeq1$ mW; $\lambda_{\textrm{detection}}=946$ nm, $P_{\textrm{detection}}\simeq30$ $\mu$W.}
\label{Spatial_dep}
\end{figure*}

\vspace{3mm} 
\textbf{General remarks about the induction experiments:} In our optical experiments (both induction and CD), light is focused to a spot of the diffraction limit ($\sim1$ $\mu$m). For the areas that are chosen for induction (see Fig.~\ref{Spatial_dep}), we have carefully studied their RCD signals as a function of $B$-field (by measuring magnetic hysteresis at $2$ K) and $T$ (upon multiple cooldowns without induction light). We found that after cooling down from $30$ K to $2$ K without induction, the RCD always reaches the maximal amplitude; only the sign is random (shown in Fig.~\ref{Cooldown_without_induction}). This suggests that, at least on the length scale of diffraction limit ($\sim1$ $\mu$m), the areas chosen for induction are always in a single-domained state. Therefore, the results of induction are ternary: (1) the induction leads to the AFM state I (positive RCD at $\lambda_{\textrm{detection}}=946$ nm); (2) the induction leads to the AFM state II (negative RCD at $\lambda_{\textrm{detection}}=946$ nm); (3) the induction cannot favor one state, so the results are random just like no induction. 

\textbf{The induction ability:} To characterize these different induction results, we define the quantity of induction ability. We define the RCD after induction from $30$ K to $2$ K with $\lambda_{\textrm{induction}}=840$ nm and $P_{\textrm{induction}}=1$ mW as the reference RCD ($\mathrm{RCD^{ref}}$). We can then perform induction at other conditions and measure RCD. The induction ability is the ratio between the measured RCD and the reference RCD. $\textrm{Induction ability}=\frac{\textrm{RCD}}{\textrm{RCD}^{\textrm{ref}}}$. Therefore, if the induction leads to the same AFM state as the reference induction, then the induction ability will be $\simeq+1$; if the induction leads to the opposite AFM state as the reference induction, then the induction ability will be $\simeq-1$. If the induction has no effect, then we perform the induction 6 consecutive inductions and take the averaged RCD value, and the induction ability will be $\simeq0$.

\vspace{3mm} 

\textbf{Spatial reproducibility:} To substantiate the Axion induction, we have performed induction experiments at multiple locations on the 8SL MnBi$_2$Te$_4$ flake (sample-S1), which gave consistent results (shown in Fig \ref{Spatial_dep}).

%Our CD measurement is a local probe ($\sim1$ $\mu$m). On this scale, the MnBi$_2$Te$_4$ 
\vspace{3mm} 

\begin{figure*}[h]
\centering
\includegraphics[width=16cm]{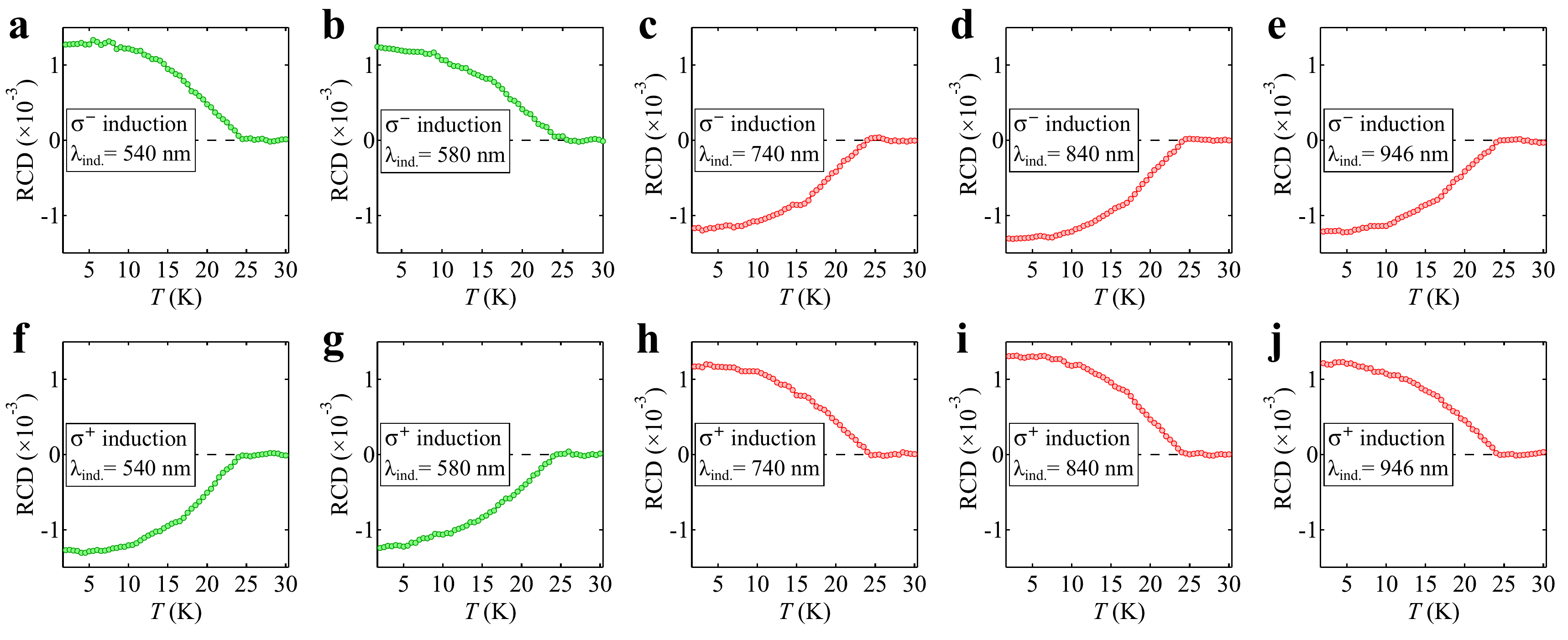}
\caption{\textbf{Detailed results of Axion induction at multiple wavelengths.} \textbf{a,} On the left is the RCD map at a particular area after induction ($\lambda_{\textrm{induction}}=540$ nm) with opposite helicity. The circle marks the laser spot for the induction light. On the right is the RCD signal as a function of temperature while warming up after induction with $\sigma^{\pm}$ light. Scale bar:2 $\mathrm{\mu m}$. \textbf{b-e,} Same as panel ($\textbf{a}$) but with the $\lambda_{\textrm{induction}}$ being $580$ nm, $740$ nm, $840$ nm and $946$ nm. Experimental parameters for data in this figure: $\lambda_{\textrm{induction}}$ noted in the figure, $P_{\textrm{induction}}\simeq1$ mW; $\lambda_{\textrm{detection}}=946$ nm, $P_{\textrm{detection}}\simeq30$ $\mu$W. }
\label{temp_dep_wave}
\end{figure*}

\textbf{Induction wavelength dependence:} In addition to the induction wavelength dependence data in the main text, Fig.~\ref{temp_dep_wave} shows additional RCD spatial maps and temperature dependence data at multiple induction wavelengths ($\mathrm{\lambda_{induction}=540\ nm}$, 580 nm, 740 nm, 840 nm and 946 nm).

\vspace{3mm} 

\textbf{Induction power dependence:} We investigate how induction depends on the optical power of the induction light. As shown in Fig.~\ref{Induct_power_dep}, we found that,  for $P_{\textrm{induction}}\gtrsim 400$ $\mu$W, the induction can effectively control the AFM state. Below this power, the induction was found to show no effect, possibly due to the influence of defects and disorder. %In addition, when  the induction power is above 400 $\mathrm{\mu W}$, the induction ability remains constant, which is also consistent with our free energy explanation.  \\

\vspace{3mm} 

\textbf{Laser heating effect:} It is important to characterize the effect of laser heating both in the induction and the CD detection processes (we use the same light source for induction and CD). In order to do so, we measured the temperature dependence of RCD while warming up at multiple laser powers, as shown in Fig.~\ref{Laser_Heating}\textbf{a}. We focus on the temperature at which the RCD signal vanishes (defined as $T^*$). Without laser heating, $T^*=T_{\textrm{N}}$. Laser heating will manifest as a decrease of $T^*$ (See detailed explanation in the caption of Fig.~\ref{Laser_Heating}). As shown in Fig.~\ref{Laser_Heating}\textbf{b}, at $P_{\textrm{detection}}=1$ mW, $T^*$ decreases by $\simeq2$ K, indicating that laser heating caused a $2$ K temperature increase locally at the sample. Therefore, we can draw the following conclusions about laser heating effect. 

%, remains at $25$ K, in agreement with the N{\'e}el temperature $T_{\textrm{N}}$. By contrast, for $P_{\textrm{detection}}\gtrsim 200$ $\mu$W, $T^*$ decreases with increasing $P_{\textrm{detection}}$

(1) For all RCD measurements, we consistently use $P_{\textrm{detection}}=30$ $\mu$W. So laser heating during RCD measurements is minimal.

(2) For the optical induction, we consistently use $P_{\textrm{induction}}=1$ mW. So laser heating caused the sample temperature to increase by $\sim2$ K. In our typical induction process, we start to shine the induction light at $30$ K and turn off the induction light at $2$ K. Therefore, this small temperature increase is unimportant for our induction.

\begin{figure*}[!htb]
\centering
\includegraphics[width=10cm]{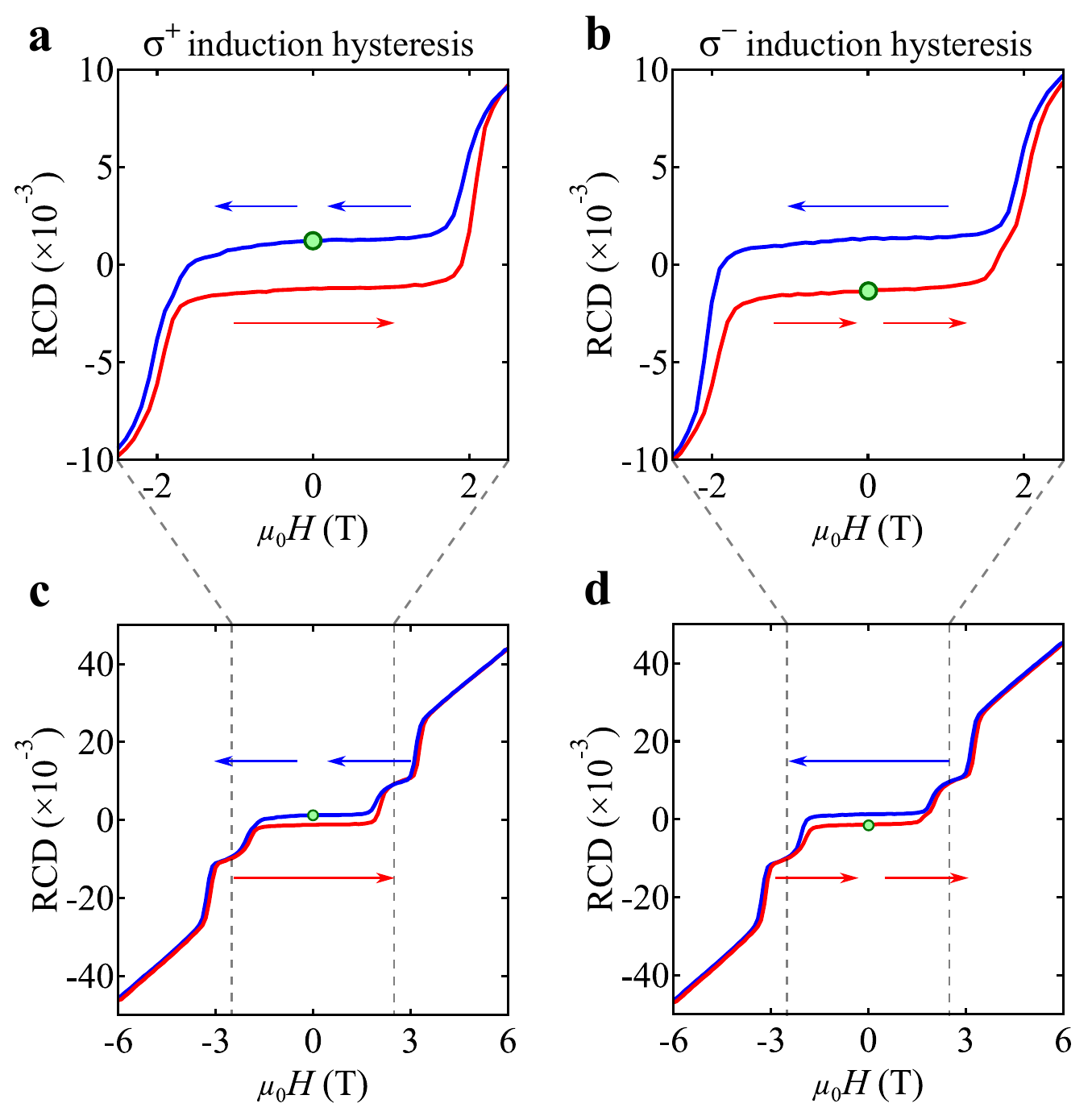}
\caption{\textbf{a-d,} Magnetic hysteresis of RCD after induction with opposite helicity in 8SL MnBi$_2$Te$_4$ of sample-S1. The starting points at $B=0$ (green dots in panels (\textbf{a,b})) are clearly on opposite hysteresis branch, again showing that the opposite induction helicity leads to opposite AFM domains. Experimental parameters for data in this figure: $\lambda_{\textrm{induction}}=840$ nm, $P_{\textrm{induction}}\simeq1$ mW; $\lambda_{\textrm{detection}}=946$ nm, $P_{\textrm{detection}}\simeq30$ $\mu$W.}
\label{Induction_hysteresis}
\end{figure*}

\begin{figure*}[!htb]
\centering
\includegraphics[width=6cm]{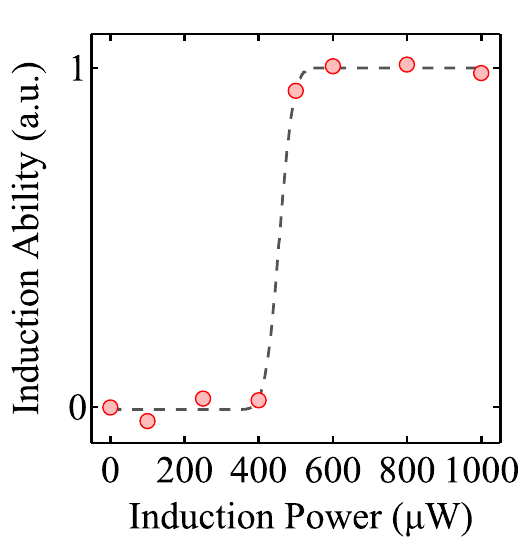}
\caption{Induction ability (see definition in the text SI.II.2) as a function of induction optical power. Experimental parameters for data in this figure: $\lambda_{\textrm{induction}}=840$ nm; $\lambda_{\textrm{detection}}=946$ nm. $P_{\textrm{detection}}= 30$ $\mu$W. The induction power is noted in the figure.}
\label{Induct_power_dep}
\end{figure*}

\begin{figure*}[!htb]
\centering
\includegraphics[width=12cm]{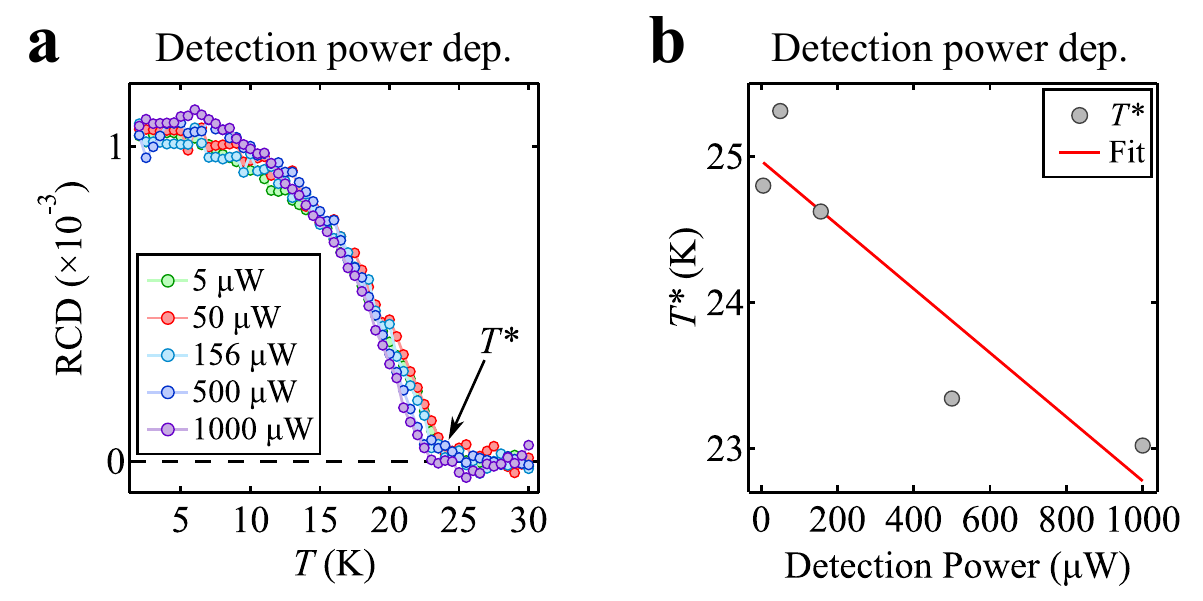}
\caption{\textbf{a,} RCD as a function of $T$ measured while warming up at different optical powers. Note that $T$ is the temperature of sample holder. In the presence of laser heating, when the sample reaches the N{\'e}el temperature $T_{\textrm{N}}$ (i.e., when RCD signal vanishes), the sample holder temperature $T^*$ is lower than $T_{\textrm{N}}$. Therefore, the difference between $T^*$ and $T_{\textrm{N}}$ is a measure of the laser-heating-induced temperature increase. \textbf{b,} $T^*$ as a function of the optical power. $T^*$ decreases with increasing $P_{\textrm{detection}}$. At $P_{\textrm{detection}}=1$ mW, $T^*$ decreases by $\simeq2$ K, indicating that laser heating caused a $2$ K temperature increase locally at the sample. $\lambda_{\textrm{detection}}=840$ nm.}
\label{Laser_Heating}
\end{figure*}

\textbf{Magnetic hysteresis after optical induction:} To further substantiate the Axion induction, we measure the magnetic hysteresis of the RCD after performing optical induction. As shown in Fig.~\ref{Induction_hysteresis}, after induction with opposite helicity, the RCD is found to start from opposite branches of the magnetic hysteresis. This further confirms that induction with opposite helicity leads to opposite AFM states at low temperatures.

\begin{figure*}[h]
\centering
\includegraphics[width=12cm]{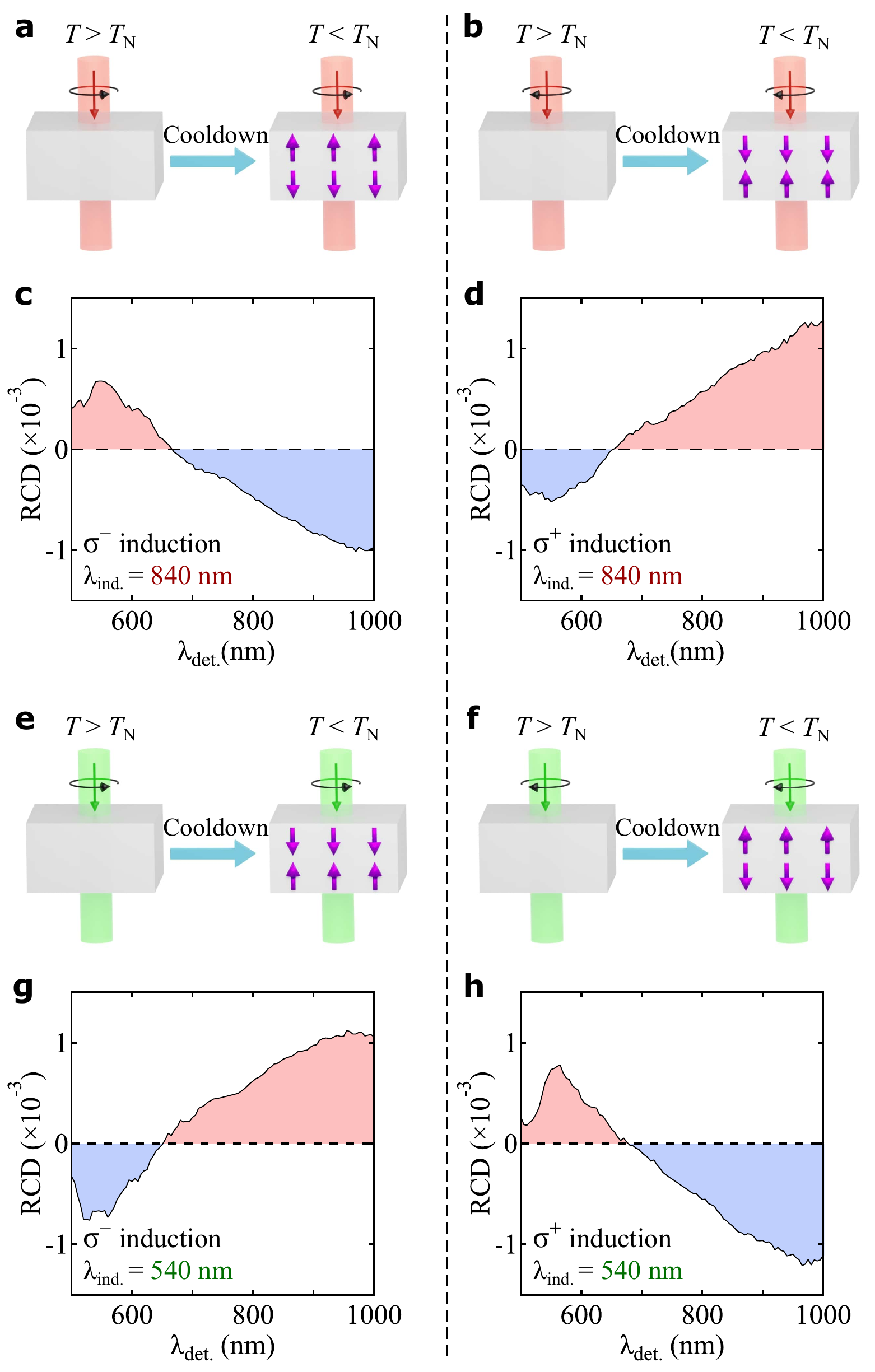}
\vspace{-5.5mm}
\caption{\textbf{RCD spectra after induction.} \textbf{a,c,} We shine  $\sigma^-$ induction light ($\lambda_{\textrm{induction}}=840$ nm, $P_{\textrm{induction}}\simeq1$ mW) on the 8SL MnBi$_2$Te$_4$ flake (sample-S1) while lowering its temperature from $T=30$ K to $2$ K (panel \textbf{a}).  Upon reaching $2$ K, we turn off the induction light, and measure the RCD's spectra (panel \textbf{c}) at $2$ K. \textbf{b,d,} Same as panels (\textbf{a,c}) except that we perform induction with $\sigma^+$. \textbf{e-h,} Same as panels (\textbf{a-d}) except that we change $\lambda_{\textrm{induction}}$ to $\lambda_{\textrm{induction}}=540$ nm.}
\label{Induction_spectra}
\end{figure*}

\textbf{RCD spectra after optical induction:} Figure~\ref{Induction_spectra} shows the RCD spectra after optical induction. The RCD spectra are opposite if we performed optical induction with opposite light helicity or different induction wavelengths (840 nm vs. 540 nm).

\color{black}
% the local sample temperature will manifest 
%
%On the other hand, we also systematically investigated the detection power dependence. As shown in Fig \ref{Induct_power_dep}\textbf{b}, the measured RCD signal ($\mathrm{\lambda_{detection}=840 \ nm}$) basically remains invariant when we vary the detection power, which shows that Axion CD is a linear optical effect. In addition, we measured the RCD as a function of temperature while warming up with multiple detection powers, as shown in Fig \ref{Induct_power_dep}\textbf{c}. The temperature dependence curves indicate a dirft of the measured N{\'e}el temperature $T_\mathrm{N}$ while we vary the detection powers. Specifically, Fig \ref{Induct_power_dep}\textbf{d} shows the extracted $T_\mathrm{N}$ linearly decreases with the increasing detection power. While we shine the detection light on the sample, it will inevitably heat up the sample due to thermal heating effect of the laser, which explains the drift of  $T_\mathrm{N}$. Additionally, since we know that the laser heating efficiency is $\mathrm{\sim 2\ K/mW}$, we could infer that the actual sample temperature is $\mathrm{\sim4\ K}$ when we remove the induction light. As a result, the sample already enters the AFM state before we remove the induction light. 

\vspace{3mm} 
\textbf{Induction initial temperature and ending temperature:} The novel coupling between circular light and the Axion insulator state allows us to lift the energy degeneracy between the opposite AFM states. In order to actually choose one AFM state over the other, we need to further overcome the potential barrier between the opposite AFM states. In the vicinity of the N{\'e}el temperature, the potential barrier is small, allowing us to achieve the control. For all induction experiments in the main text, the induction initial and ending temperatures were kept at $30$ K and $2$ K. 

\begin{figure*}[!htb]
\centering
\includegraphics[width=14cm]{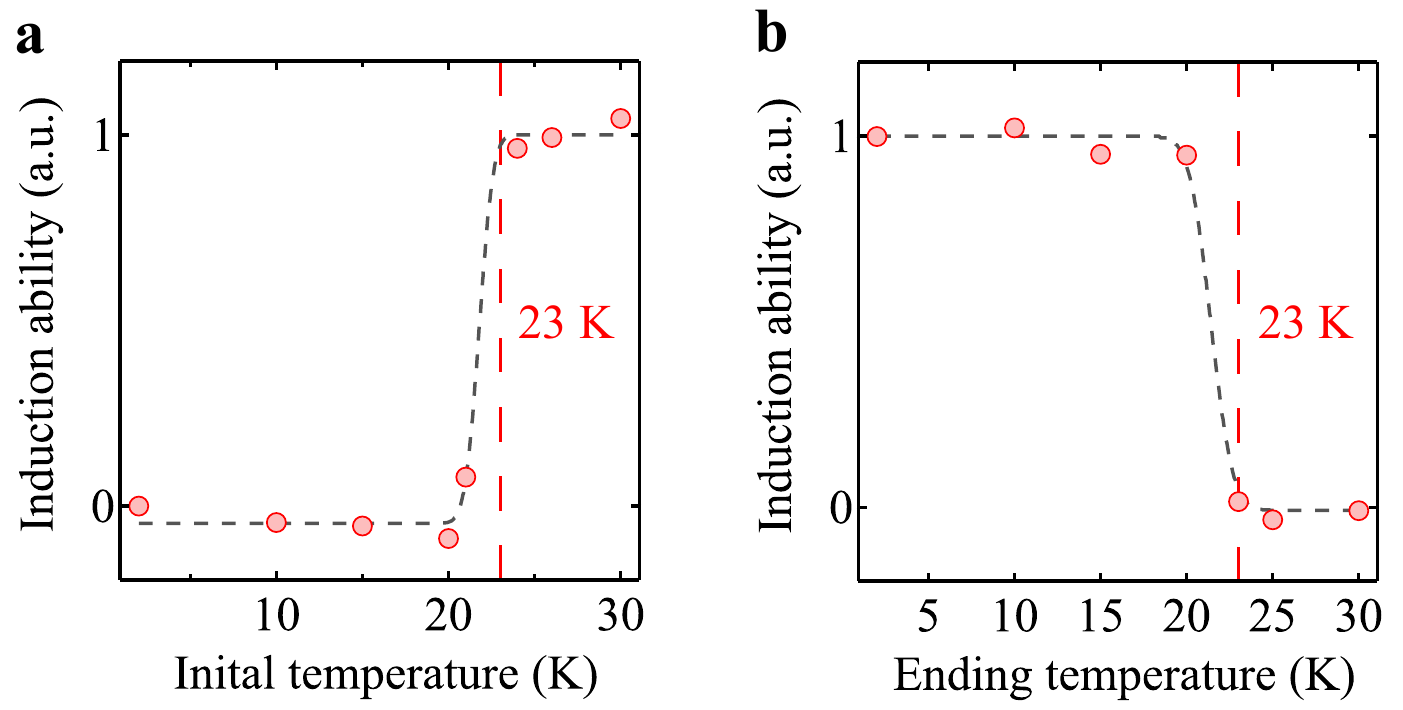}
\caption{\textbf{Dependence of induction initial temperature and ending temperature}. The initial (ending) temperature is the temperature at which we turn on (off) the induction light. \textbf{a,} We keep the ending temperature at $2$ K and measure the induction ability as a function of the initial temperature. \textbf{b,} We keep the initial temperature at $30$ K and measure the induction ability as a function of the ending temperature. Experimental parameters for data in this figure: $\lambda_{\textrm{induction}}=840$ nm, $P_{\textrm{induction}}\simeq1$ mW; $\lambda_{\textrm{detection}}=946$ nm, $P_{\textrm{detection}}\simeq30$ $\mu$W.}
\label{Induct_temp_dep}
\end{figure*}

\begin{figure*}[!htb]
\centering
\includegraphics[width=16cm]{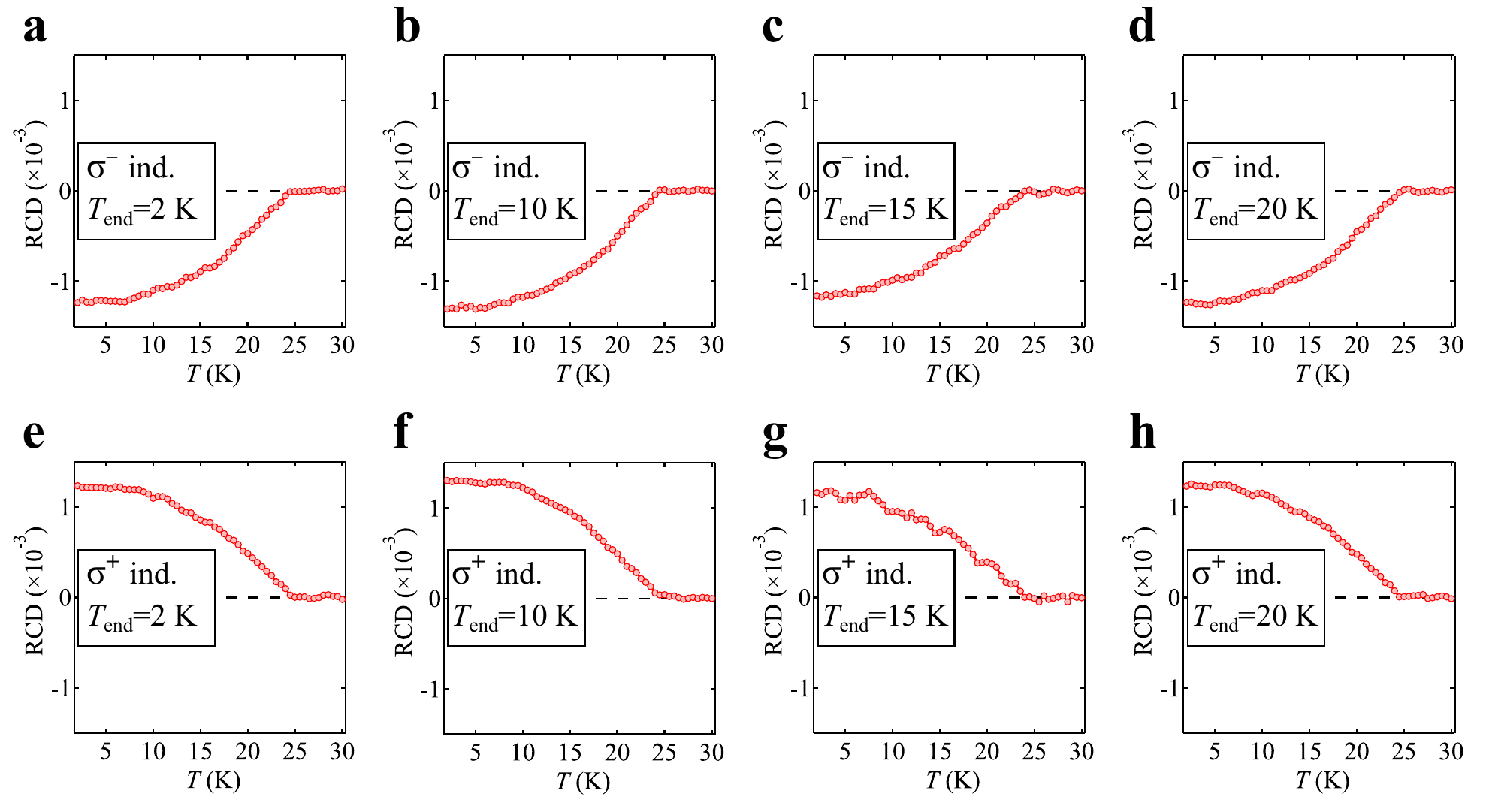}
\caption{\textbf{a-d,} We performed induction with $\sigma^{-}$ helicity by keeping the initial temperature at $30$ K and varying the ending temperature $T_\mathrm{{end}}$. After induction, we cool the sample to $2$ K. We then measure RCD as a function of temperature while warming up. \textbf{e-h,} Same as panels (\textbf{a-e}) but induction was performed with $\sigma^{+}$ helicity. This figure is in supplementary to Fig.~\ref{Induct_temp_dep}\textrm{b}. Experimental parameters for data in this figure: $\lambda_{\textrm{induction}}=840$ nm, $P_{\textrm{induction}}\simeq1$ mW; $\lambda_{\textrm{detection}}=946$ nm, $P_{\textrm{detection}}\simeq30$ $\mu$W.}
\label{End_temp_dep}
\end{figure*}

We now investigated how the induction initial and ending temperatures influence the induction result. First, we keep the ending temperature at $2$ K while varying the initial temperature. As shown in Fig.~\ref{Induct_temp_dep}\textbf{a}, for initial temperature higher than $23$ K, the induction can effectively control the AFM state; by contrast, for initial temperature below $23$ K, the induction ability approaches zero, suggesting that the potential barrier between opposite AFM states at low temperature is too strong to overcome. Second, we keep the initial temperature at $30$ K while varying the ending temperature. As shown in Fig \ref{Induct_temp_dep}\textbf{b} and Fig~\ref{End_temp_dep}, for ending temperature lower than $23$ K, the induction can effectively control the AFM state.

\begin{figure*}[h]
\centering
\includegraphics[width=12cm]{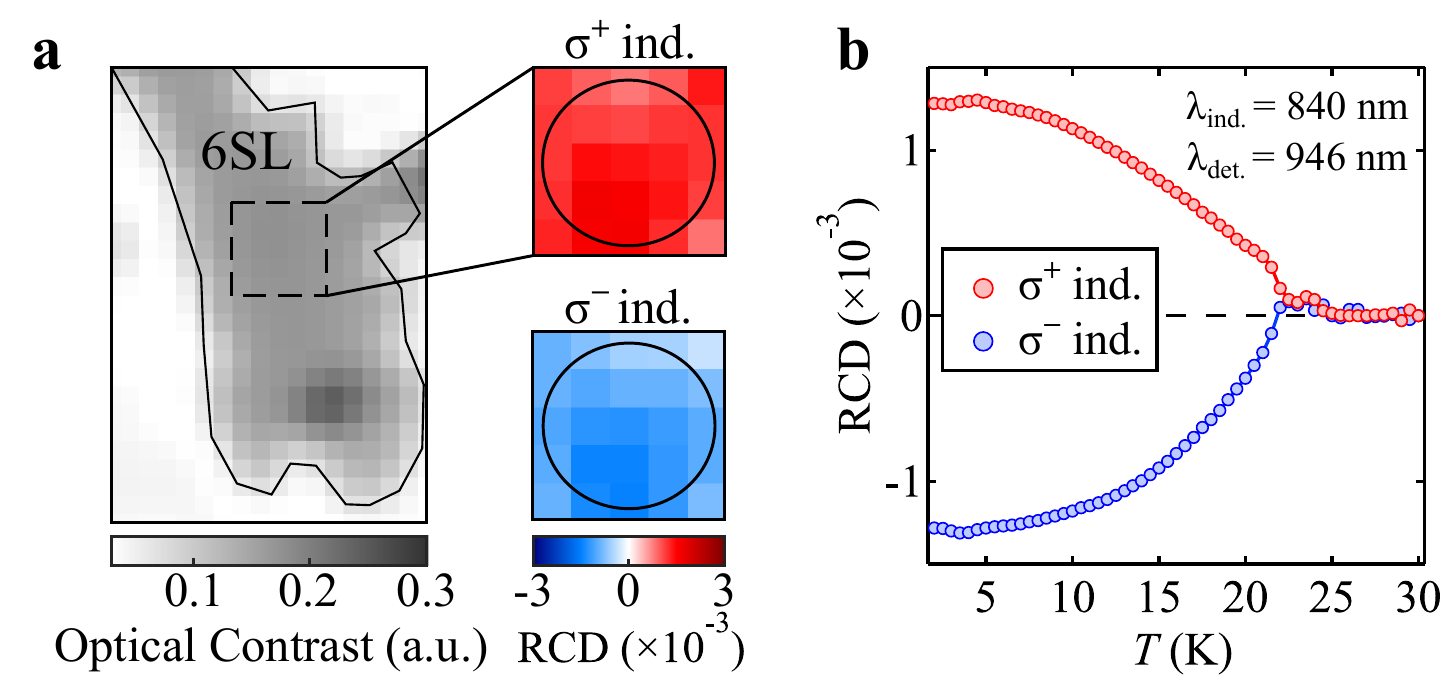}
\vspace{-5.5mm}
\caption{\textbf{a,} Left: Optical contrast of the 6SL flake. Right: RCD upon induction using $\sigma^{\pm}$. \textbf{b,} Temperature-dependence of RCD after induction using $\sigma^{\pm}$.}
\label{6SL_Induction}
\end{figure*}

\vspace{3mm} 
\textbf{Induction of 6SL:} In Fig.~\ref{6SL_Induction}, we show the induction results of 6SL sample.

\begin{figure*}[h]
\centering
\includegraphics[width=17cm]{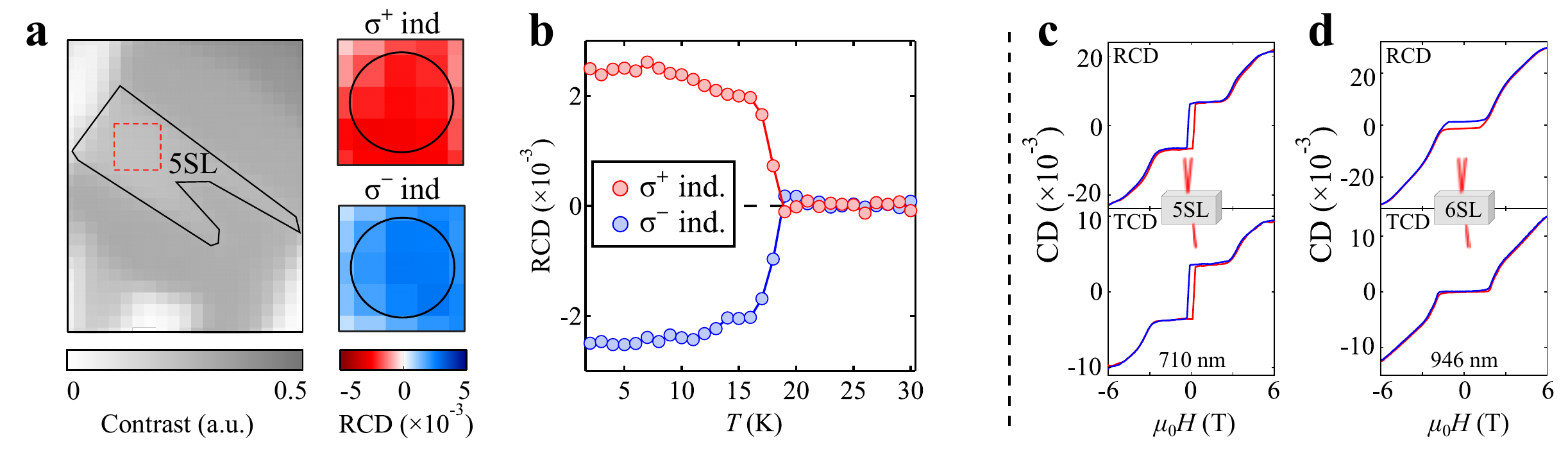}
\vspace{-5.5mm}
\caption{\textbf{a,} Left: Optical contrast of the 5SL flake. Right: RCD upon induction using $\sigma^{\pm}$. \textbf{b,} Temperature-dependence of RCD after induction using $\sigma^{\pm}$. \textbf{c,d,} Simultaneous RCD TCD measurements for 5SL and 6SL.}
\label{5SL_Induction}
\end{figure*}

\vspace{3mm} 
\textbf{Induction of 5SL:} In Figs.~\ref{5SL_Induction}\textbf{a,b}, we show the induction results of 5SL sample. Because odd-layered samples has an obvious $M$ (like a ferromagnet), the primary interaction is between $M$ and circularly-polarized light (confirmed by our simultaneous RCD TCD measurements in 5SL Figs.~\ref{5SL_Induction}\textbf{c,d}). Therefore, the induction of odd-layered sample arises from the helicity-dependent optical control of M, which has been demonstrated previously in a range of ferromagnets. %By contrast, the induction of even-layered sample ($M=0$) is qualitatively different. 

\color{black}
%\pagebreak

%In our optical experiments (both induction and CD), light is focused to a spot of the diffraction limit ($\sim1$ $\mu$m). For the areas that are chosen for induction

\pagebreak

\pagebreak

\pagebreak

%$T^*$ is the temperature at which the RCD signal vanishes (defined as $T^*$). Without laser heating, $T^*=T_{\textrm{N}}$. Laser heating will manifest as a drift of $T^*$ (See detailed explanation in the caption of Fig.~\ref{Induct_power_dep}\textbf{XXX}). As shown in Fig.~\ref{Induct_power_dep}\textbf{XXX}, for $P_{\textrm{detection}}\lesssim 200$ $\mu$W, $T^*$ remains at $25$ K, in agreement with the N{\'e}el temperature $T_{\textrm{N}}$. By contrast, for $P_{\textrm{detection}}\gtrsim 200$ $\mu$W, $T^*$ decreases with increasing $P_{\textrm{detection}}$. At $P_{\textrm{detection}}=1$ mW, $\delta T^*\simeq2$ K, indicating that laser heating caused a $2$ K temperature increase locally at the sample. Therefore, we can draw the following conclusions about laser heating effect.
%\textbf{b}, RCD signal as a function of detection power, which indicates the Axion CD is a first-order effect. The initial state of this measurement is obtained by the $\sigma^{+}$ light induction with $\mathrm{\lambda_{induct}=840 \ nm}$. 

\pagebreak

%We perform After induction with $\sigma^{-}$ light ($P_\mathrm{{induction}}\mathrm{\simeq 1\ mW, \ \lambda_{induction}=840\ nm}$), RCD signal as a function of temperature  with $T_\mathrm{{end}}=$ 2 K, 10 K, 15 K and 20 K, respectively.  \textbf{e-h}, Same as panels (\textbf{a-e}) but with opposite induction light helicity $\sigma^{+}$. This figure is in supplementary to fig \ref{Induct_temp_dep}. 

\clearpage
 
\subsection*{II.3. Additional double induction data}

\begin{figure*}[h]
\centering
\includegraphics[width=12cm]{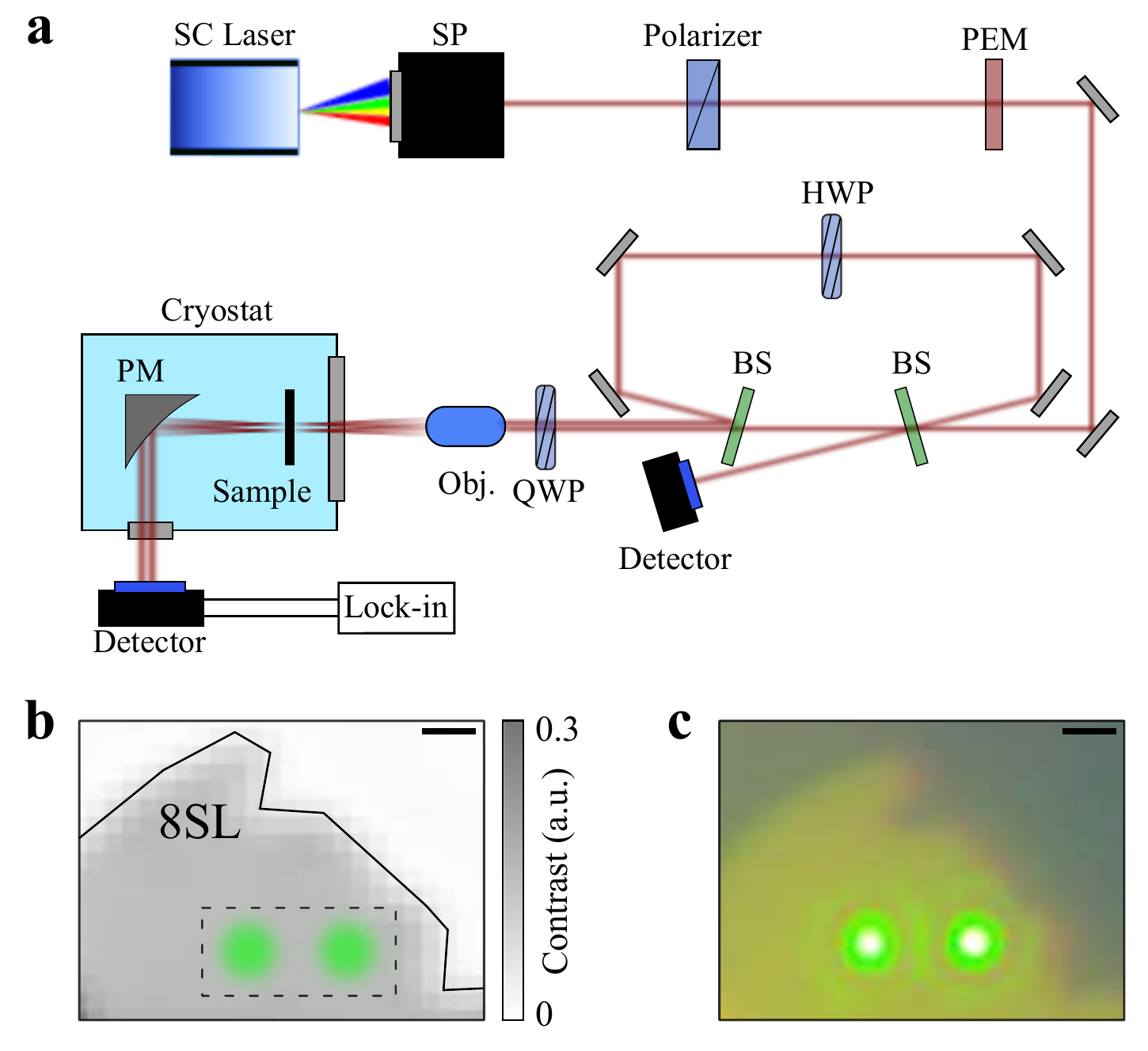}
\vspace{-5.5mm}
\caption{\textbf{a}, Schematic drawing of optical set-up for double induction experiments. SC Laser: super-continuum Laser, SP: spectrometer, BS: beamsplitter,  HWP: half-waveplate, QWP: quarter-waveplate, Obj: Objective, PM: parabolic mirror. \textbf{b},  Optical contrast of the 8SL flake. \textbf{c}, Microscope image of the 8SL flake with double induction beams. Scale bar: $2$ $\mathrm{\mu m}$. }
\label{double_ind_setup}
\end{figure*}

\hspace{5mm}\textbf{Optical set-up for the double induction experiments:} As shown in Fig.~\ref{double_ind_setup}\textbf{a}, starting from a single beam, we constructed two collinear, spatially separated beams by the two beam splitters. Their polarizations were controlled separately by the combination of the half-waveplate and quarter-waveplate. Figures~\ref{double_ind_setup}\textbf{b-c} show the optical contrast map and microscope image of double induction beams. 

\begin{figure*}[h]
\centering
\includegraphics[width=12cm]{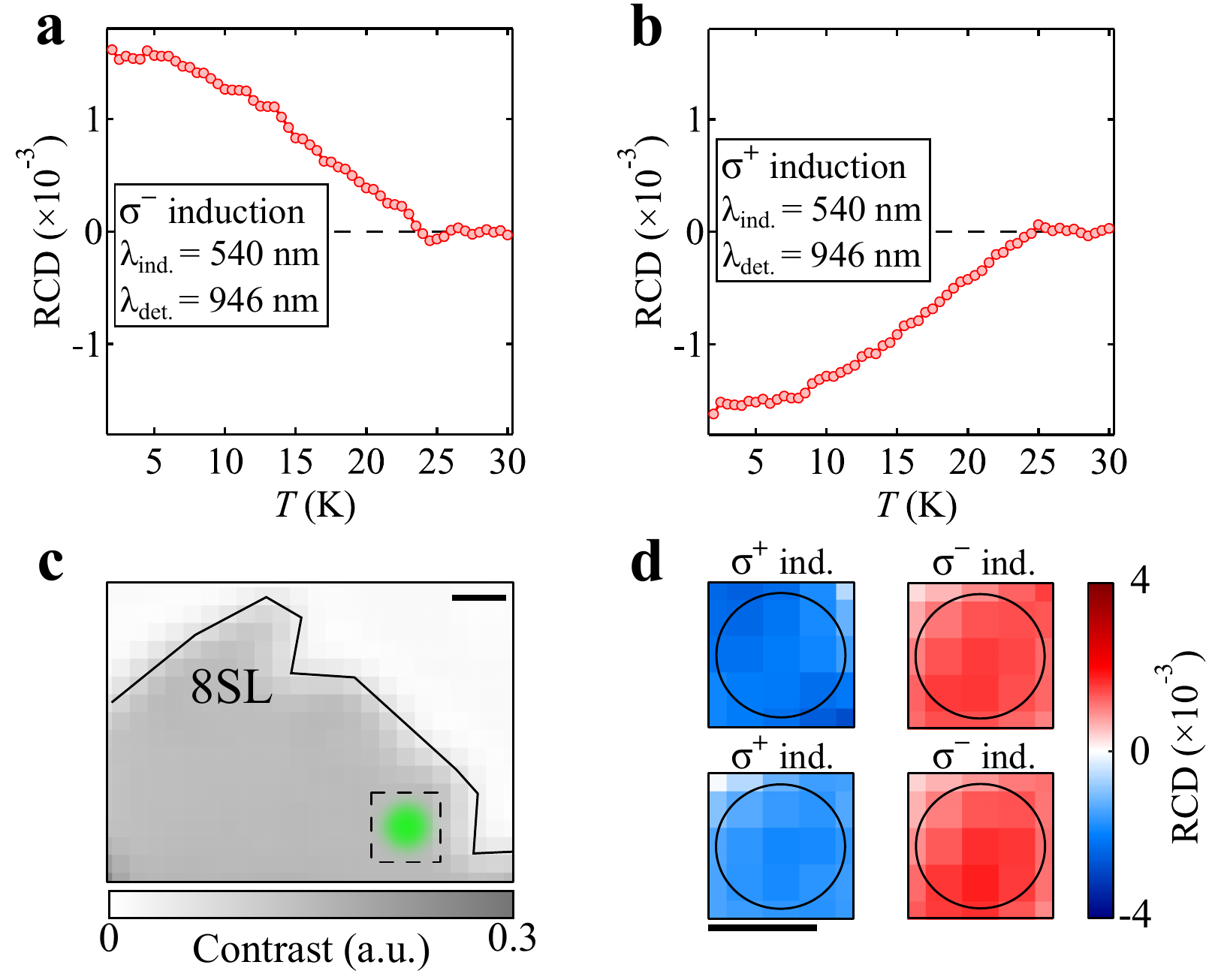}
\vspace{-5.5mm}
\caption{\textbf{a,b,} Temperature dependence of RCD after induction using $\sigma^{\pm}$ (with $\lambda_{\mathrm{induction}}=540$ nm). \textbf{c,} Optical contrast of the 8SL flake. \textbf{d,} RCD map upon induction using $\sigma^{\pm}$. }
\label{double_ind_single}
\end{figure*}

\textbf{Single optical induction experiments on sample-S5:} The double induction experiments were conducted on a new sample (S5). For consistency, we conducted single induction experiments as shown in Fig.~\ref{double_ind_single}. We observed consistent optical induction results as before, which provide the basis for the double induction experiments.

\begin{figure*}[h]
\centering
\includegraphics[width=14cm]{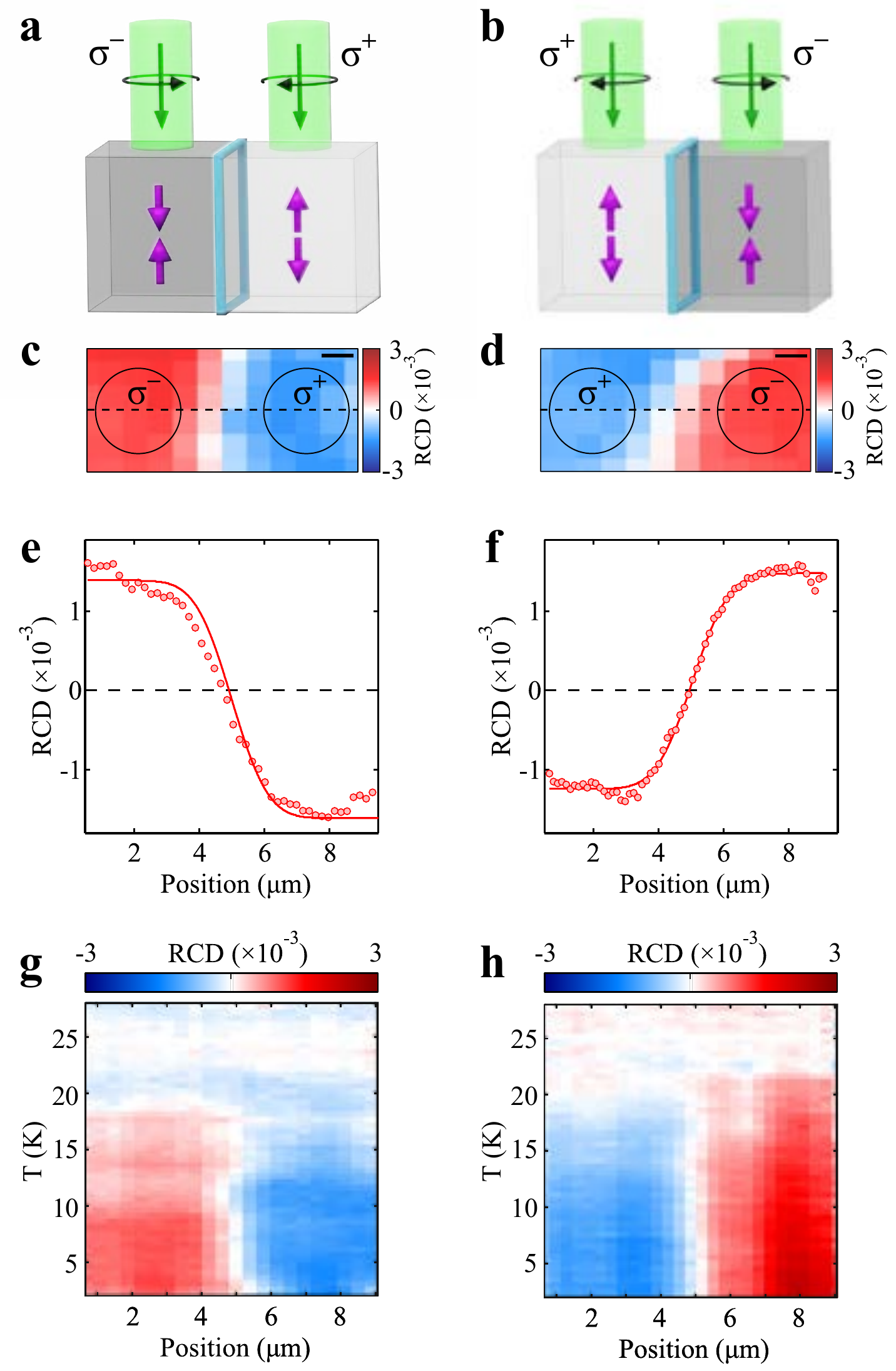}
\vspace{-5.5mm}
\caption{\textbf{a-b} Schematics of the double induction. \textbf{c-d,} RCD map after double induction. \textbf{e-f,} RCD line-cut plot along the dashed lines in panels \textbf(c,d). \textbf{g-h,} Temperature dependence of the line-cut plot. }
\label{double_ind_line}
\end{figure*}

\textbf{Detailed investigation into the AFM domain wall:} In the main text, we showed that double induction beams with opposite helicity could create an AFM domain wall (also shown in Figs.~\ref{double_ind_line}\textbf{a-d}). Here, we scan a line-cut plot across the AFM domain wall (Figs.~\ref{double_ind_line}\textbf{e-f}). In contrast to both domains, the domain wall itself showed zero RCD signal, consistent with either N\'eel type or Bloch type AFM domain walls. In Figs.~\ref{double_ind_line}\textbf{g-h}, the temperature dependence of the RCD line-cut further supports our conclusion.

\color{black}

\clearpage
 
\subsection*{II.4. Additional ultrafast switching data}

\begin{figure*}[h]
\centering
\includegraphics[width=12cm]{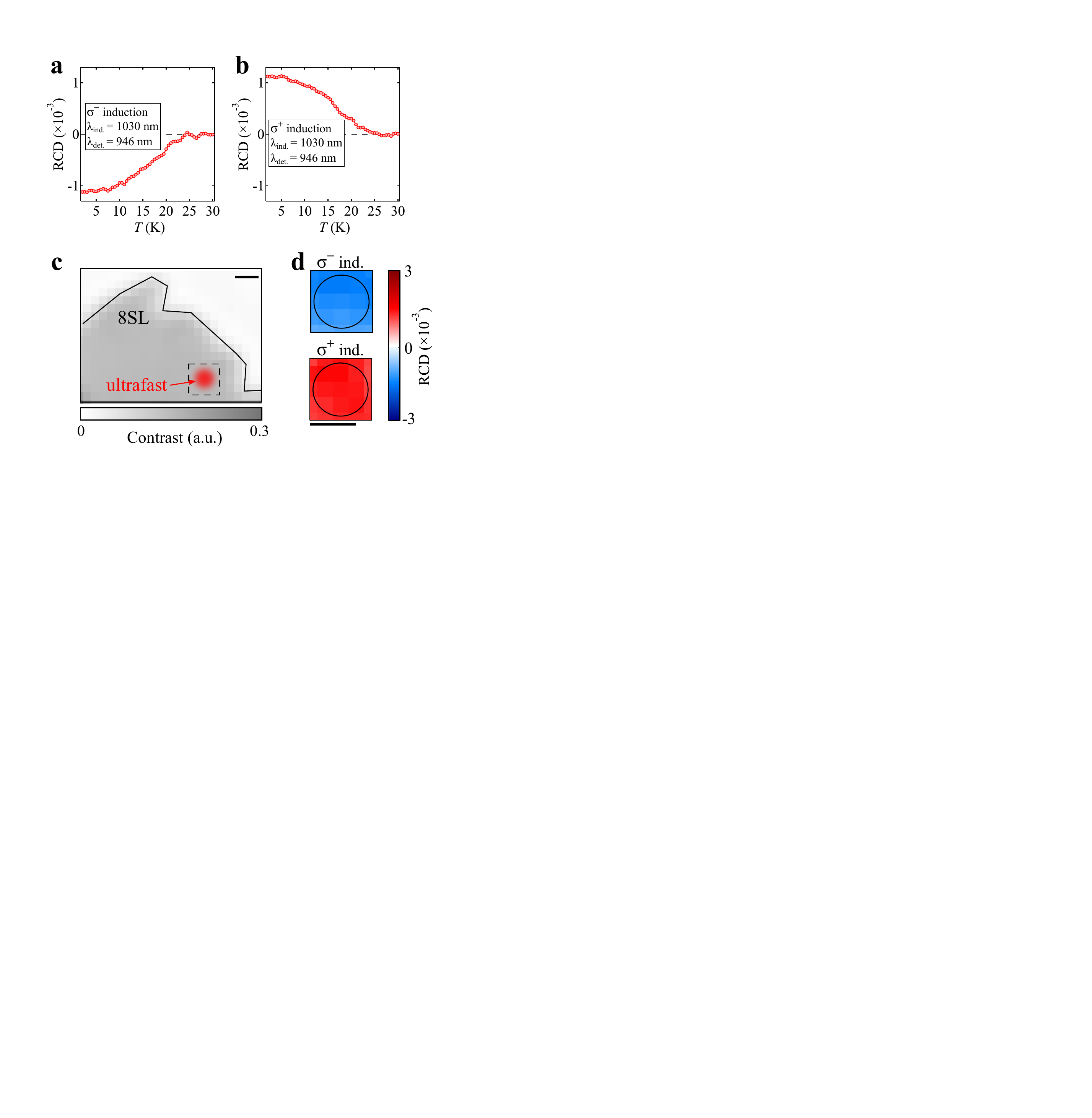}
\vspace{-5.5mm}
\caption{\textbf{Axion induction with the circularly-polarized ultrafast light.} \textbf{a-b,} Temperature-dependence of RCD after induction using $\sigma^{\pm}$ ultrafast light (with $\lambda_{\mathrm{induction}}=1030$ nm). \textbf{c,} Optical contrast of the 8SL flake. \textbf{d,} RCD map upon induction using $\sigma^{\pm}$. }
\label{ultrafast_switch_ind}
\end{figure*}

\hspace{5mm}\textbf{Optical induction experiments with ultrafast laser:} In the main text, we showed direct switching by ultrafast pulses while the sample temperature was kept at 18 K. Before doing that, we also conducted optical induction experiments with ultrafast pulses, i.e., we shone the ultrafast pulsed light while cooling the entire sample from 30 K to 2K. We found that the ultrafast pulses can also achieve optical induction with similar results as the super-continuum laser (Fig.~\ref{ultrafast_switch_ind}). 

\begin{figure*}[h]
\centering
\includegraphics[width=16cm]{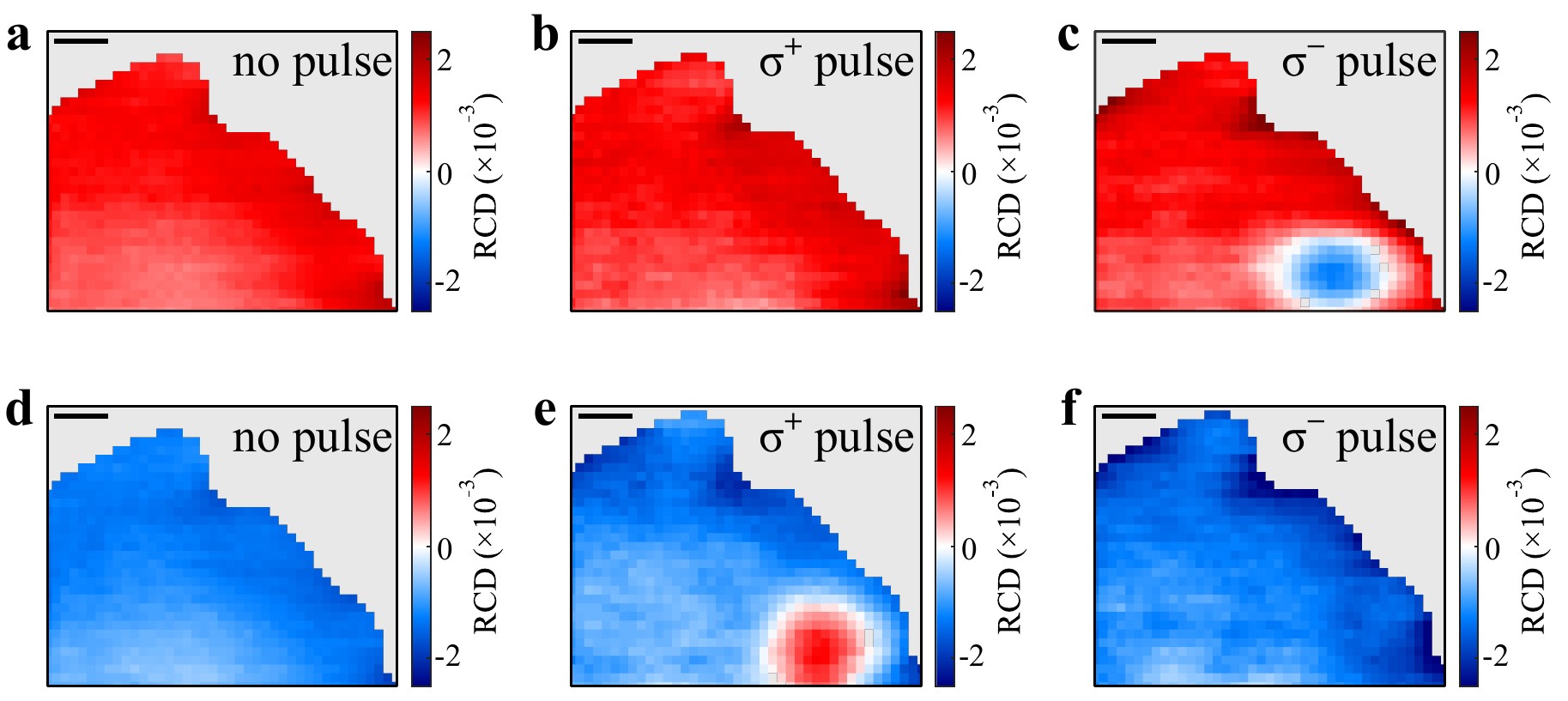}
\vspace{-5.5mm}
\caption{\textbf{a,} RCD map of the single domain state prepared by sweeping the B field from $+7$ T to $0$ T. \textbf{b-c,} RCD map after optical induction with $\sigma^{\pm}$ ultrafast pulse. \textbf{d-f,} Same as panel (\textbf{a-c}) but with the single domain state prepared by sweeping the B field from $-7$ T to $0$ T.}
\label{ultrafast_switch_map}
\end{figure*}

\textbf{Comparison of RCD map before and after ultrafast switching:} Figures~\ref{ultrafast_switch_map}\textbf{a,d} show the RCD map of the prepared single domain state before ultrafast switching, which is in clear contrast to RCD map after ultrafast switching (shown in Figs.~\ref{ultrafast_switch_map}\textbf{b-c,e-f}).

\textbf{Reproducibility of ultrafast switching:} To further substantiate the ultrafast switching results, we repeated the switching experiments multiple times for each helicity and domain type (as shown in Fig.~\ref{ultrafast_switch_reproduce}), which all yielded consistent results. 

\begin{figure*}[h]
\centering
\includegraphics[width=6cm]{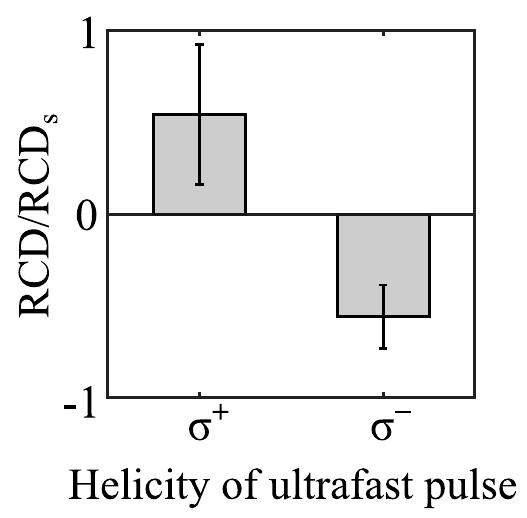}
\vspace{-5.5mm}
\caption{\textbf{Ultrafast switching statistics.} RCD value after illumination with ultrafast light, weighed against the saturated value RCD$_{\textrm{s}}$.}
\label{ultrafast_switch_stats}
\end{figure*}

\begin{figure*}[h]
\centering
\includegraphics[width=16cm]{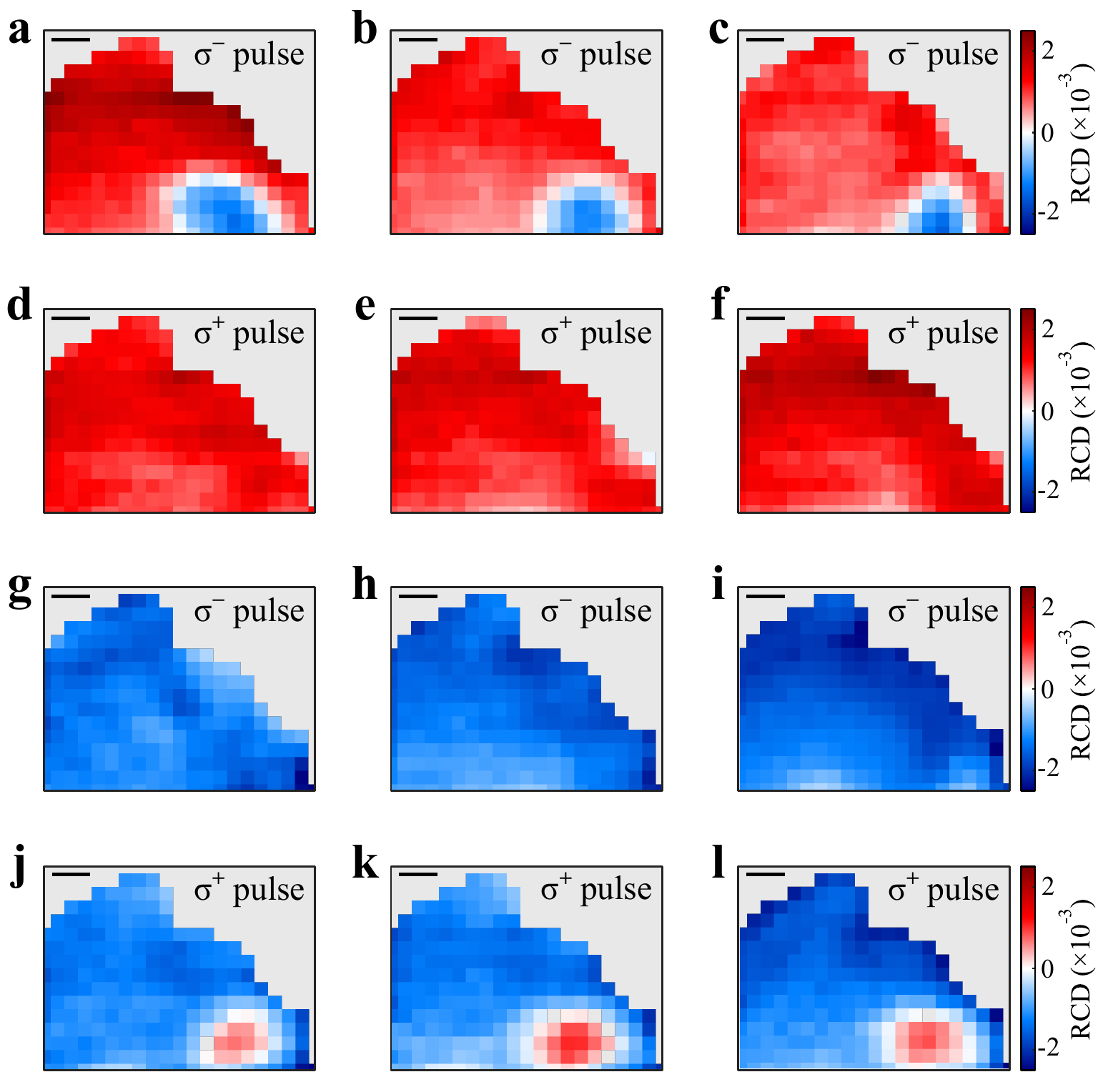}
\vspace{-5.5mm}
\caption{\textbf{Reproducibility of switching with circularly-polarized ultrafast light.} \textbf{a-c}, RCD maps after shining $\sigma^-$  ultrafast pulsed light (we first prepare the single domain state by sweeping the B field from $+7$ T to $0$ T). \textbf{d-f}, The same as panels (\textbf{a-c}) but with $\sigma^+$  ultrafast pulsed light. \textbf{g-i}, Same as panels (\textbf{a-f}) but with the single domain state prepared by sweeping the B field from $-7$ T to $0$ T. Scale bar: 2 $\mu$m.}
\label{ultrafast_switch_reproduce}
\end{figure*}

\textbf{Statistics of ultrafast switching:} As shown in Fig.~\ref{ultrafast_switch_stats}, we measured the statistics of ultrafast switching. While we can definitely switch the sign of the RCD based on helicity, we cannot reach the saturated RCD value every time. We note that our preliminary results can be further improved by detailed explorations in the future. With optimized conditions (e.g. wavelength, rep. rate, pulse energy, etc.), it may be possible to switch to a fully-saturated state. 

\color{black}

\clearpage
\subsection*{II.5. Additional electrical transport data}

\begin{figure*}[!htb]
\centering
\includegraphics[width=10cm]{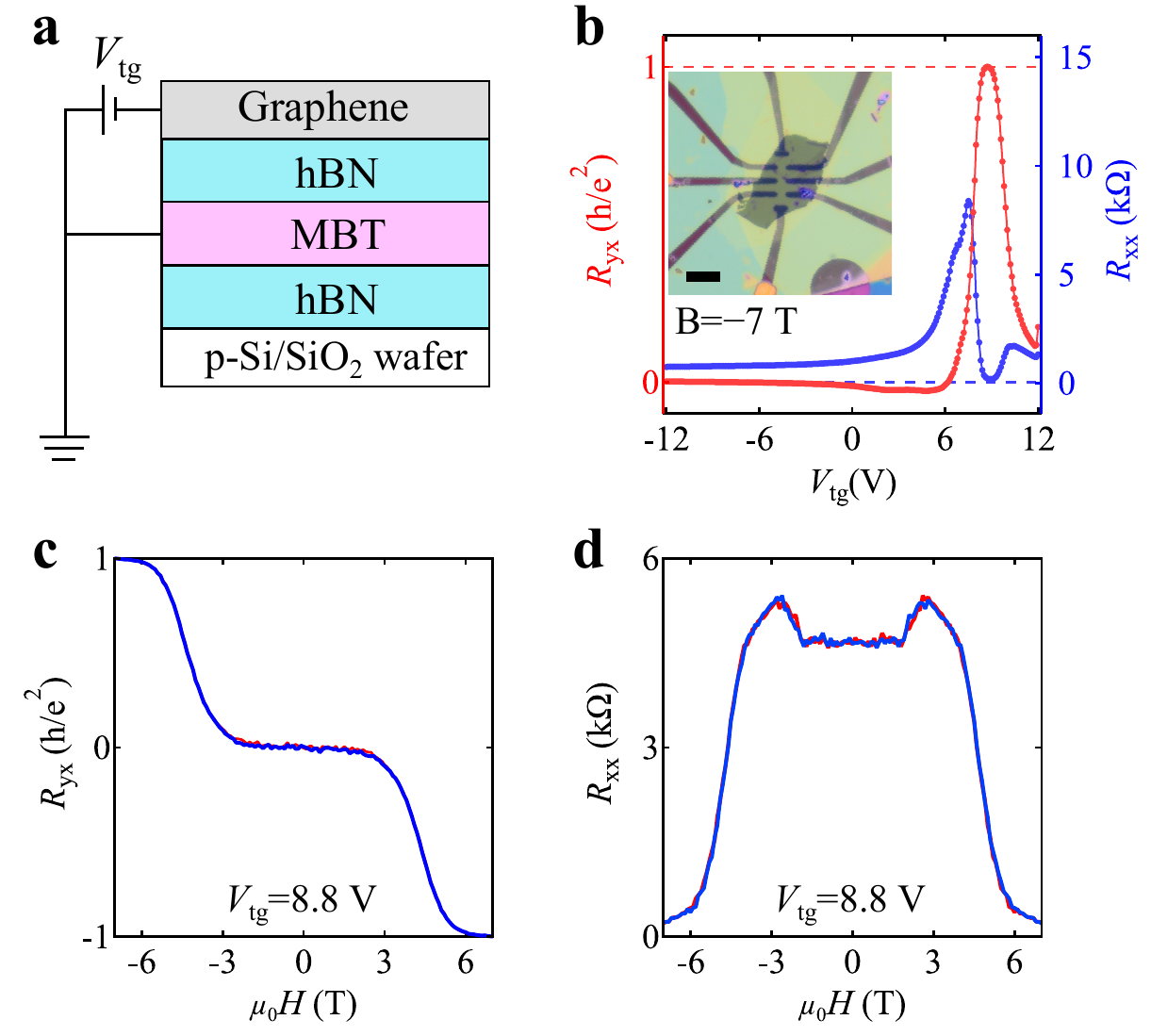}
\caption{\textbf{a,} Schematic layout for the 6SL MnBi$_2$Te$_4$ sample (Sample-S5) used for electrical transport. The transport experiments were performed in the same cryostat as the optical CD. \textbf{b,} Longitudinal ($R_{xx}$) and Hall ($R_{yx}$) responses as a function of the gate voltage in the FM phase at $B=-7$ T. Inset: Optical image of the sample. Scale bar: $10$ $\mu$m. \textbf{c,d,} $R_{yx}$ and $R_{xx}$ as a function of $B$ with the gate voltage tuned to the charge neutrality. }
\label{Transport_data_dev87}
\end{figure*}

\hspace{5mm}The entire device fabrication process was finished in an Ar-filled glovebox without exposure to air, chemicals, or heat. Using a similar fabrication process, we have also fabricated a 6SL MnBi$_2$Te$_4$ device with electrical contacts and performed transport experiments in the same cryostat as the optical CD measurements (Fig.~\ref{Transport_data_dev87}). In the FM phase at $B=-7$ T, we observed clear topological Chern insulator state with fully quantized $R_{yx}$ and zero $R_{xx}$.

\clearpage
\section*{III. Symmetry analysis of the CD}

\subsection*{III.1. General principles}

\hspace{4mm} We start by outlining the general principles for the symmetry analysis of the CD. We need to specify a system (i.e., a chiral crystal, a FM, or an AFM) and enumerate all the symmetries of that particular system. We also need to specify a particular CD process, e.g. RCD or TCD. Once both the system and the CD process are specified, then the general principles for the symmetry analysis are as follows:

\textbf{(1) To prove the CD is zero:} One needs to identify a symmetry which can keep the system and the light path invariant but flip the light helicity. 

\textbf{(2) To prove the CD is allowed:} One needs to exhaustively examine all symmetries, to show that there is no symmetry that can achieve \textbf{(1)}.

%\subsection*{III.1. General principles}
%
%\hspace{4mm} We need to specify a system (i.e., a chiral crystal, a FM, or an AFM) and enumerate all the symmetries of that particular system. We also need to specify a particular CD process, e.g. RCD or TCD. Once both the system and the CD process are specified, then the general principles for the symmetry analysis are as follows
%
%\textbf{(1) To prove the CD is zero:} One needs to identify a symmetry which can keep the system and the light path invariant but flip the light helicity. 
%
%\textbf{(2) II. To prove the CD is allowed:} One exhaustively goes through all symmetries of the system but cannot find a symmetry that can achieve \textbf{(1)}.
%

\subsection*{III.2. Three material classes and their symmetries}

\begin{figure*}[h]
\centering
\includegraphics[width=12cm]{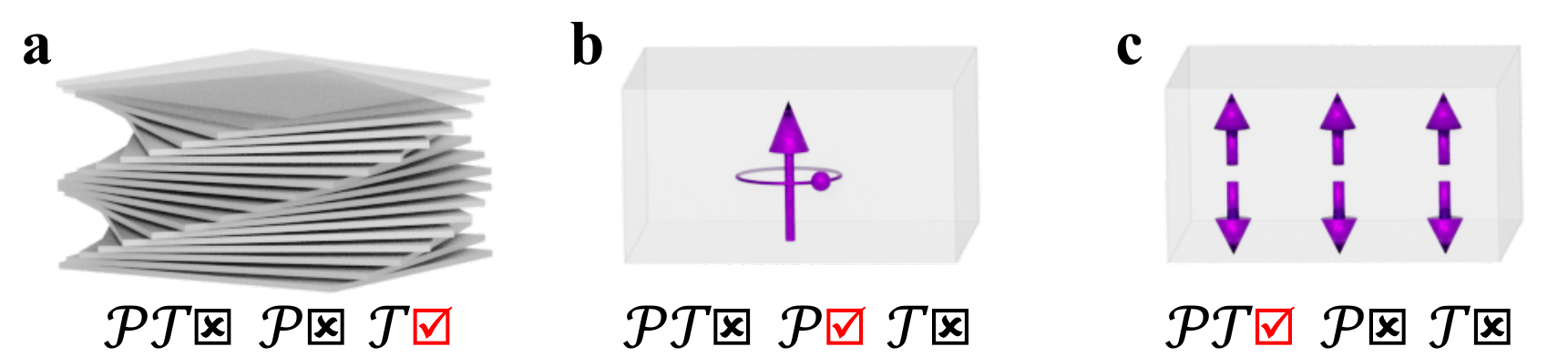}
\caption{\textbf{Three material classes (Chiral crystals, FM and AFM) and their respective symmetry properties.} Note that the symmetries described here are for the typical cases. Strictly speaking, it is possible for FMs to be in a noncentrosymmetric lattice ($\mathcal{P}$-breaking), or chiral crystals to have magnetism ($\mathcal{T}$-breaking). But most FMs are not considered to be associated with inversion symmetry breaking and most spatially chiral systems are not considered to be associated with time-reversal symmetry breaking. When all symmetries are broken, then all kinds of CD would be allowed in the system (in some sense, the distinction between different material classes is lost when all symmetries are broken, because phases of matter are only classified based on the symmetry).}
\label{AFM_Chiral_FM_symmetry}
\end{figure*}

%Similar to the main text, we focus on three material classes, i.e., chiral crystals, ferromagnets and $\mathcal{PT}$-symmetric AFMs. They have distinct symmetries, as shown in Fig.~\ref{AFM_Chiral_FM_symmetry}.

\subsection*{III.3. Application to AFM and other systems}

\hspace{4mm} We now carry out symmetry analysis for CD in these three material classes. 

\begin{figure*}[t]
\centering
\includegraphics[width=16cm]{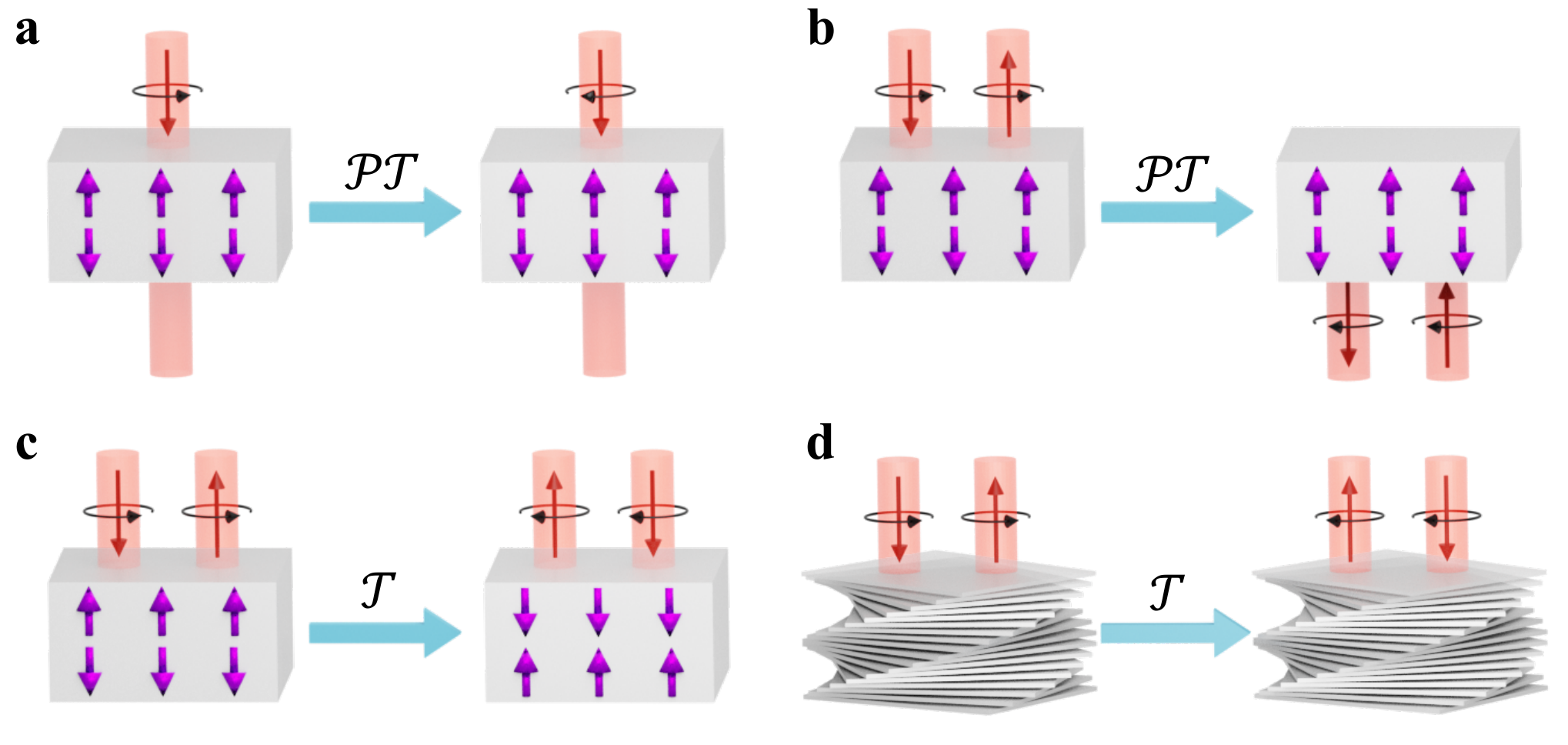}
\caption{Symmetry transformation for different CD experiments on different material systems (see detailed analysis in SI III.3).}
\label{AFM_Chiral_PT_T_symmetry}
\end{figure*}

We start from the $\mathcal{PT}$-symmetric AFMs. In the main text, we have shown that $\mathcal{PT}$ enforces $\textrm{TCD}=0$. So we will be brief: As shown in Fig.~\ref{AFM_Chiral_PT_T_symmetry}\textbf{a}, upon $\mathcal{PT}$, both the AFM order and light path remain invariant, but light helicity is reversed. As such, $\mathcal{PT}$ enforces the transmission coefficients for $\sigma^{\pm}$ to be identical, which means $\textrm{TCD} =0$. We can also show that RCD is allowed, because there is no symmetry that can keep the AFM order and reflection light path invariant but only flip the light helicity. For instance, $\mathcal{PT}$ changes the reflection to the bottom surface (Fig.~\ref{AFM_Chiral_PT_T_symmetry}\textbf{b}). On the other hand, $\mathcal{T}$ flips the AFM order (Fig.~\ref{AFM_Chiral_PT_T_symmetry}\textbf{c}). 

We now study the chiral crystals. We show that $\mathcal{T}$ enforces $\textrm{RCD}=0$. As shown in Fig.~\ref{AFM_Chiral_PT_T_symmetry}\textbf{d}, upon $\mathcal{T}$, the chiral crystal remains invariant, the reflection light path also stays the same, but light helicity is reversed. As such, $\mathcal{T}$ enforces the reflection coefficients for $\sigma^{\pm}$ to be identical, which means $\textrm{RCD} =0$ in chiral crystals. 

Using the same method, we can show that chiral crystals allow TCD, and that FMs allow both RCD and TCD.

\clearpage

\begin{wrapfigure}{r}{0.20\textwidth}
  \vspace{2.5ex}
  \includegraphics[width=0.20\textwidth]{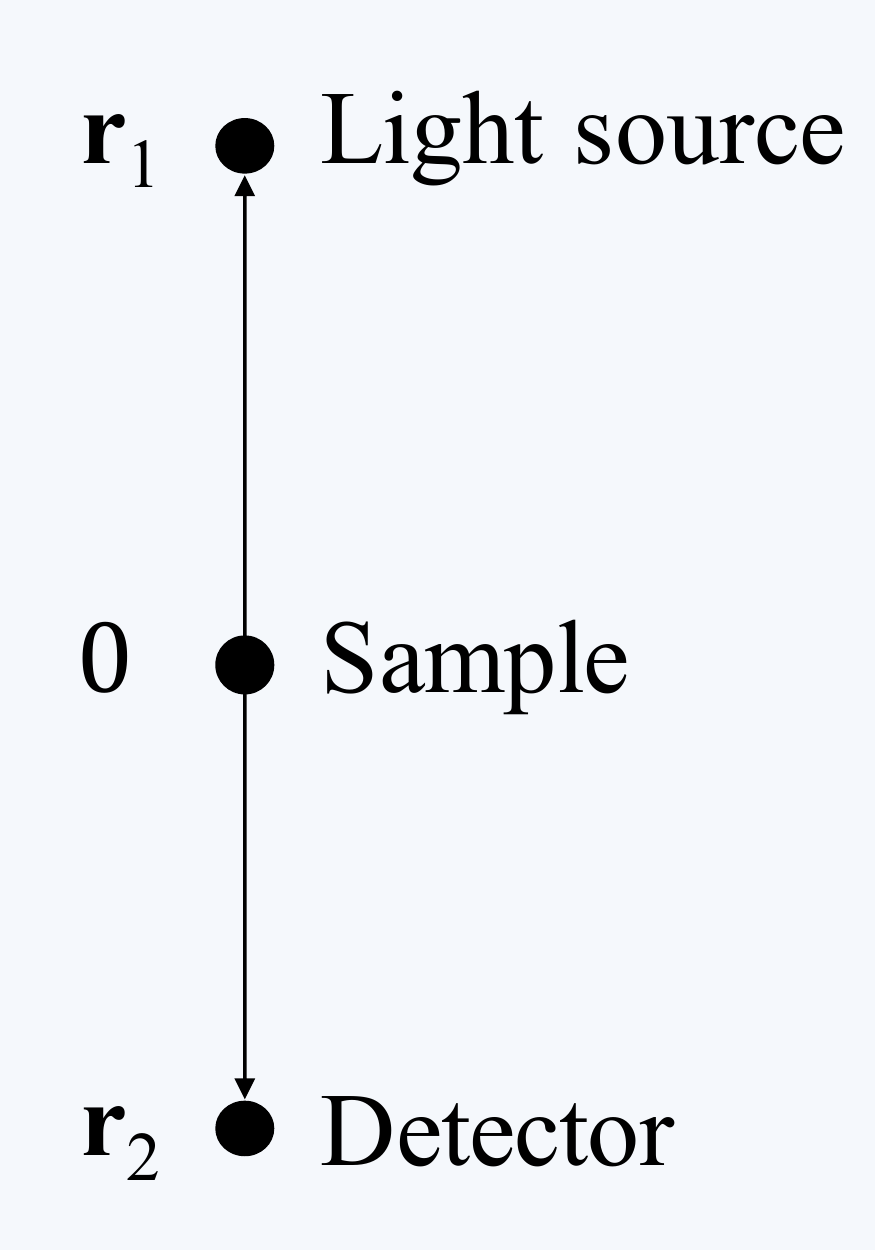}
   \vspace{-4.ex}
  \caption{\small Schematics layout.}
    \vspace{-10.ex}
  \label{Reciprocal}
\end{wrapfigure}
\leavevmode

\subsection*{III.4. Math derivation for symmetry analysis} 

\hspace{4mm} We now provide mathematical derivation for the above symmetry analysis. We will be brief in this section because the detailed derivation can be found in previous works by Halperin and others \cite{Halprin1992,fried2014relationship,armitage2014constraints,hosur2015kerr}. We consider a generic experimental layout (Fig.~\ref{Reciprocal}), which consists of the sample at the origin, the light source at position $\mathbf{r}_1$, and the detector at position $\mathbf{r}_2$. The propagator $\chi_{ij}(\mathbf{r}_1, \mathbf{r}_2)$ (the Green's function) contains all information about how the light starts from the light source at $\mathbf{r}_1$, interacts with sample at the origin, and reaches the detector at $\mathbf{r}_2$. This propagator is defined as follows

\begin{align}
\braket{\hat{A}_a(t_1,{\bf r}_1)}
=\braket{\hat{A}_a(t_1,{\bf r}_1)}_{J=0}+
\int dt_2d{\bf r}_2\chi_{ab}(t_1,{\bf r}_1;t_2,{\bf r}_2)J_b(t_2,{\bf r}_2),
\end{align} where $\hat{A}$ is the vector potential, $J_a(t',{\bf r}')$ is the source current, and
\begin{align}
\chi_{ab}(t_1,{\bf r}_1;t_2,{\bf r}_2)
&=\braket{\hat{A}_a(t_1,{\bf r}_1)\hat{A}_b(t_2,{\bf r}_2)}
=\sum_n\rho_n\braket{\hat{A}_a(t_1,{\bf r}_1)\psi_n|\hat{A}_b(t_2,{\bf r}_2)\psi_n}
\end{align}
is the Green's function for the vector potential $\hat{A}$, where $\psi_n$ is the eigenstate of the density operator $\hat{\rho}$ of the whole medium, including the sample and the vacuum while excluding the detector and source, where $\rho_n$ is the corresponding eigenvalue.

Therefore, we can perform symmetry analysis for the Green's function $\chi_{ij}(\mathbf{r}_1, \mathbf{r}_2)$. In particular, the circular dichroism with light propagating along the $z$ direction is given by the antisymmetric part of the Green's function
\begin{align}
\textrm{circular dichroism}\propto \chi_{xy}-\chi_{yx}
\end{align}

1. \textbf{Reflection CD:} For reflection CD, the key feature is that the light source and the detector are on the same side of the sample, so we can make them spatially overlap, i.e., $\mathbf{r}_2=\mathbf{r}_1$. Therefore, we have
\begin{align}
\textrm{Reflection CD}\propto \chi_{xy}(\mathbf{r}_1, \mathbf{r}_1)-\chi_{yx}(\mathbf{r}_1, \mathbf{r}_1).
\end{align}

One can show that time-reversal symmetry $\mathcal{T}$ enforces 
\begin{align}
\chi_{xy}(\mathbf{r}_1, \mathbf{r}_1)\equiv\chi_{yx}(\mathbf{r}_1, \mathbf{r}_1) \qquad \textrm{under }\mathcal{T}\textrm{ symmetry}
\end{align}
As such, we see that RCD identically vanishes under $\mathcal{T}$.\\

2. \textbf{Transmission CD:} For transmission CD, the key feature is that the light source and the detector are on the opposite side of the sample, so we can require, i.e., $\mathbf{r}_2=-\mathbf{r}_1$. Therefore, we have
\begin{align}
\textrm{Transmission CD}\propto \chi_{xy}(\mathbf{r}_1, -\mathbf{r}_1)-\chi_{yx}(\mathbf{r}_1, -\mathbf{r}_1).
\end{align}

One can show that space-time symmetry $\mathcal{PT}$ enforces 
\begin{align}
\chi_{xy}(\mathbf{r}_1, -\mathbf{r}_1)\equiv\chi_{yx}(\mathbf{r}_1, -\mathbf{r}_1) \qquad \textrm{under }\mathcal{PT}\textrm{ symmetry}
\end{align}
As such, we see that TCD identically vanishes under $\mathcal{PT}$.

\clearpage
\section*{IV. Theoretical studies}
\subsection*{IV.1. Theoretical expressions for calculating the Axion CD}

\begin{figure*}[h]
\centering
\includegraphics[width=14cm]{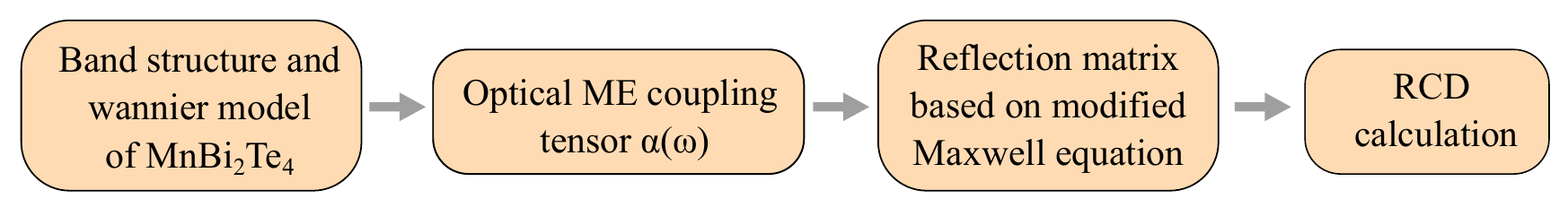}
\caption{\textbf{Key steps for the theoretical calculations of the Axion CD }}
\label{theory_flow}
\end{figure*}

\hspace{4mm}In this section, we provide the theoretical expressions that are needed to compute the RCD based on the optical Axion electrodynamics. The detailed derivations for these expressions can be found in Ref. \cite{ahn2022Axion} (Please note the same ME coefficient is denoted as $\alpha$ in our paper but $G^{(z)}$ in Ref. \cite{ahn2022Axion}). An illustration of the key steps is shown in Fig.~\ref{theory_flow}. 

\subsubsection*{IV.1.1. Band Structure and the Wannier model}

\hspace{4mm}First we calculated the band structure and Wannier functions of 2D even-layer $\mathrm{MnBi_2Te_4}$ flakes using first-principles DFT calculations.

\subsubsection*{IV.1.2. The Optical ME coupling}

\hspace{4mm} Using the first-principles electronic structure, we can calculate the optical ME coupling. Based on the detailed theoretical studies in Ref. \cite{ahn2022Axion}, the expression for optical ME coupling is as follows:
\begin{align}
\label{eq:Gij-quasi2D}
\alpha_{xx}(\omega) &= \frac{e^2 }{\hbar L}\sum_{\textrm{o,u}}\int d^2\textbf{k}\frac{\varepsilon_{\textrm{ou}}(\textbf{k})}{\varepsilon_{\textrm{uo}}(\textbf{k})-\hbar\omega}\ {\rm Im}[\frac{\hbar^2\braket{\textrm{o}(\textbf{k})| \hat{v}^x|\textrm{u}(\textbf{k})}\braket{\textrm{u}(\textbf{k})|-\frac{1}{2}\left(\hat{v}^y\hat{r}^z+\hat{r}^z\hat{v}^y\right)+\hat{m}^{\rm s}_x|\textrm{o}(\textbf{k})}}{\varepsilon_{\textrm{uo}}^2(\textbf{k})}], 
\end{align}
$\mathrm{o}(\textbf{k})$ and $\mathrm{u}(\textbf{k})$ are the occupied and unoccupied Bloch states and $\varepsilon_{\textrm{uo}}(\textbf{k})$ is their energy difference. As explained in the main text, the ME coupling can be decomposed into the spin and orbital (Berry curvature) contributions as follows:
\begin{align}
\alpha_{xx}(\omega)
&=\alpha^{\mathrm{Berry}}_{xx}(\omega)+\alpha^{\mathrm{spin}}_{xx}(\omega),\label{eq:Gij-quasi2D2}\\
\alpha_{xx}^{\mathrm{Berry}}(\omega) &=\frac{e^2 }{ \hbar L} \sum_{\mathrm{o, u}} \int d^2\textbf{k} \frac{\varepsilon_{\textrm{uo}}(\textbf{k})}{\varepsilon_{\textrm{uo}}(\textbf{k})-\hbar\omega} \ {\rm Im}[\frac{\hbar^2\braket{\textrm{o}(\textbf{k})| \hat{v}^x|\textrm{u}(\textbf{k})}\braket{\textrm{u}(\textbf{k})|\hat{v}^y\hat{r}^z|\textrm{o}(\textbf{k})}}{\varepsilon_{\textrm{uo}}^2(\textbf{k})}],\label{eq:Gij-quasi2D3}\\
\alpha_{xx}^{\mathrm{spin}}(\omega) &=\frac{e^2}{ \hbar L} \sum_{\textrm{o, u}} \int d^2\textbf{k} \frac{\varepsilon_{\mathrm{ou}}(\textbf{k})}{\varepsilon_{\mathrm{uo}}(\textbf{k})-\hbar\omega} \ {\rm Im}[\frac{\hbar^2\braket{\mathrm{o}(\textbf{k})| \hat{v}^x|\mathrm{u}(\textbf{k})}\braket{\mathrm{u}(\textbf{k})|\hat{m}^{\rm s}_x|\mathrm{o}(\textbf{k})}}{\varepsilon_{\textrm{uo}}^2(\textbf{k})}]\label{eq:Gij-quasi2D4},
\end{align}

\begin{figure*}[htb]
\centering
\includegraphics[width=10cm]{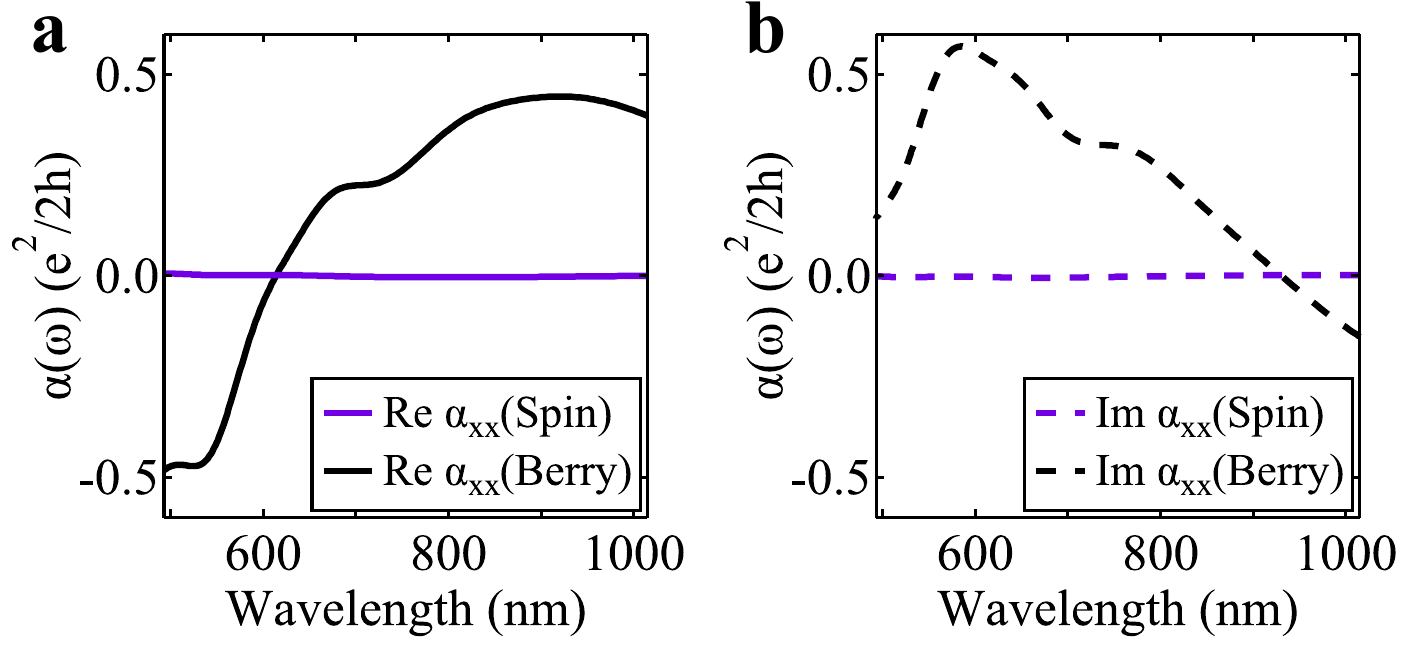}
\caption{Real and imaginary parts of $\mathrm{\alpha_{xx}^{spin}}$ and $\mathrm{\alpha_{xx}^{Berry}}$.}
\label{Theory_alpha_spin_ReIm}
\end{figure*}

Here we have assumed that $\hat{r}^z$ commutes with $\hat{v}^y$ (we have confirmed that this is a good approximation for our first-principles calculated DFT electronic structure). Equation~\ref{eq:Gij-quasi2D3} is the Berry curvature induced optical ME coupling we used in the main text. Equation~\ref{eq:Gij-quasi2D4} is the spin induced optical ME coupling. Figure~\ref{Theory_alpha_spin_ReIm} shows the real and imaginary components of $\alpha_{xx}(\omega)$ for both spin and orbital (Berry curvature) contributions. 

% The first term in \ref{eq:Gij-quasi2D} is the berry curvature induced optical ME coupling, and the second term originates from the magnetic spins. In the tight-binding model of $\mathrm{MnBi_2Te_4}$ where $\hat{r}^z$ commutes with $\hat{v}^y$, we can simplify the expression of $\alpha_{xx}(\omega)$
%The definition of the inter-band berry curvature is:
%\begin{align}
%\label{eq:Gij-quasi2D}
%\Omega^z_{ou}(\textbf{k})&=
%i\hbar^2\epsilon^{ijz}\frac{\braket{o(\textbf{k})| \hat{v}^i|u(\textbf{k})}\braket{u(\textbf{k})|\hat{v}^j|o(\textbf{k})}}{\varepsilon_{uo}(\textbf{k})^2}
%=\hbar^2 {\rm Im}[\frac{\braket{o(\textbf{k})| \hat{v}^x|u(\textbf{k})}\braket{u(\textbf{k})|\hat{v}^y|o(\textbf{k})}}{\varepsilon_{uo}(\textbf{k})^2}]
%\end{align}

\vspace{5mm}
In addition, we clarify the relationship between the optical ME coupling coefficients and Axion angle $\theta(\omega)$.

\begin{align}
\theta(\omega) &= \pi \frac{2h}{e^2}\frac{1}{3}\sum_{i=x,y,z}\alpha_{ii}(\omega), \label{eq:theta-definition}
\end{align}
\begin{align}
\alpha_{xx}(\omega) &= \frac{e^2 }{\hbar L}\sum_{\textrm{o,u}}\int d^2\textbf{k}\frac{\varepsilon_{\textrm{ou}}(\textbf{k})}{\varepsilon_{\textrm{uo}}(\textbf{k})-\hbar\omega}\ {\rm Im}[\frac{\hbar^2\braket{\textrm{o}(\textbf{k})| \hat{v}^x|\textrm{u}(\textbf{k})}\braket{\textrm{u}(\textbf{k})|-\frac{1}{2}\left(\hat{v}^y\hat{r}^z+\hat{r}^z\hat{v}^y\right)+\hat{m}^{\rm s}_x|\textrm{o}(\textbf{k})}}{\varepsilon_{\textrm{uo}}^2(\textbf{k})}], \\
\alpha_{yy}(\omega) &= \frac{e^2 }{\hbar L}\sum_{\textrm{o,u}}\int d^2\textbf{k}\frac{\varepsilon_{\textrm{ou}}(\textbf{k})}{\varepsilon_{\textrm{uo}}(\textbf{k})-\hbar\omega}\ {\rm Im}[\frac{\hbar^2\braket{\textrm{o}(\textbf{k})| \hat{v}^y|\textrm{u}(\textbf{k})}\braket{\textrm{u}(\textbf{k})|+\frac{1}{2}\left(\hat{v}^x\hat{r}^z+\hat{r}^z\hat{v}^x\right)+\hat{m}^{\rm s}_y|\textrm{o}(\textbf{k})}}{\varepsilon_{\textrm{uo}}^2(\textbf{k})}],\\
\alpha_{zz}(\omega) &= \frac{e^2 }{\hbar L}\sum_{\textrm{o,u}}\int d^2\textbf{k}\frac{\varepsilon_{\textrm{ou}}(\textbf{k})}{\varepsilon_{\textrm{uo}}(\textbf{k})-\hbar\omega}\ {\rm Im}[\frac{\frac{\hbar^2}{2}(\braket{\textrm{o}(\textbf{k})| \hat{r}^z|\textrm{u}(\textbf{k})}\braket{\textrm{u}(\textbf{k})| \hat{v}^x\hat{v}^y|\textrm{o}(\textbf{k})}-\braket{\textrm{o}(\textbf{k})|  \hat{r}^z\hat{v}^x|\textrm{u}(\textbf{k})}\braket{\textrm{u}(\textbf{k})|\hat{v}^y |\textrm{o}(\textbf{k})}}{\varepsilon_{\textrm{uo}}^2(\textbf{k})}\nonumber\\
& \qquad\qquad\quad\qquad\qquad\qquad\qquad\qquad\quad\quad\ \ \frac{-(x\leftrightarrow y))+\hbar^2\braket{\textrm{o}(\textbf{k})|\hat{v}^z|\textrm{u}(\textbf{k})}\braket{\textrm{u}(\textbf{k})| \hat{m}^{\rm s}_z |\textrm{o}(\textbf{k})}}{\varepsilon_{\textrm{uo}}^2(\textbf{k})}]
\end{align}

%-\braket{\textrm{o}(\textbf{k})| \hat{r}^z|\textrm{p}(\textbf{k})}\braket{\textrm{p}(\textbf{k})| \hat{v}^y|\textrm{u}(\textbf{u})}\braket{\textrm{o}(\textbf{k})| \hat{v}^x|\textrm{o}(\textbf{k})}

%\alpha_{zz}(\omega) &= \frac{e^2 }{\hbar L}\sum_{\textrm{o,u}}\int d^2\textbf{k}\frac{\varepsilon_{\textrm{ou}}(\textbf{k})}{\varepsilon_{\textrm{uo}}(\textbf{k})-\omega}\ {\rm Im}[\frac{\frac{1}{2}\left(\hat{r}^z_{nm}\hat{v}^y_{mp}\hat{v}^y_{pm}-\hat{r}^z_{np}\hat{v}^x_{pm}\hat{v}^y_{mn}-(x\leftrightarrow y)\right)+\hat{m}^{\rm s}_z}{\varepsilon_{\textrm{uo}}^2(\textbf{k})}],

We see that $\theta(\omega)$ is proportional to the trace of the optical ME tensor. For our CD experiments, $\alpha_{xx}$ and $\alpha_{yy}$ are relevant since the light's $E$ and $B$ fields are inside the 2D plane. Moreover, $\alpha_{xx}\equiv\alpha_{yy}$ because of $\mathrm{MnBi_2Te_4}$'s three-fold rotational symmetry. As such, $\alpha_{xx}(\omega)$ is sufficient for the CD calculations, so in the main text we omitted the subscript $xx$. In fact, our calculations (Fig.~\ref{Theory_all_alpha_xyz}) show that $\alpha_{xx}$, $\alpha_{yy}$ and $\alpha_{zz}$ have very similar amplitude. Therefore, we have 

\begin{align}
\theta(\omega) &= \pi \frac{2h}{e^2}\frac{1}{3}\sum_{i=x,y,z}\alpha_{ii}(\omega) \simeq \pi \frac{2h}{e^2}\alpha_{xx}(\omega)
\end{align}

%We will focus on $\alpha_{xx}$ because $\alpha_{xx}=\alpha_{yy}$. Due to the three fold symmetry (C3) of $\mathrm{MnBi_2Te_4}$, $\alpha_{yy}(\omega)$ is equivalent to $\alpha_{xx}(\omega)$. Since the light always incidents along the z-axis in our experiments, the light electric field is always in-plane. 

\begin{wrapfigure}{r}{0.25\textwidth}
  \vspace{5.5ex}
  \includegraphics[width=0.25\textwidth]{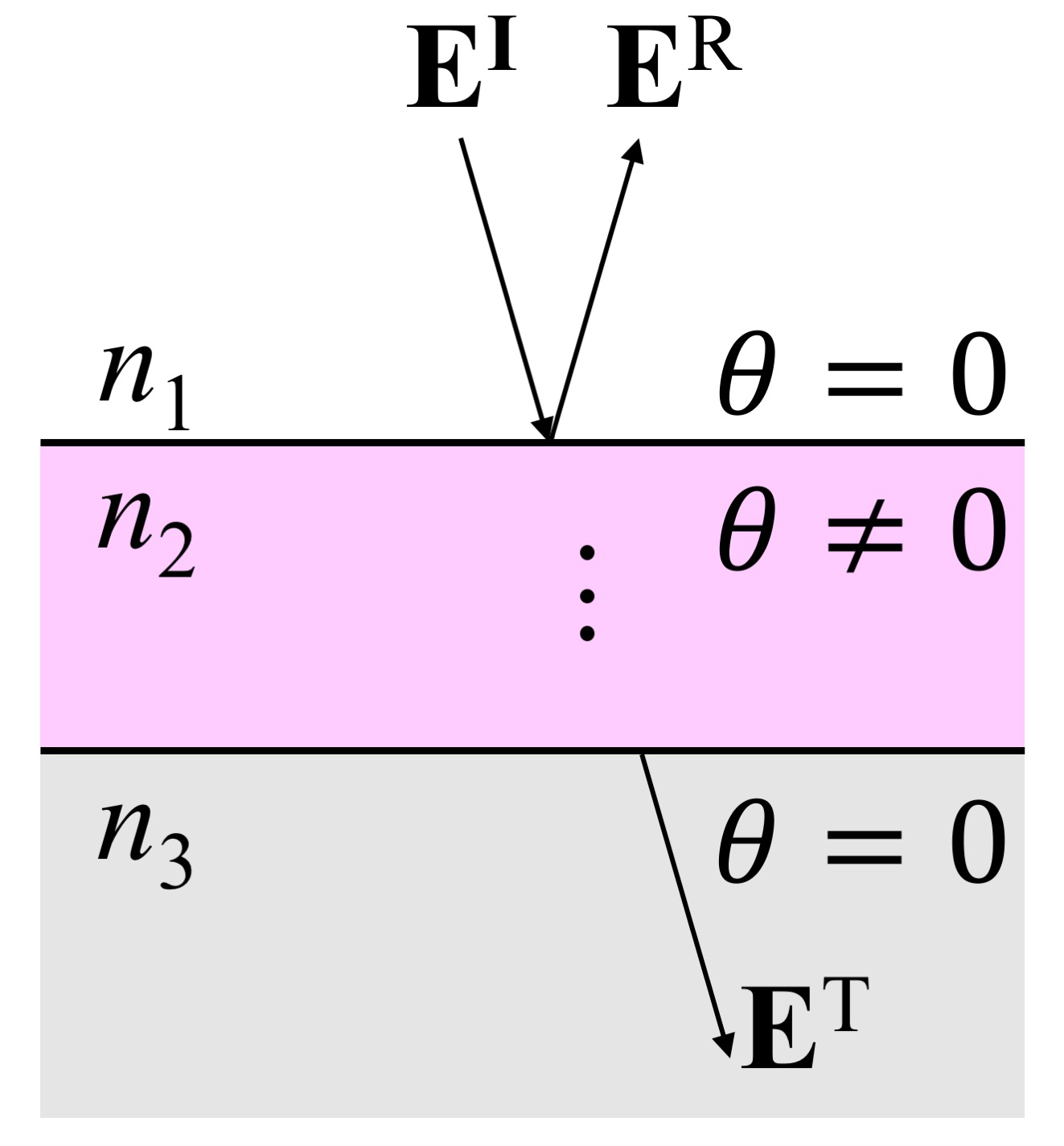}
   \vspace{-4.ex}
  \caption{\small Schematics of reflection and transmission.}
    \vspace{-10.ex}
  \label{Theory_medium_plot}
\end{wrapfigure}
\leavevmode

\subsubsection*{IV.1.3. Reflection matrix based on modified Maxwell equations}

\hspace{4mm} We now aim to analyze the light reflection based on the optical Axion electrodynamics. As shown in Fig.~\ref{Theory_medium_plot}, we consider an Axion insulator film (medium 2) sandwiched by two dielectric media (1 and 3). Suppose the incident electric field is $\mathbf{E}^{\textrm{I}}=(E^{\textrm{I}}_x,E^{\textrm{I}}_y)$. Then the reflected electric field can be related by a reflection matrix $\mathbf{E}^{\textrm{R}}=R\mathbf{E}^{\textrm{I}}$, and the reflection matrix $R$ is defined in Eq.~\ref{eq:Reflection_matrix}

%The reflection ($r$) and transmission ($t$) matrices are defined by considering light incident from medium 1 to medium 2 (Fig.~\ref{Theory_medium_plot}\textbf{a}). The electric field in medium $1$ consists of incident and reflected light while that in medium $2$ only consists of the transmitted light.
%\begin{align}
%\label{eq:Er_Et}o
%{\bf E}_1
%&={\bf E}^i+{\bf E}^r
%\equiv (1+r){\bf E}^i,\notag\\
%{\bf E}_2
%&={\bf E}^t
%\equiv t{\bf E}^i,
%\end{align}

\begin{figure*}[htb]
\centering
\includegraphics[width=16cm]{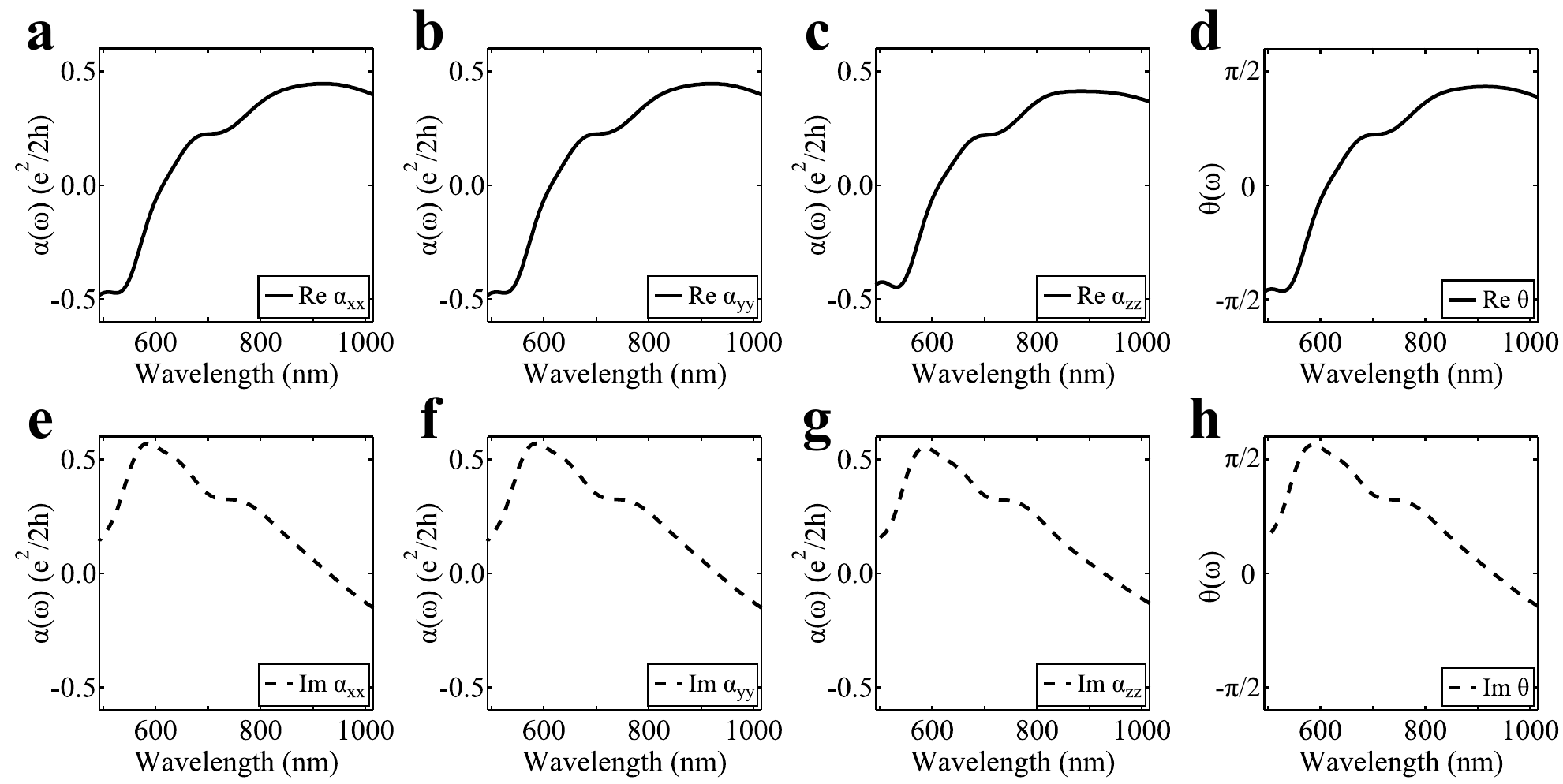}
\caption{Real and imaginary parts of $\alpha$ and $\theta$. Note that we only show the orbital (Berry curvature) contribution here, since the spin contribution is vanishingly small.}
\label{Theory_all_alpha_xyz}
\end{figure*}

\begin{align}
\label{eq:Reflection_matrix}
R
=\begin{pmatrix}
R_{xx}&R_{xy}\\
-R_{xy}&R_{xx}
\end{pmatrix}
%,\quad
%t
%=
%1+r
\end{align}

%\begin{figure*}[h]
%\centering
%\includegraphics[width=8cm]{Theory_medium_plot_v2}
%\caption{Schematics of reflection and transmission for two media (\textbf{a}) and three media (\textbf{b}).}
%\label{Theory_medium_plot}
%\end{figure*}

For the film with finite thickness considered here (Fig.~\ref{Theory_medium_plot}\textbf{b}), the final reflected electric field $\mathbf{E}^{\textrm{R}}$ consists of contributions from both the top (t) and the bottom (b) surfaces, which can be described as follows:

\begin{align}
\label{eq:rT_rB}
(R_\mathrm{t})_{xx}
&=\frac{(n_1-n_2)(n_1+n_2)-(\mu_0 c\alpha_{xx})^2}{(\mu_0 c\alpha_{xx})^2+(n_1+n_2)^2},\notag\\
(R_\mathrm{t})_{xy}
&=\frac{-2n_1(\mu_0 c\alpha_{xx})}{(\mu_0 c\alpha_{xx})^2+(n_1+n_2)^2},\notag\\
(R'_\mathrm{t})_{xx}
&=\frac{-(n_1-n_2)(n_1+n_2)-(\mu_0 c\alpha_{xx})^2}{(\mu_0 c\alpha_{xx})^2+(n_1+n_2)^2},\notag\\
(R'_\mathrm{t})_{xy}
&=\frac{-2n_2(\mu_0 c\alpha_{xx})}{(\mu_0 c\alpha_{xx})^2+(n_1+n_2)^2},\notag\\
(R_\mathrm{b})_{xx}
&=\frac{(n_2-n_3)(n_2+n_3)-(\mu_0 c\alpha_{xx})^2}{(\mu_0 c\alpha_{xx})^2+(n_2+n_3)^2},\notag\\
(R_\mathrm{b})_{xy}
&=\frac{2n_2(\mu_0 c\alpha_{xx})}{(\mu_0 c\alpha_{xx})^2+(n_2+n_3)^2}.
\end{align}

%\begin{align}
%\label{eq:rT_rB}
%(r_\mathrm{T})_{xx}
%&=\frac{(n_1-n_2)(n_1+n_2)-\bar{m}_2^2}{\bar{m}_2^2+(n_1+n_2)^2},\notag\\
%(r_\mathrm{T})_{xy}
%&=\frac{-2n_1\bar{m}_2}{\bar{m}_2^2+(n_1+n_2)^2},\notag\\
%(r'_\mathrm{T})_{xx}
%&=\frac{-(n_1-n_2)(n_1+n_2)-\bar{m}_2^2}{\bar{m}_2^2+(n_1+n_2)^2},\notag\\
%(r'_\mathrm{T})_{xy}
%&=\frac{-2n_2\bar{m}_2}{\bar{m}_2^2+(n_1+n_2)^2},\notag\\
%(r_\mathrm{B})_{xx}
%&=\frac{(n_2-n_3)(n_2+n_3)-\bar{m}_2^2}{\bar{m}_2^2+(n_2+n_3)^2},\notag\\
%(r_\mathrm{B})_{xy}
%&=\frac{2n_2\bar{m}_2}{\bar{m}_2^2+(n_2+n_3)^2}.
%\end{align}

where $R_\textrm{t}$ and $R_\textrm{t}'$ are reflections of the top surface from medium 1 to 2 and from medium 2 to 1, $R_\textrm{b}$ is the reflection of the bottom surface from medium 2 to 3, and $n_i$ is the refractive index of the medium $i$. We see that $\alpha_{xx}$ appears in the expressions. This is because to solve for  $R_\textrm{t}$, $R_\textrm{t}'$ and $R_\textrm{b}$, we needed to use the optical Axion electrodynamics based on the modified Maxwell equations. The detailed derivations are described in Ref.~\cite{ahn2022Axion}. 

\vspace{3mm}
We then plug in $R_\textrm{t}$, $R_\textrm{t}'$ and $R_\textrm{b}$ to the following equation. The final reflection matrix $R$ is then obtained by solving this equation:

%At last, by plugging the value of $r_\textrm{T}, r_\textrm{T}', r_\textrm{B}$ into equation \ref{eq:r_rT_rB}, we could obtain the final reflection matrix for $\mathrm{MnBi_2Te_4}$ flakes.
%In the case of the $\mathrm{MnBi_2Te_4}$ flake with a finite thickness, we then need to consider the case of three mediums (shown in Fig.~\ref{Theory_medium_plot}\textbf{b}). As a result,  we need to include the reflection from both the top surface and the bottom surface. Nevertheless, we can still use the reflection matrix to characterize the reflection from $\mathrm{MnBi_2Te_4}$ flakes, which just requires some modifications. 
\begin{align}
\label{eq:r_rT_rB}
R=R_\mathrm{t}+e^{2i\phi}(1+R'_\mathrm{t})R_\mathrm{b}\left(1-e^{2i\phi}R'_\mathrm{t}R_\mathrm{b}\right)^{-1}(1+R_\mathrm{t}),
\end{align}
Here, $\phi=n_2\omega L/c$ is the complex-valued phase obtained by the one-way propagation through the sample ($L$: sample thickness). 

%In addition, the subscripts of $\mathrm{T}$ and $\mathrm{B}$ indicate the top and bottom surface of the sample, and the prime sign indicates the reversed light propagation process (following the notations in Ref.~\cite{tse2010giant}). \\

\subsubsection*{IV.1.4.  RCD calculation}

\hspace{4mm} Based on the reflection matrix obtained above, we then calculate the RCD by:
\begin{align}
\label{eq:RCD_eq}
{\rm RCD}
&=\frac{|R_{++}|^2-|R_{--}|^2}{|R_{++}|^2+|R_{--}|^2}=\frac{2{\rm Im}[R_{xx}^*R_{xy}]}{|R_{xx}|^2+|R_{xy}|^2},
\end{align} where $R_{++}=2(R_{xx}-iR_{xy})$ and $R_{--}=2(R_{xx}+iR_{xy})$.

\subsubsection*{IV.1.5.  TCD calculation}

\hspace{4mm} Similarly, we could also express the transmission matrix $T$ as follows
\begin{align}
\label{eq:T_rT_rB}
T=e^{i\phi}(1+R_\mathrm{b})\left(1-e^{2i\phi}R'_\mathrm{t}R_\mathrm{b}\right)^{-1}(1+R_\mathrm{t})=\begin{pmatrix}
T_{xx}&T_{xy}\\
-T_{xy}&T_{xx}
\end{pmatrix}.
\end{align}
Based on the transmission matrix obtained above, we then calculate the TCD by:
\begin{align}
\label{eq:TCD_eq}
{\rm TCD}
&=\frac{|T_{++}|^2-|T_{--}|^2}{|T_{++}|^2+|T_{--}|^2}=\frac{2{\rm Im}[T_{xx}^*T_{xy}]}{|T_{xx}|^2+|T_{xy}|^2}
\end{align} where $T_{++}=2(T_{xx}-iT_{xy})$ and $T_{--}=2(T_{xx}+iT_{xy})$.
\\

\textbf{(1) No TCD with $\mathcal{PT}$ symmetry} 

When we consider the 2D even-layered $\mathrm{MnBi_2Te_4}$ flakes alone or hBN-encapsulated $\mathrm{MnBi_2Te_4}$ flakes, $\mathcal{PT}$ symmetry is strictly preserved and TCD is strictly prohibited. Indeed, using Eqs.~\ref{eq:T_rT_rB} and~\ref{eq:TCD_eq}, one can check that

\begin{align}
\label{eq:TCD_PT}
{\rm TCD} & \equiv 0 \qquad \textrm{under }\mathcal{PT}
\end{align}

\textbf{(2) Very small TCD with asymmetric dielectrics} 

%In the case of hBN-encapsulated $\mathrm{MnBi_2Te_4}$ flakes (hBN/$\mathrm{MnBi_2Te_4}$/hBN),  $\mathcal{PT}$ symmetry is still preserved, which means TCD is also prohibited.  On the other hand, f

In the experiments, the $\mathrm{MnBi_2Te_4}$ flakes are interfaced with different dielectric materials. So we need to reconsider the symmetry of the whole system. For the hBN/$\mathrm{MnBi_2Te_4}$/diamond structure, since the refractive indices of hBN and diamond are slightly different ($n^\mathrm{{hBN}=2.2}$, $n^\mathrm{{diamond}=2.4}$ at 1 eV \cite{lee2019refractive, zaitsev2013optical}), the $\mathcal{PT}$ symmetry of the whole system is slightly broken, which in principle could allow TCD (see detailed discussion in Ref. \cite{ahn2022Axion}). This TCD signal should be very small, since the degree of $\mathcal{PT}$ symmetry breaking is very weak. Main Fig.~4\textbf{f} shows the calculated results for the hBN/$\mathrm{MnBi_2Te_4}$/diamond structure. Indeed, our calculations show that the TCD in this case is negligibly small.

\vspace{5mm}
\subsection*{IV.2. Band structure of MnBi$_2$Te$_4$ and additional calculations}

\hspace{4mm} In this subsection, we provide an introduction of the electronic structure of 2D MnBi$_2$Te$_4$. We hope this serves as useful information for readers who are not familiar with this new material.

%Otrokov2019a, Deng2020, Liu2020a, Ge2020, Liu2020b, Deng2020a, yang2021odd, Ovchinnikov2020,gao2021layer,cai2021electric, li2021nonlocal,tai2021polarity
%\begin{wrapfigure}{r}{0.4\textwidth}
%  \vspace{-3.ex}
%  \includegraphics[width=0.4\textwidth]{ConvMag_TopoMag_v4.pdf}
%  \vspace{-5ex}
%  \caption{\small Known 2D magnets \textit{vs.} the MnBi$_2$Te$_4$.}
%  \label{ConvMag_TopoMag}
%  \vspace{-3ex}
%\end{wrapfigure}

\subsubsection*{IV.2.1. Localized magnetic ions versus low-energy itinerant electrons}

\begin{figure*}[h]
\centering
\includegraphics[width=7cm]{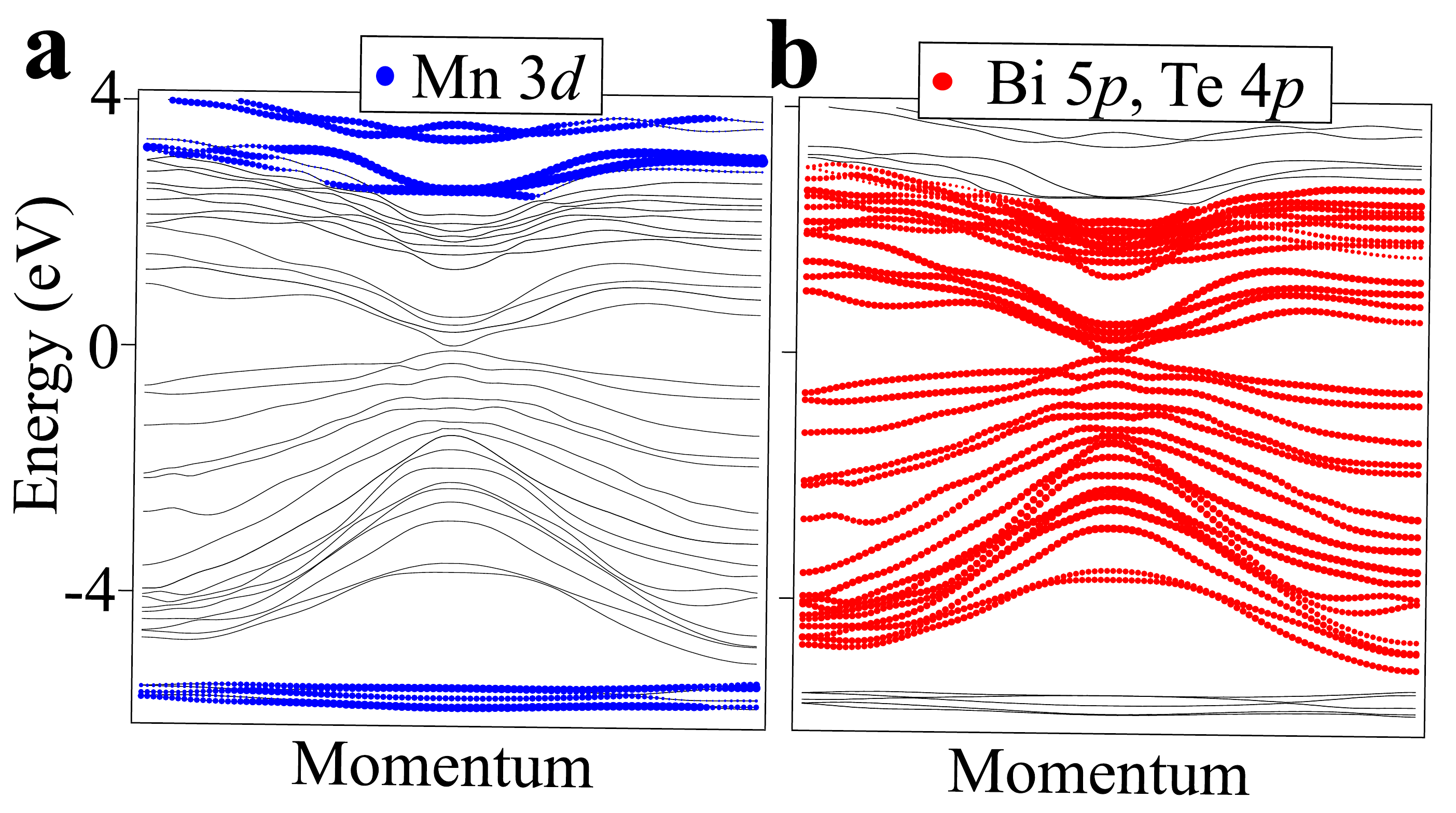}
\vspace{-4mm}
\caption{\textbf{a,} First-principles calculated electronic structure of MnBi$_2$Te$_4$ with the Mn $3d$ orbitals highlighted. \textbf{b,} Same as panel (\textbf{a}) but with the Bi $5p$ and Te $4p$ orbitals highlighted. }
\label{Mn_d}
\end{figure*}

The recent discoveries of the vdW magnets \cite{huang2017layer,gong2017discovery,burch2018magnetism} have attracted great interest as they enable us to explore magnetism in the ultra-2D limit. MnBi$_2$Te$_4$ \cite{Otrokov2019a, Deng2020, Liu2020a, Ge2020, Liu2020b, deng2021high, yang2021odd, Ovchinnikov2020,gao2021layer,cai2022electric, li2021nonlocal,tai2021polarity} provides novel possibilities: Known 2D magnets such as CrI$_3$ are wide-gap magnetic insulators whose physics is dominated by the localized magnetic ions. By contrast,  in addition to localized magnetic ions, MnBi$_2$Te$_4$ features low-energy itinerant electrons. These low-energy electronic states arise from the delocalized Bi and Te $p$ orbitals: They make up the bulk conduction and valance bands, with the topological surface states (TSSs) in between. Figure~\ref{Mn_d} shows the first-principles calculated electronic structure of MnBi$_2$Te$_4$ over a large energy window. Indeed, we see that the Mn $3d$ bands are located at very high energies. Therefore, in our optical experiments (photon energy $1.2$ eV to $2.4$ eV), they are not expected to directly contribute. This is consistent with our calculations showing that the spin contribution is negligibly small in the ME coupling.% ($\alpha^{\textrm{spin}}({\omega})\ll\alpha^{\textrm{Berry}}({\omega})$ in Fig.~\ref{Theory_alpha_spin_ReIm}). %Because of such electronic structure, we expect two distinct contributions for the ME coupling in MnBi$_2$Te$_4$: (1) Localized spin contribution from Mn ions (similar to other magnetoelectric or multiferroic insulators) and (2) Berry curvature contribution from the low-energy Bi and Te electrons, i.e., $\alpha(\omega)=\alpha^{\textrm{spin}}({\omega})+\alpha^{\textrm{Berry}}({\omega})$. 

%\begin{figure*}[h]
%\centering
%\vspace{-4mm}
%\includegraphics[width=12cm]{Magnetic_TI.pdf}
%\vspace{-4mm}
%\caption{Schematic illustration of how a nonmagnetic band insulator evolves into a topological insulator with Dirac surface states via inversion between bulk bands, and then to a magnetic topological insulator via introducing magnetism (the Dirac surface states open up a gap due to magnetism). }
%\label{Magnetic_TI}
%\end{figure*}

%2. Low-energy electronic structures of 2D MnBi$_2$Te$_4$
%\vspace{5mm}
%
%As a reference, Fig.~\ref{Layer_bands} shows the first-principles calculated low-energy electronic structures of 2D MnBi$_2$Te$_4$ as a function of thickness. We note the following points. (1) The odd layers are quantum anomalous Hall insulators (except 1SL). They have a nonzero $M$ and they break $\mathcal{PT}$ symmetry. Their bands are singly-degenerates. (2) The even layers are Axion insulators. They are fully-compensated ($M=0$) and they respect $\mathcal{PT}$ symmetry. Their bands are doubly-degenerates. (3) The number of bands increases with the number of layers. (4) The lowest two bands (1 conduction + 1 valence) are the Dirac surface states. The higher bands arise from inner layers, which will evolve into the true 3D bulk bands as we further increase the thickness.
\vspace{5mm}
\subsubsection*{IV.2.2. Low-energy electronic structures of 2D MnBi$_2$Te$_4$}

\begin{figure*}[h]
\centering
\includegraphics[width=13cm]{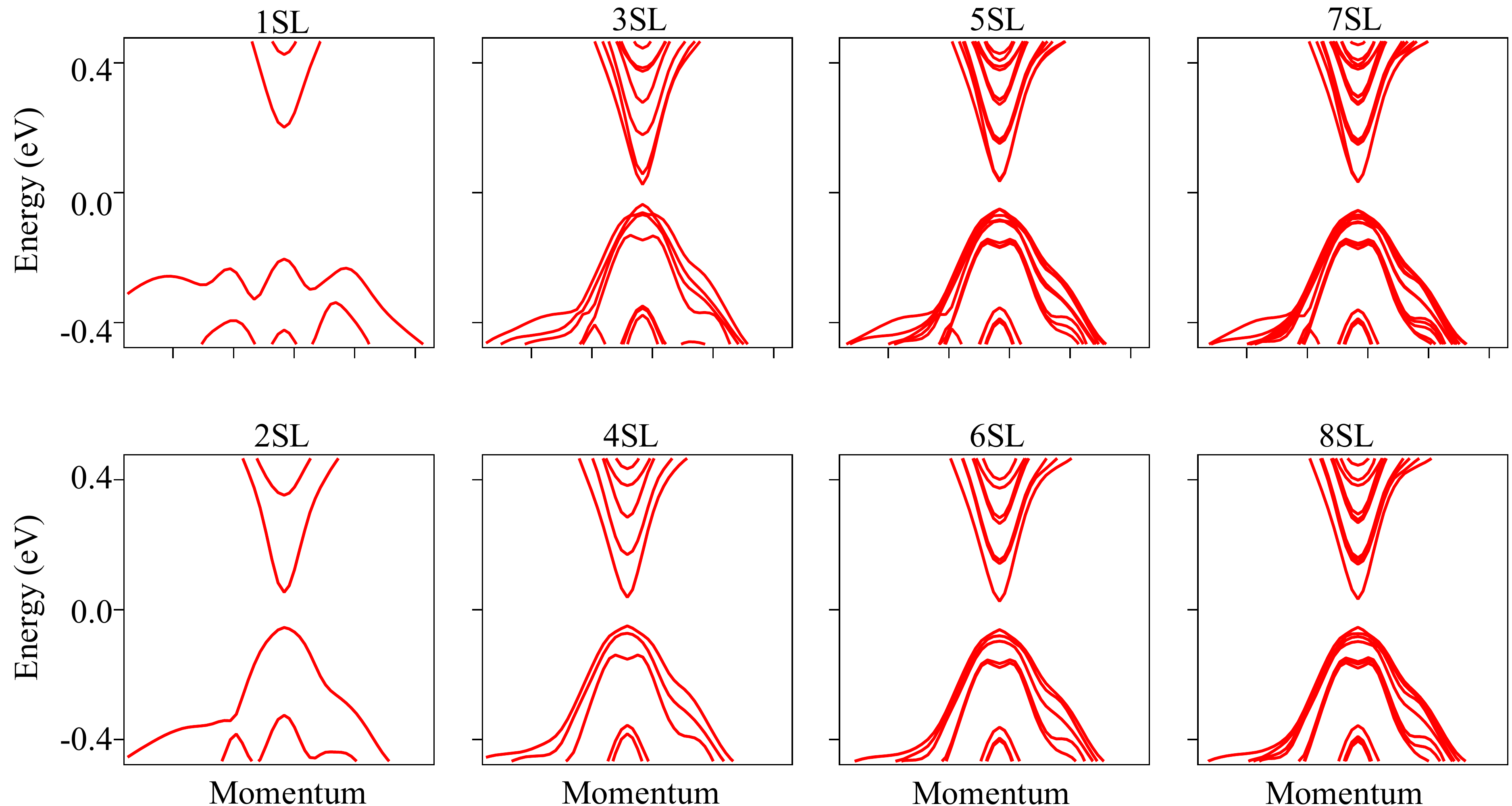}
\vspace{-4mm}
\caption{First-principles calculated band structures of 2D MnBi$_2$Te$_4$.}
\label{Layer_bands}
\end{figure*}

As a reference, Fig.~\ref{Layer_bands} shows the first-principles calculated low-energy electronic structures of 2D MnBi$_2$Te$_4$ as a function of thickness. We note the following aspects: (1) The lowest conduction band plus the highest valence band make up the topological surface states (TSSs). (2) The higher bands are the bulk bands. 

%The bulk bands are also crucial for the nontrivial topology. As shown in Fig.~\ref{Magnetic_TI}, the system changes from a normal insulator to a topological insulator because bulk bands go through a band inversion. 

\vspace{5mm}
\subsubsection*{IV.2.3. Band-resolved contributions for optical ME coupling}

\begin{figure*}[h]
\centering
\includegraphics[width=6cm]{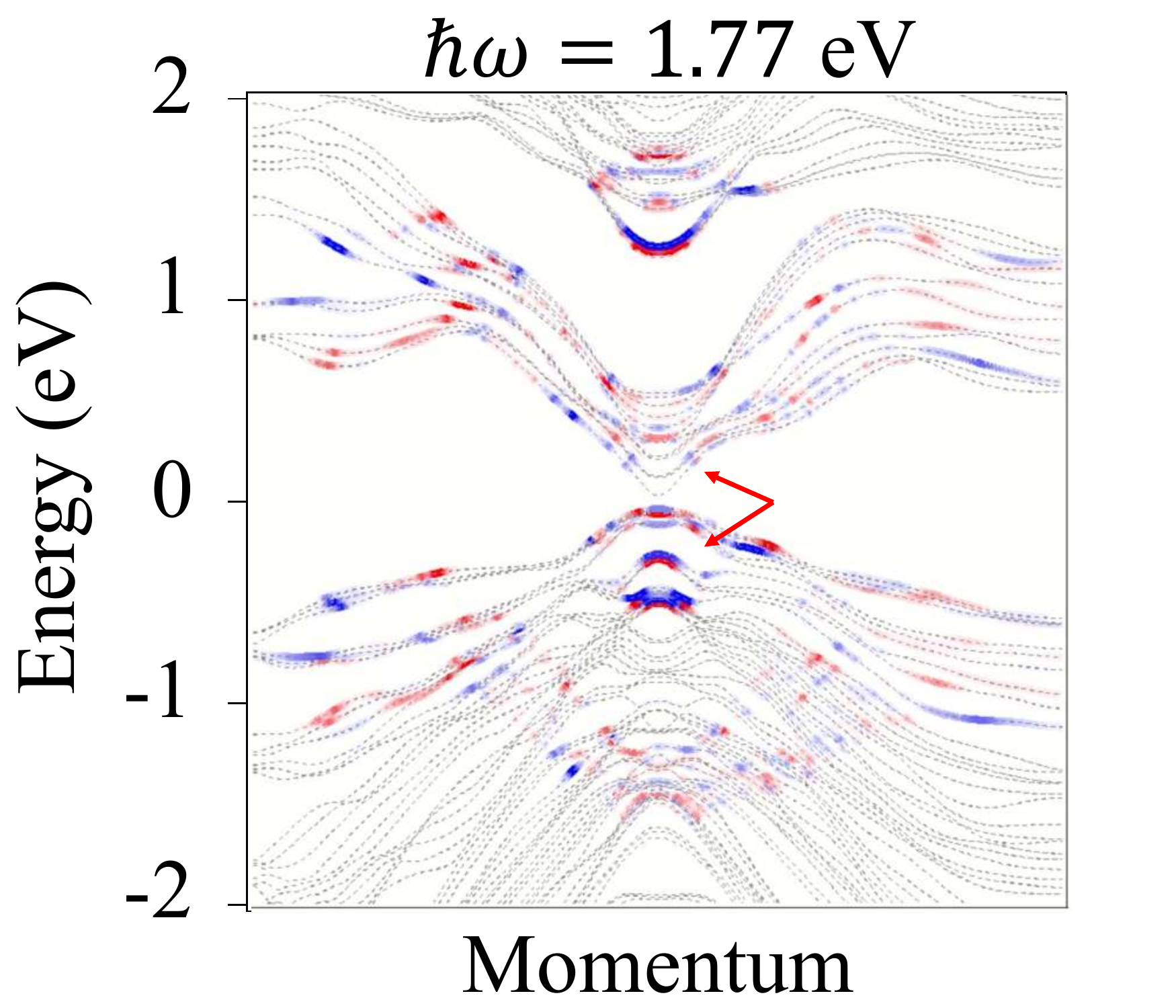}
\vspace{-1mm}
\caption{\textbf{Band-resolved contribution to the optical ME coupling at $\hbar\omega=1.77$ eV (700 nm).} }
\label{Band_resolved}
\end{figure*}

As described above, the Mn magnetic spins are not expected to contribute significantly to our optical ME coupling, because the Mn bands are located at very high energies (Fig.~\ref{Mn_d}). By contrast, the main contribution is from the Bi and Te electronic bands. Figure~\ref{Band_resolved} shows the band-resolved contribution for the optical ME coupling at $\hbar\omega=1.77$ eV.

\begin{figure*}[h]
\centering
\includegraphics[width=12cm]{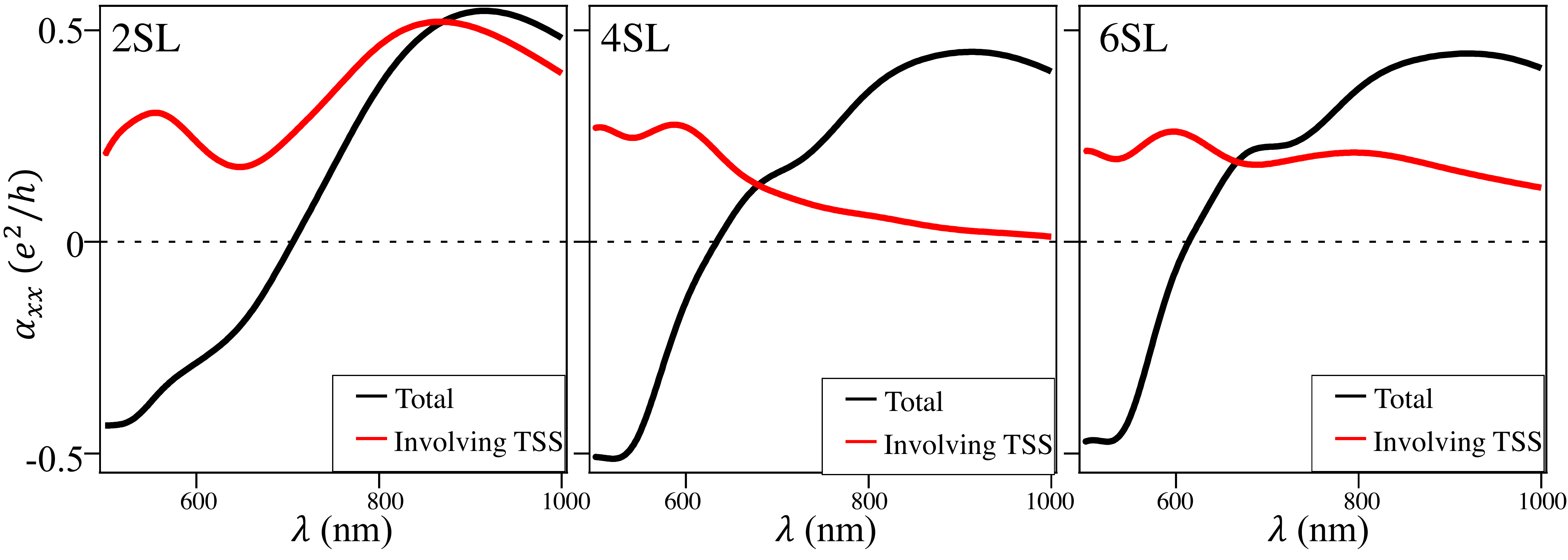}
\vspace{-1mm}
\caption{Calculated $\alpha(\omega)$ for 2, 4, 6 SL from all possible contributions (black) and$\alpha(\omega)$ from contributions only involving the lowest CB or highest VB.}
\label{TSS_thickness}
\end{figure*}

In Fig.~\ref{TSS_thickness}, we compare the total $\alpha$ and  the $\alpha$ from TSS contributions (i.e., inter-band transitions where the initial state or final state or both are the TSSs) for 2, 4, and 6SL.  While the TSSs give finite contribution, the contribution from higher bands is generally larger. For 2SL, the TSS contribution is quite significant, which is possibly because there are fewer higher bulk bands in 2SL. 

%\begin{figure*}[h]
%\centering
%\includegraphics[width=16cm]{./SI_Figures/Reply_figure/SI_fig8_12_Txx_Axion_CD.pdf}
%\caption{\textbf{a,} Calculated $\mathrm{T_{xx}}$ and $\theta$.  \textbf{a,} RCD calculated from the $\mathrm{T_{xx}}$ term and  Axion $\theta$ term. \textbf{c. } Experimentally %measured RCD. }
%\label{Theta_Txx_calc}
%\end{figure*}

\begin{figure*}[h]
\centering
\includegraphics[width=10cm]{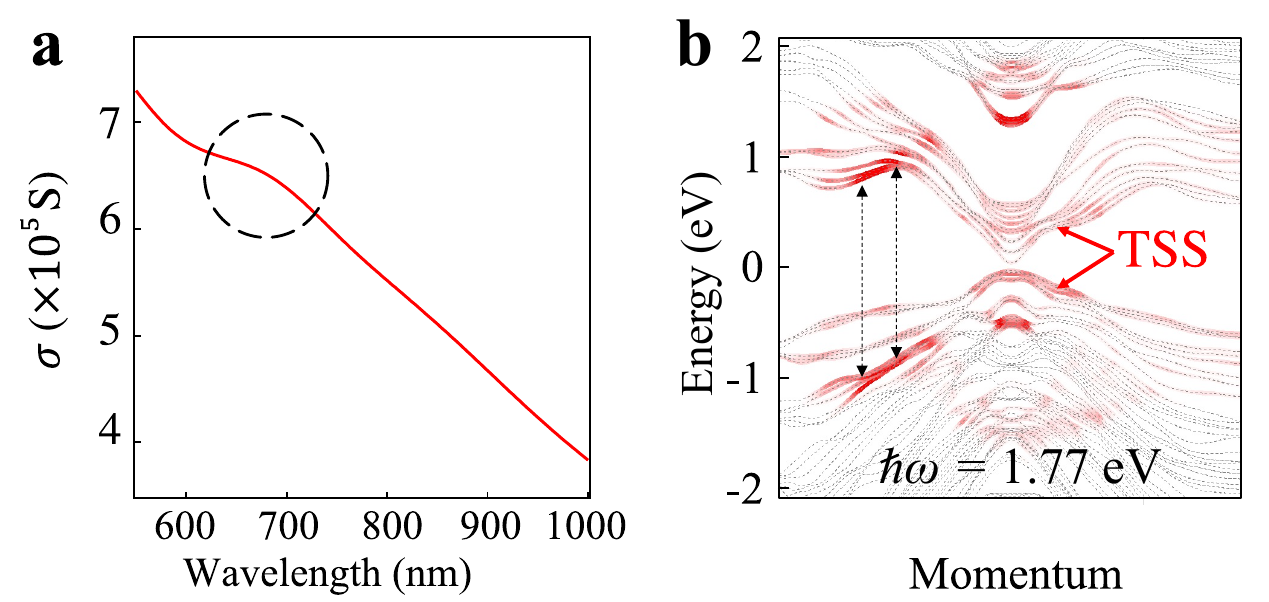}
\vspace{-1mm}
\caption{\textbf{a,} Calculated optical conductivity $\sigma(\omega)$ for 6SL. \textbf{b,} Band resolved contribution for the calculated optical conductivity $\sigma(\omega)$. The lowest conduction and highest valence bands are the topological surface states (TSS).}
\label{Optical_conductivity}
\end{figure*}

In Fig.~\ref{Optical_conductivity}\textbf{a}, we compute the optical conductivity $\sigma(\omega)$ from $550-1000$ nm for 6SL MnBi$_2$Te$_4$. We observe a small bump near $700$ nm ($1.77$ eV). In Fig.~\ref{Optical_conductivity}\textbf{b}, we show a band resolved contribution for $\sigma(\omega)$. This small bump may arise from the transitions shown by the black arrows shown in Fig.~\ref{Optical_conductivity}\textbf{b}, where the conduction and valence bands turn out to be roughly ''parallel''. 
\color{black}

%
%\begin{align}
%\label{eq:Gij-quasi2D}
%\alpha_{xx}(\omega) &= \frac{e^2 }{\hbar L}\sum_{\textrm{o,u}}\int d^2\textbf{k}\frac{\varepsilon_{\textrm{ou}}(\textbf{k})}{\varepsilon_{\textrm{uo}}(\textbf{k})-\hbar\omega+i\hbar/\tau}\ {\rm Im}[\frac{\braket{\textrm{o}(\textbf{k})| \hat{v}^x|\textrm{u}(\textbf{k})}\braket{\textrm{u}(\textbf{k})|-\frac{1}{2}\left(\hat{v}^y\hat{r}^z+\hat{r}^z\hat{v}^y\right)+\hat{m}^{\rm s}_x|\textrm{o}(\textbf{k})}}{\varepsilon_{\textrm{uo}}^2(\textbf{k})}], 
%\end{align}

%This is consistent with our calculations showing that the spin contribution is negligibly small in the ME coupling ($\alpha^{\textrm{spin}}({\omega})\ll\alpha^{\textrm{Berry}}({\omega})$ in Fig. 4\textbf{e}).Our optical experiments (photon energy $1.2$ eV to $2.4$ eV), they are not expected to directly contribute. This is consistent with our calculations showing that the spin contribution is negligibly small in the ME coupling ($\alpha^{\textrm{spin}}({\omega})\ll\alpha^{\textrm{Berry}}({\omega})$ in Fig. 4\textbf{e}). Therefore, in MnBi$_2$Te$_4$, the itinerant electrons from Bi and Te orbitals dominate the low-energy optical and transport physics, whereas the localized magnetic ions provide the background magnetism.

%As such, in MnBi$_2$Te$_4$, we not only have localized ions providing the underlying magnetism, but also low-energy topological electrons responsible for the emergent transport. This allows us to ask exciting questions inaccessible in known 2D magnets: How does the magnetism affects topology? Can topological electrons inversely influence the magnetism? What happens when considering layer and stacking degrees of freedom? By explore these questions, we can (1) realize ultra-tunable 2D magnetism, which is of great interest given the rareness of 2D magnets. (2) provide the first demonstration of the topological magnetoelectric effect arising from the topological $\theta$ term, and (3) understand the magnetism in 2D MnBi$_2$Te$_4$, which sets the foundation for its nontrivial topology. 

\clearpage
\subsection*{IV.3. Mathematical derivation of RCD/Kerr under different symmetry condition}
Below, we show that by considering the general wave equation and proper boundary conditions, we can show in a unifying scheme that RCD/Kerr is zero under $\mathcal{T}$-symmetry; but RCD/Kerr is nonzero with $\mathcal{T}$-breaking (especially for our MnBi$_2$Te$_4$ system). In this derivation, we used $G$ for the optical ME coupling.

\vspace{4mm}
\textbf{1. Wave equation}

  {The wave equation up to electric quadrupole/magnetic dipole takes the following form~\cite{raab2004multipole}:
\begin{align}
\left[\delta_{ij}+\epsilon_0^{-1}\tilde{\chi}_{ij}
+i\mu_0nc\sum_k\kappa_k\tilde{\sigma}_{ijk}
+n^2(\kappa_i\kappa_j-\delta_{ij})\right]E_j=0
\end{align}
where $\tilde{\chi}_{ij}=\chi_{ij}-i\chi'_{ij}$ satisfying $\chi_{ij}=\chi_{ji}$ and $\chi'_{ij}=-\chi'_{ji}$, $\kappa_i=k_i/|{\bf k}|$ is the propagation direction of light, $n$ is the refractive index, $\tilde{\sigma}_{ijk}=\sigma_{ijk}^{\textrm{S}}-i\sigma_{ijk}^{\textrm{A}}$ is the complex bulk conductivity coefficient defined by $\tilde{\sigma}_{ij}({\bf q})=\tilde{\sigma}_{ij}(0)+\tilde{\sigma}_{ijk}q_k+\hdots$, and 
\begin{align}
\sigma_{ijk}
&=i\left[
\epsilon_{ikl}G_{jl}+\epsilon_{jkl}G_{il}
-\frac{1}{2}\omega(a_{ijk}'+a_{jik}')\right]
=\sigma_{jik},\notag\\
\sigma'_{ijk}
&=i\left[-\epsilon_{ikl}G'_{jl}+\epsilon_{jkl}G'_{il}
+\frac{1}{2}\omega(a_{ijk}-a_{jik})\right]
=-\sigma'_{jik}.
\end{align}
}

  {Let us assume $C_{3z}$ and $C_{2x}$ symmetries for simplicity. We further impose that the bulk Hall response is zero, i.e., $\chi'_{ij}=0$, in order to focus on the magneto-electric and electric-quadrupole effects. For ${\mathbf{\kappa}}=\pm\hat{z}$, the wave equation is
\begin{align}
\begin{pmatrix}
1+\epsilon_0^{-1}\chi_{xx}-n^2& n\kappa_z\mu_0c\sigma'_{xyz}\\
-n\kappa_z\mu_0c\sigma'_{xyz}&1+\epsilon_0^{-1}\chi_{xx}-n^2
\end{pmatrix}
\begin{pmatrix}
E_x\\
E_y
\end{pmatrix}=0.
\end{align}
The refractive index satisfying the wave equation is given by
\begin{align}
\label{eq:npm}
n_{\pm}^{\kappa_z}=\sqrt{1+\epsilon_0^{-1}\chi_{xx}+(i\mu_0c\sigma'_{xyz}/2)^2}\mp i\kappa_z\mu_0c\sigma'_{xyz}/2,
\end{align}
for circular polarization $\hat{\pm}=\hat{x}+i\hat{y}$.}

\vspace{4mm}
\textbf{2. Reflection and transmission from a single interface}

  {We consider the interface of medium 1 ($z>0$) and medium 2 ($z<0$) with the surface normal $\hat{z}$.
For normal incidence, in the circularly polarized basis,
\begin{align}
\begin{pmatrix}
H_+\\
H_-
\end{pmatrix}
=\frac{1}{\mu_0c}
\begin{pmatrix}
-\mu_0c\tilde{T}^{\mu}_{xx}-in^{\kappa_z}_{\mu+}\kappa_z &0\\
0& -\mu_0c\tilde{T}^{\mu}_{xx}+in^{\kappa_z}_{\mu-}\kappa_z
\end{pmatrix}
\begin{pmatrix}
E_+\\
E_-
\end{pmatrix}
\end{align}
within the media $\mu=1$ or $2$, where $n_{\pm}$ depends on the sign $\kappa_z$, and $\kappa_z=-1$ for incident and transmitted light, while $\kappa_z=1$ for reflected light.
Here,
\begin{align}
\tilde{T}_{xx} &= T_{xx} - iT_{xx}'\notag\\
&=\frac{1}{3}(G_{xx}-G_{zz})-\frac{1}{6}\omega(a'_{yzx}-a'_{zyx})
-i\left[G'_{xx}-\frac{1}{2}\omega(a_{yzx}-a_{zyx})\right]\notag\\
&=
\frac{i}{3}\sigma_{zxy}
-\frac{1}{2}\sigma'_{xyz}.
\end{align}
As we consider light incident from medium $1$ to medium $2$, the electric field in medium $1$ consists of incident and reflected fields while that in medium $2$ is the transmitted field.
\begin{align}
{\bf E}_1
&={\bf E}^i+{\bf E}^r
\equiv (1+r){\bf E}^i,\notag\\
{\bf E}_2
&={\bf E}^t
\equiv t{\bf E}^i,
\end{align}
where
\begin{align}
r
=
\begin{pmatrix}
r_{++}&r_{++}\\
r_{-+}&r_{--}
\end{pmatrix}
=\begin{pmatrix}
r_{++}&0\\
0&r_{--}
\end{pmatrix}
,\quad
t
=
1+r
\end{align}
by $C_{3z}$ symmetry and the continuity of ${\bf E}$ at the interface.}

  {The ${\bf B}$ field satisfies the boundary condition as follows
\begin{align}
{\bf B}^t
&={\bf B}^i+{\bf B}^r+\mu_0\hat{z}\times{\bf j}_s,
\end{align}
where ${\bf j}_s
=[(e^2/2\pi h)(\theta^{(z)}_2-\theta^{(z)}_1)
+(\tilde{T}^{2}_{xx}-\tilde{T}^{1}_{xx})]{\bf E}\times\hat{z}$ is the total two-dimensional surface current density.
By solving the boundary condition, we obtain
\begin{align}
r_{++}
&=\frac{n_{1\textrm{L}}-n_{2\textrm{L}}-i\mu_0c\sigma^s_{xy}}{n_{1\textrm{R}}+n_{2\textrm{L}}+i\mu_0c\sigma^s_{xy}},\notag\\
r_{--}
&=\frac{n_{1\textrm{R}}-n_{2\textrm{R}}+i\mu_0c\sigma^s_{xy}}{n_{1\textrm{L}}+n_{2\textrm{R}}-i\mu_0c\sigma^s_{xy}},\notag\\
r_{+-}
&=r_{-+}=0,
\end{align}
where $n_{\mu \textrm{L}}=n_{\mu+}^-=n_{\mu-}^+$ is the refractive index for left circularly polarization, and $n_{\mu R}=n_{\mu-}^-=n_{\mu+}^+$ is the refractive index for the right circularly polarization, where the expression of $n_{\mu+}^{\kappa_z}$ is given by Eq.~\ref{eq:npm}, and
\begin{align}
\sigma^s_{xy}
=(e^2/2\pi h)(\theta^{(z)}_2-\theta^{(z)}_1)
+(\tilde{T}^{2}_{xx}-\tilde{T}^{1}_{xx})
\end{align}
is the two-dimensional surface conductivity.
From the expressions of $r_{++}$ and $r_{--}$ and Eq.~\eqref{eq:npm}, we obtain the Kerr angle
\begin{align}
\varphi_K
&=\tan^{-1}\frac{r_{xy}}{r_{xx}}=\tan^{-1}\left[\frac{-i(r_{++}-r_{--})}{r_{++}+r_{--}}\right].
\end{align}}

\vspace{4mm}
\textbf{3. Cancellation and absence of Kerr in nonmagnetic systems}

  {In a nonmagnetic system, the $\mathcal{T}$-symmetry dictates $G_{ij}=a'_{ijk}=T_{xx}=\theta^{(z)}=0$, $G'_{ij}, a_{ijk}, T'_{xx}\neq0$, Therefore, we found
\begin{align}
r_{++}-r_{--} & \propto[2i\mu_0c(T'^{\mu=2}_{xx}-T'^{\mu=1}_{xx})]-[n_{\textrm{1\textrm{R}}}-n_{\textrm{1\textrm{L}}}-(n_{\textrm{2\textrm{R}}}-n_{\textrm{2\textrm{R}}})]\notag\\
&=[i\mu_0c(\sigma^{\textrm{A}, \mu=2}_{xyz}-\sigma^{\textrm{A}, \mu=1}_{xyz})]-[n_{\textrm{1\textrm{R}}}-n_{\textrm{1\textrm{L}}}-(n_{\textrm{2\textrm{R}}}-n_{\textrm{2\textrm{R}}})]\notag\\
&=0
\end{align}
Again, we have $n_{\mu \textrm{L}}=n_{\mu+}^-=n_{\mu-}^+$ is the refractive index for left circularly polarization, and $n_{\mu R}=n_{\mu-}^-=n_{\mu+}^+$ is the refractive index for the right circularly polarization, where the expression of $n_{\mu+}^{\kappa_z}$ is given by $n_{\pm}^{\kappa_z}=\sqrt{1+\epsilon_0^{-1}\chi_{xx}+(i\mu_0c\sigma'_{xyz}/2)^2}\mp i\kappa_z\mu_0c\sigma'_{xyz}/2$. We also have $\tilde{T}^{\mu}_{xx}=T'^{\mu}_{xx}=\sigma^{\textrm{A}, \mu=2}_{xyz}/2$ because $T^{\mu}_{xx}=0$ with $\mathcal{T}$ symmetry (Eq.~\ref{eq:npm}). }

  {We can understand this cancellation as a compensation between bulk and surface responses. 
\begin{itemize}
\item The refractive indices part $``n_{\textrm{1\textrm{R}}}-n_{\textrm{1\textrm{L}}}-(n_{\textrm{2\textrm{R}}}-n_{\textrm{2\textrm{R}}})''$ are responsible for circular birefringence (the natural optical activity) in the bulk.
\item $\sigma^{\textrm{A}, \mu=2}_{xyz}-\sigma^{\textrm{A}, \mu=1}_{xyz}$ is responsible for the surface current that leads to the jump of $B$ field at the surface.
\end{itemize}}

  {Their effects cancel such that there is no net polar Kerr rotation (i.e., no Kerr rotation at normal incidence), compatible with results from previous papers including Refs. \cite{agranovich1973phenomenological, Halperin_book, hosur2015erratum}.}

\vspace{4mm}
\textbf{4. No cancellation and nonzero Kerr in MnBi$_2$Te$_4$}

  {By contrast, in $\mathcal{PT}$-symmetric AFMs, the  $\mathcal{PT}$-symmetry and the $\mathcal{T}$-breaking dictates $G'_{ij}=a_{ijk}=T'_{xx}=0$, $G_{ij}, a'_{ijk}, T_{xx}, \theta^{(z)}\neq0$. Correspondingly, we got\begin{align}
r_{xy}
&=\frac{1}{2i}(r_{++}-r_{--})
=-\frac{2n_1\mu_0c[\frac{e^2}{2\pi h}(\theta^{(z)}_2-\theta^{(z)}_1)
+(\tilde{T}^{2}_{xx}-\tilde{T}^{1}_{xx})]}{(\mu_0c[\frac{e^2}{2\pi h}(\theta^{(z)}_2-\theta^{(z)}_1)
+(\tilde{T}^{2}_{xx}-\tilde{T}^{1}_{xx})])^2+(n_1+n_2)^2}.
\end{align}}

  {Here, we see that the Kerr rotation has two origins: $(\tilde{T}^{2}_{xx}-\tilde{T}^{1}_{xx})$ is the gyrotropic birefringence (which has been known previous) and $(\theta^{(z)}_2-\theta^{(z)}_1)$ is the Axion contribution (our new results). 
\begin{itemize}
\item In MnBi$_2$Te$_4$, $\tilde{T}^{2}_{xx}-\tilde{T}^{1}_{xx}=0$ because the bulk MnBi$_2$Te$_4$ respects inversion symmetry. So the dominant contribution is the optical Axion electrodynamics. 
\item In Cr$_2$O$_3$, both contributions exist.
\end{itemize}}

  {In any case, there is no cancellation hence the Kerr rotation is always nonzero in $\mathcal{PT}$-symmetric AFM including even-layered MnBi$_2$Te$_4$.}

\clearpage
\section*{V. Additional discussion}

\subsection*{V.1. Additional discussion about Berry curvature real space dipole}

\hspace{4mm} In the main text, we noted that the $\alpha_{xx}^{\mathrm{Berry}}(\omega)$ can be understood as the Berry curvature real space dipole. We provide some additional discussion about this point.

\begin{wrapfigure}{r}{0.25\textwidth}
  \vspace{-6.5ex}
  \includegraphics[width=0.25\textwidth]{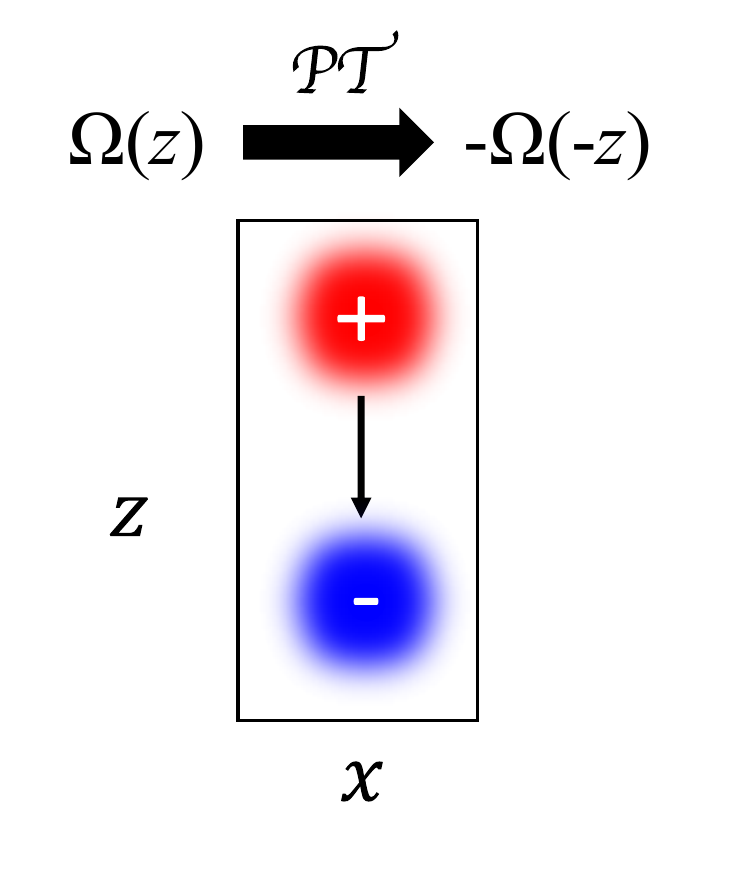}
   \vspace{-7.ex}
  \caption{\small $\mathcal{PT}$ symmetry.}
    \vspace{-1.ex}
  \label{PTsymmetry}
\end{wrapfigure}
\leavevmode

  \vspace{1.ex}
\subsubsection*{V.1.1. Berry curvature real space dipole is dictated by $\mathcal{PT}$ symmetry}

$\mathcal{PT}$ maps real space location $z$ to $-z$; $\mathcal{PT}$ also flips $\Omega$ to $-\Omega$. Therefore, we have the following relationship
\leavevmode

\begin{equation}
\Omega (z) \stackrel{\mathcal{PT}}{=}  - \Omega (-z),
\label{eq:PT}
\end{equation}
\leavevmode

which means that the Berry curvature $\Omega$ at location $z$ is always opposite to that at $-z$. Therefore,  $\mathcal{PT}$ dictates (1) the total Berry curvature summed up all over the space identically vanishes but (2) there is a real space dipole of Berry curvature. 

As a side note, using the same philosophy, we can study spatially-chiral materials, where time-reversal symmetry $\mathcal{T}$  dictates $\Omega (k) \stackrel{\mathcal{T}}{=}  - \Omega (-k)$. So in noncentrosymmetric materials (1) total Berry curvature summed up all over the $k$ space identically vanishes but (2) there is a $k$ space dipole of Berry curvature. We can also study FMs, where space-inversion symmetry $\mathcal{P}$  dictates $\Omega (k) \stackrel{\mathcal{P}}{=}  \Omega (-k)$. So the total Berry curvature in FMs is nonzero.

  \vspace{3.ex}
\subsubsection*{V.1.2. The ME effect can be visualized by the Berry curvature real space dipole}

\begin{figure*}[h]
\centering
\includegraphics[width=12cm]{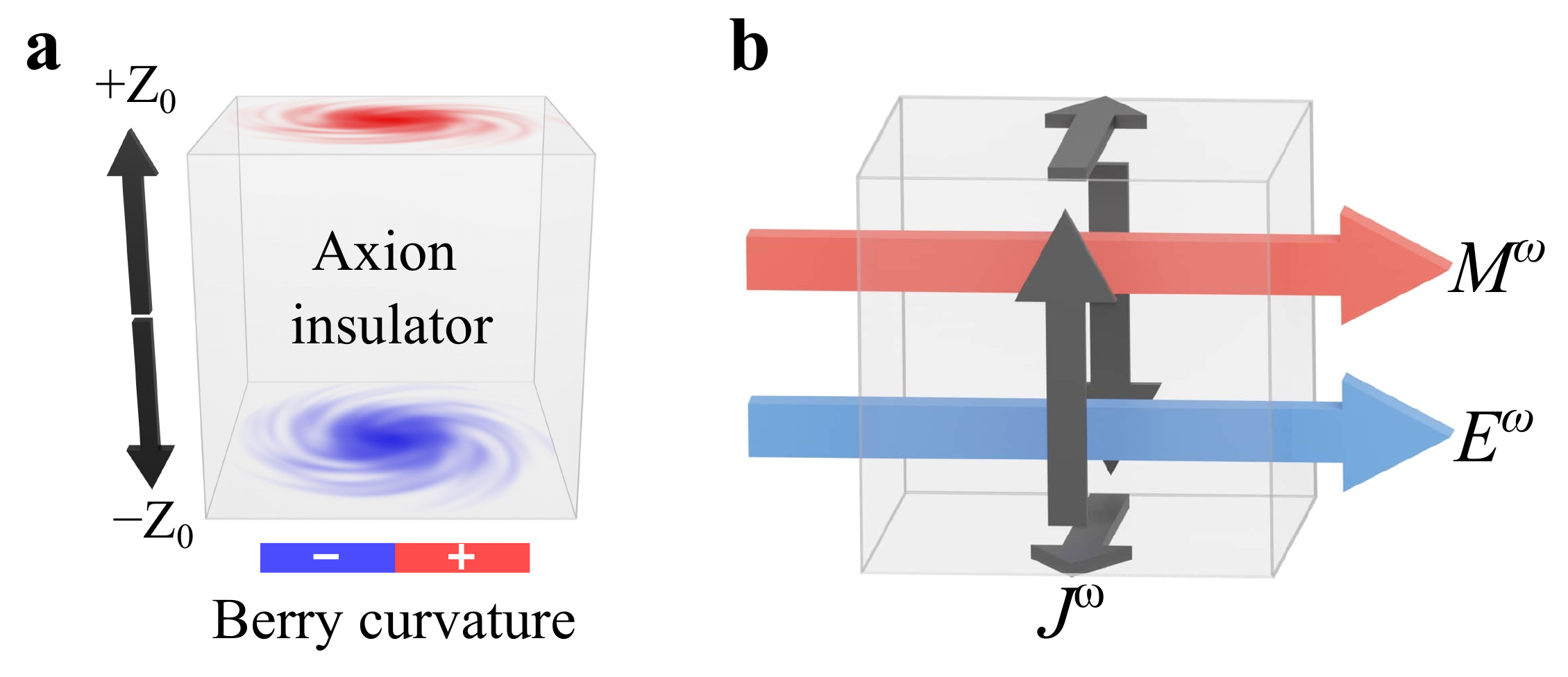}
\caption{\textbf{The ME effect can be visualized by the Berry curvature real space dipole.}}
\label{BCME}
\end{figure*}

Next, we show that the ME effect can be visualized by the Berry curvature real space dipole. Figure~\ref{BCME}\textbf{a} shows the Berry curvature real space dipole. It is known that Berry curvature leads to an anomalous velocity, i.e., a deflection of electron trajectory. Because the Berry curvature at $\pm z$ is opposite, upon applying in-plane $E$ field, electrons at $\pm z$ deflect toward opposite directions. As shown in Figure~\ref{BCME}\textbf{b}, this can be visualized as an itinerant circulation, which in turn leads to a magnetization $M$. Therefore,  because of the Berry curvature real space dipole, $E$ can generate $M$, i.e., an ME effect. %If $E$ is a static $E$ field, then Berry curvature leads to a static ME effect. If $E$ is the oscillating $E^{\omega}$ of light, then Berry curvature leads to an optical ME effect. 

%As a side note, the explanation above mainly focuses on Berry curvature at $\pm z$ (e.g., the top and bottom surfaces). One might wonder what happens to the side surfaces. Ideally, the side surfaces should also develop Berry curvature. This requires the magnetic moments to have finite component perpendicular to the side surfaces. For the layered AFM we consider here, this means that the side surfaces should be tilted. In our real experiments, the side surfaces of MnBi$_2$Te$_4$ flakes are rough and uncontrolled. Nevertheless, because the side surfaces are so small in our atomically thin flakes, their contribute to the total current is negligible. As supporting evidence, our transport shows that the 6SL sample is highly insulating (Fig.~\ref{Transport_data_dev87}) at charge neutrality. This would be impossible if the side surfaces were highly conductive. 
  \vspace{3.ex}
\textbf{3. Mathematical derivations}

As explained above, qualitatively, the ME effect is a natural consequence of the Berry curvature real space dipole. We further establish the mathematical relationship between the ME coefficient $\alpha_{xx}^{\mathrm{Berry}}(\omega)$ and the Berry curvature real space dipole. As shown in Eq.~\ref{eq:Dipole}, we have
\begin{align}
\alpha_{xx}^{\mathrm{Berry}}(\omega) &=\frac{e^2 }{ \hbar L} \sum_{\mathrm{o, u}} \int d^2\textbf{k} \frac{ \varepsilon_{\textrm{uo}}(\textbf{k})}{\varepsilon_{\textrm{uo}}(\textbf{k})-\hbar\omega} \ {\rm Im}[\frac{\hbar^2\braket{\textrm{o}(\textbf{k})| \hat{v}^x|\textrm{u}(\textbf{k})}\braket{\textrm{u}(\textbf{k})|\hat{v}^y\hat{r}^z|\textrm{o}(\textbf{k})}}{\varepsilon_{\textrm{uo}}^2(\textbf{k})}]\nonumber\\
&\simeq\frac{e^2}{\hbar L}\sum_{\textrm{o,u}}\int d^2\mathbf{k} \ \frac{\varepsilon_{\textrm{uo}}(\mathbf{k})}{\varepsilon_{\textrm{uo}}(\mathbf{k})-\hbar\omega}\  \langle\hat{r}^z\rangle \textrm{Im}[\frac{\hbar^2\langle\textrm{o}(\mathbf{k})|\hat{v}^x|\textrm{u}(\mathbf{k})\rangle\langle\textrm{u}(\mathbf{k})|\hat{v}^y|\textrm{o}(\mathbf{k})\rangle}{\varepsilon_{\textrm{uo}}^2(\mathbf{k})}]\nonumber\\
&=\frac{e^2}{2\hbar L}\sum_{\textrm{o,u}}\int d^2\mathbf{k} \ \frac{\varepsilon_{\textrm{uo}}(\mathbf{k})}{\varepsilon_{\textrm{uo}}(\mathbf{k})-\hbar\omega}\  \langle\hat{r}^z\rangle \Omega_{\textrm{uo}},
\label{eq:Dipole}
\end{align}
where $\Omega_{\textrm{uo}}=2\textrm{Im}[\frac{\hbar^2\langle\textrm{o}(\mathbf{k})|\hat{v}^x|\textrm{u}(\mathbf{k})\rangle\langle\textrm{u}(\mathbf{k})|\hat{v}^y|\textrm{o}(\mathbf{k})\rangle}{\varepsilon_{\textrm{uo}}^2(\mathbf{k})}]$ is the inter-band Berry curvature. The key approximation in Eq.~\ref{eq:Dipole} is that $\hat{r}^z$ is moved out of the inner product. We provide additional discussions/clarifications about this point. This is a good approximation when the wavefunction of the electronic states is concentrated at a particular height $z$ (i.e., in a particular layer). To see that, we insert a complete Hilbert space into the inner product as shown in Eq.~\ref{eq:Appro}. When wavefunction of the electronic states is concentrated in a particular layer, then we have $\braket{\textrm{p}(\textbf{k})|\hat{r}^z|\textrm{o}(\textbf{k})}\simeq\delta_{\textrm{po}}\langle\hat{r}^z\rangle$. As a result, we can move $\hat{r}^z$ out of the inner product and Eq.~\ref{eq:Appro} becomes the Berry curvature real space dipole. In the limit of decoupled layer systems, then the wavefunctions are indeed purely localized at each layers. In the presence of interlayer coupling, the wavefunction will have finite weight on multiple layers. Nevertheless, for van der Waals materials, the interlayer coupling is expected to be relatively weak. Hence, the wavefunction of the electronics states is relatively localized.

%When $|\textrm{o}(\textbf{k})\rangle$ is an eigenstate of $\hat{r}^z$, then $\braket{\textrm{p}(\textbf{k})|\hat{r}^z|\textrm{o}(\textbf{k})}=\delta_{po}\langle\hat{r}^z\rangle$, where $\langle\hat{r}^z\rangle$ is the $\hat{r}^z$ eigenvalue of $|\textrm{o}(\textbf{k})\rangle$. In this case, the approximation becomes strictly true. 

\begin{align}
\alpha_{xx}^{\mathrm{Berry}}(\omega) &=\frac{e^2 }{ \hbar L} \sum_{\mathrm{o, u}} \int d^2\textbf{k} \frac{ \varepsilon_{\textrm{uo}}(\textbf{k})}{\varepsilon_{\textrm{uo}}(\textbf{k})-\hbar\omega} \ {\rm Im}[\frac{\hbar^2\braket{\textrm{o}(\textbf{k})| \hat{v}^x|\textrm{u}(\textbf{k})}\braket{\textrm{u}(\textbf{k})|\hat{v}^y\hat{r}^z|\textrm{o}(\textbf{k})}}{\varepsilon_{\textrm{uo}}^2(\textbf{k})}]\nonumber\\
&=\frac{e^2 }{ \hbar L} \sum_{\mathrm{o, u}} \int d^2\textbf{k} \frac{ \varepsilon_{\textrm{uo}}(\textbf{k})}{\varepsilon_{\textrm{uo}}(\textbf{k})-\hbar\omega} \ {\rm Im}[\frac{\hbar^2\braket{\textrm{o}(\textbf{k})| \hat{v}^x|\textrm{u}(\textbf{k})}\sum_{\mathrm{p}}\braket{\textrm{u}(\textbf{k})|\hat{v}^y|\textrm{p}(\textbf{k})}\braket{\textrm{p}(\textbf{k})|\hat{r}^z|\textrm{o}(\textbf{k})}}{\varepsilon_{\textrm{uo}}^2(\textbf{k})}]\nonumber\\
&\simeq\frac{e^2 }{ \hbar L} \sum_{\mathrm{o, u}} \int d^2\textbf{k} \frac{ \varepsilon_{\textrm{uo}}(\textbf{k})}{\varepsilon_{\textrm{uo}}(\textbf{k})-\hbar\omega} \ \langle\hat{r}^z\rangle{\rm Im}[\frac{\hbar^2\braket{\textrm{o}(\textbf{k})| \hat{v}^x|\textrm{u}(\textbf{k})}\sum_{\mathrm{p}}\braket{\textrm{u}(\textbf{k})|\hat{v}^y|\textrm{p}(\textbf{k})}\delta_{\textrm{po}}}{\varepsilon_{\textrm{uo}}^2(\textbf{k})}]\nonumber\\
&=\frac{e^2 }{ \hbar L} \sum_{\mathrm{o, u}} \int d^2\textbf{k} \frac{ \varepsilon_{\textrm{uo}}(\textbf{k})}{\varepsilon_{\textrm{uo}}(\textbf{k})-\hbar\omega} \ \langle\hat{r}^z\rangle{\rm Im}[\frac{\hbar^2\braket{\textrm{o}(\textbf{k})| \hat{v}^x|\textrm{u}(\textbf{k})}\braket{\textrm{u}(\textbf{k})|\hat{v}^y|\textrm{o}(\textbf{k})}}{\varepsilon_{\textrm{uo}}^2(\textbf{k})}]\nonumber\\
&=\frac{e^2}{2\hbar L}\sum_{\textrm{o,u}}\int d^2\mathbf{k} \ \frac{\varepsilon_{\textrm{uo}}(\mathbf{k})}{\varepsilon_{\textrm{uo}}(\mathbf{k})-\hbar\omega}\  \langle\hat{r}^z\rangle \Omega_{\textrm{uo}}
\label{eq:Appro}
\end{align}

\subsection*{V.2. Additional discussion about optical control of AFM}

In this section, we briefly discuss previous pioneering works on optical control of AFM \cite{manz2016reversible, kharchenko2005odd,liou2019deterministic,kavspar2021quenching,kimel2009inertia,reichlova2019imaging,ostler2012ultrafast,higuchi2016control,dannegger2021ultrafast}, which are of great interest especially given the difficulty of this task. This discussion also justifies that helicity-dependent optical control of fully-compensated AFM order has not been achieved before. Especially, we discuss three representative works:

1. \textbf{Optical control through electric polarization $P$ in TbMnO$_3$:} In the multiferroic TbMnO$_3$, the AFM order $L$ is connected to $P$. Therefore, by reversing $P$, one can reverse $L$. In Ref. \cite{manz2016reversible}, a single $P$ domain was first prepared by electric-field in a TbMnO$_3$ sample; a small area was optically heated above $T_{\textrm{c}}$; when the laser was turned off, the small area went to opposite $P$ (because of the stray electric field exerted by the surrounding single $P$ domain \cite{hadni1973localized}) and therefore the opposite $L$. In this way, AFM order is controlled by controlling $P$. The optical induced thermal effect is the driving force for the order parameter reversal.

2. \textbf{Optical control by $B$ field and linear dichroism in MnF$_2$:} Upon applying $B$ field,  MnF$_2$ shows a linear dichroism (LD) \cite{kharchenko2005odd} (different absorption when light polarization is along $[110]$ and $[\bar{1}10]$). Moreover, the sign of LD is opposite for the opposite AFM states. Therefore, by applying $B=0.5$ T, Ref. \cite{higuchi2016control} achieved control of AFM order using linearly polarized light.

3. \textbf{Optical control of ferrimagnet:} Ferrimagnet has $M$ and $L$ that are connected. $M$ can directly couple to circularly-polarized light like in a ferromagnet. Therefore, by reversing $M$, one also reverses $L$, as demonstrated by Ref. \cite{ostler2012ultrafast}. 
 
\subsection*{V.3. Additional discussion about previous theoretical works on MnBi$_2$Te$_4$}
Previous theoretical works have comprehensively studied the electronic, magnetic and topological properties of MnBi$_2$Te$_4$ bulk and thin films. Here we have made a table to summarize their important pioneering findings:
\begin{center}
\centering
\begin{tabular}{ | m{3.5cm} | m{3.5cm}| m{3.5cm} | m{4cm}| } 
  \hline
    & Magnetization & Topological phase & Topological phase at high $B$ field  \\ 
  \hline
  MnBi$_2$Te$_4$ monolayer & $M\neq0$ & Trivial insulator\cite{Otrokov2019unique} & Trivial insulator\cite{Otrokov2019unique} \\ 
  \hline
  MnBi$_2$Te$_4$ even layer  & $M=0$ & Axion insulator\cite{Otrokov2019unique,Li2019a,Liu2020} & Chern insulator\cite{Otrokov2019unique} \\ 
  \hline
  MnBi$_2$Te$_4$ odd layer  & $M\neq0$ & QAH insulator\cite{Otrokov2019unique,Li2019a}  & Chern insulator\cite{Otrokov2019unique}   \\ 
  \hline
  MnBi$_2$Te$_4$ bulk* & $M=0$ & AFM TI \cite{Otrokov2019a, Li2019a, Zhang2019a} & Weyl semi-metal\cite{Li2019a, Zhang2019a}\\ 
  \hline
\end{tabular}
\end{center}
*In addition, the ‘M$ \mathrm{\ddot o}$bius insulator’ phase is also reported in the canted AFM phase of MnBi$_2$Te$_4$ bulk\cite{Zhang2020}.
\color{black}

\subsection*{V.4. Additional general discussion about thickness dependence}

The Axion CD, induction and the underlying optical Axion ME coupling can persist in the thick sample limit (which is indeed what we observed in Fig.~\ref{RCD_thickness}). Below, we explain how we can understand the physics in the thick limit, focusing on two aspects: (1) symmetry constraints and (2) finite optical penetration depth. Because the Axion CD and induction arises from the Axion optical ME coupling (Fig. R13). We will focus on understanding the Axion optical ME coupling in thick samples.

 (1) Symmetry: Bulk MnBi$_2$Te$_4$ restores the space inversion symmetry $\mathcal{P}$ and the time-reversal symmetry $\mathcal{T}$ (precisely $\mathcal{T}$ multiples a fractional lattice translation). This leads to a problem: symmetry dictates the ME coupling to vanish under $\mathcal{P}$ or $\mathcal{T}$, so it seems that bulk should not support an Axion ME coupling. The key is to differentiate a periodic sample with no surface and a finite thick sample.  
 
\begin{figure*}[h]
\centering
\includegraphics[width=14cm]{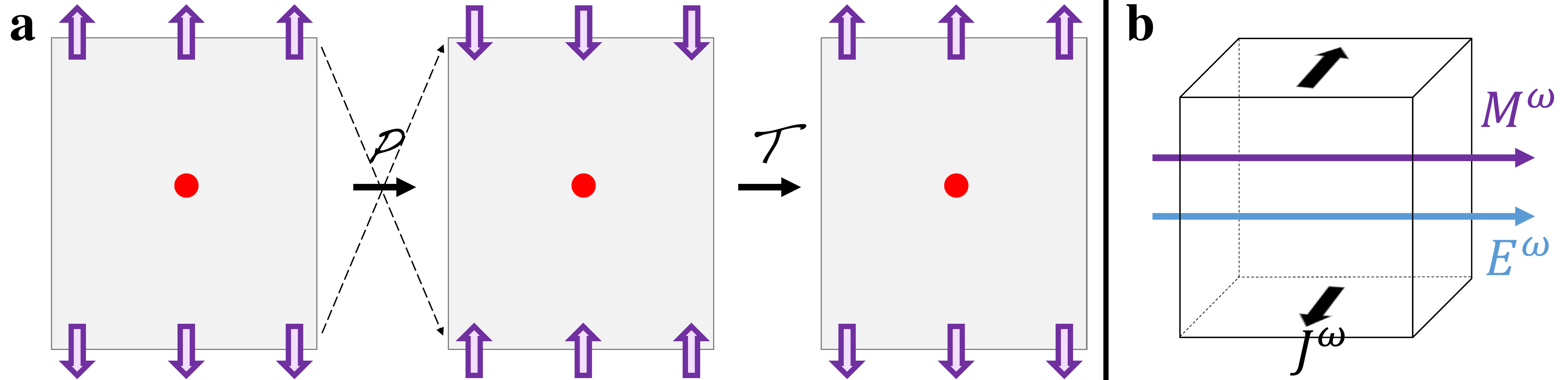}
\vspace{-1mm}
\caption{\textbf{a,} A finite even-layered thick sample (no matter how thick) breaks $\mathcal{P}$ and $\mathcal{T}$ macroscopically due to the opposite magnetization at the surface. The red circle denote the macroscopic inversion center. \textbf{b,} The Axion ME coupling manifests as surface itinerant current $J^\omega \perp E^\omega$. }
\label{Bulk_surface_Axion}
\end{figure*}

 \textbf{For a finite thick sample:} $\mathcal{P}$ and $\mathcal{T}$ symmetries are still broken in the macroscopic sense. As shown in Fig.~\ref{Bulk_surface_Axion}\textbf{a}, the interior of a thick sample respects $\mathcal{P}$ and $\mathcal{T}$ . But the surfaces strongly break $\mathcal{P}$ and $\mathcal{T}$ . This is crucial because the Axion ME coupling manifests as a surface response (surface itinerant current perpendicular to the applied $E$ field). Therefore, even though the interior respects $\mathcal{P}$ and $\mathcal{T}$, a thick sample still hosts Axion ME coupling. 
 
 \textbf{For a periodic system:} both $\mathcal{P}$ and $\mathcal{T}$ symmetries are preserved. It has no surfaces, therefore the Axion ME coupling cannot be defined. 
 
 \vspace{4mm}
(2) Light penetration depth: We now try to understand how the light penetration depth influences the Axion optical ME coupling (and hence the Axion CD). The Axion optical ME coupling can be visualized by the surface Hall currents: As shown in Fig.~\ref{Bulk_surface_Axion}\textbf{b}, light electric field $E^\omega$ leads to a Hall current $J^\omega$ at the surface. Because the top and bottom surfaces have opposite Berry curvature, they have opposite $J^\omega$ Fig.~\ref{Bulk_surface_Axion}\textbf{b}). Globally, we can visualize the  $J^\omega$ at the surfaces as a circulating current, which naturally leads to an  $M^\omega$.
 
 In a thin flake, light penetrates the sample, so the top and bottom surfaces have similarly large $E^\omega$ and hence similarly large $J^\omega$, which leads to an $M^\omega$ as explained above.  
 
In a thick sample, light strongly decays. So only the top surface has a large $E^\omega$ and hence a large $J^\omega$. Nevertheless, this still leads to a nonzero $M^\omega$.

\clearpage
\vspace{0.5cm}

\end{document}